\documentclass[iop,revtex4]{emulateapj}
\usepackage{epstopdf}
\usepackage{graphicx}
\usepackage{epsfig}
\usepackage{amsmath}
\usepackage{wasysym}
\usepackage{txfonts}
\pdfoutput=1
\usepackage[pdftex]{color,xcolor}
\usepackage[backref]{hyperref}
\usepackage{hyperref}
\hypersetup{pdfauthor=Paul Molliere}
\hypersetup{backref=true, pagebackref=true, hyperindex=true, breaklinks=true,colorlinks=true,urlcolor=blue, linkcolor=blue,  citecolor=blue,pagecolor=red, bookmarks=true, bookmarksopen=true}
\usepackage{epstopdf}
\hypersetup{backref=true}
\usepackage{lipsum}
\usepackage{natbib}
\usepackage{mathtools}

\bibpunct{(}{)}{;}{a}{}{,}

\def\rsun{{\rm R}_\odot}

\def\f1{f_{\rm I}}
\def\mj{{\rm M}_{\textrm{\tiny \jupiter }}}

\newcommand{\rj}{{\rm R}_{\textrm{\tiny \jupiter}}}

\def\beq{\begin{equation}}
\def\eeq{\end{equation}}

\def\t2{\tau_{\rm II}}

\def\sigmas0{\Sigma_{\rm s,0}}

\newcommand{\rch}[1]{{#1}}

\def\apjl{ApJ}                % Astrophysical Journal, Letters
\def\apjs{ApJS}               % Astrophysical Journal, Supplement
\def\ao{Appl.Optics}          % Applied Optics
\def\aap{A\&A}                % Astronomy and Astrophysics
\def\({\left(}
\def\){\right)}
\def\<{\left<}
\def\>{\right>}

\shorttitle{Irradiated exoplanets: The influence of stellar parameters, metallicity, and the C/O ratio}
\shortauthors{P. Molli\`ere et al.}

\begin{document}

\title{Model atmospheres of irradiated exoplanets: \\
The influence of stellar parameters, metallicity, and the C/O ratio}

\author{P. Molli\`{e}re\altaffilmark{1},  R. van Boekel\altaffilmark{1}, C. Dullemond\altaffilmark{3}, Th. Henning\altaffilmark{1}, C. Mordasini\altaffilmark{2}}

\altaffiltext{1}{Max-Planck-Institut f\"ur Astronomie, K\"onigstuhl 17, D-69117 Heidelberg, Germany}
\altaffiltext{2}{Physikalisches Institut, Universit\"at Bern, Sidlerstrasse 5, CH-3012 Bern, Switzerland}
\altaffiltext{3}{Institut f\"ur Theoretische Astrophysik, Universit\"at Heidelberg, Albert-Ueberle-Stra\ss e 2, D-69120, Germany}
\email{molliere@mpia.de}
\begin{abstract}
Many parameters constraining the spectral appearance of exoplanets are still poorly understood. We therefore study the properties of irradiated exoplanet atmospheres over a wide parameter range including metallicity, C/O ratio and host spectral type. We calculate a grid of 1-d radiative-convective atmospheres and emission spectra. We perform the calculations with our new \textbf{P}r\textbf{e}ssure-\textbf{T}emperature \textbf{It}erator and Spectral Emission Calculator for Planetary Atmospheres (\emph{PETIT}) code\rch{, assuming chemical equilibrium}. The atmospheric structures and spectra are made available online. We find that atmospheres of planets with C/O ratios $\sim$ 1 and $T_{\rm eff}$ $\gtrsim$ 1500 K can exhibit inversions due to heating by the alkalis because the main coolants CH$_4$, H$_2$O and HCN are depleted. \rch{Therefore, temperature inversions possibly occur without the presence of additional absorbers like TiO and VO.
At low temperatures we find that the pressure level of the photosphere strongly influences whether the atmospheric opacity is dominated by either water (for low C/O) or methane (for high C/O), or both (regardless of the C/O). For hot, carbon-rich objects this pressure level governs whether the atmosphere is dominated by methane or HCN. Further we find that host stars of late spectral type lead to planetary atmospheres which have shallower, more isothermal temperature profiles. In agreement with prior work we find that} for planets with $T_{\rm eff}$ $<$ 1750 K the transition between water or methane dominated spectra occurs at C/O $\sim$ 0.7, instead of $\sim$ 1, because condensation preferentially removes oxygen.
\end{abstract}

\section{Introduction}
\label{sect:intro}
In a number of existing studies the range of possible C/O ratios in protoplanetary disks and
the resulting implications for the C/O ratios in the gaseous envelopes of extrasolar planets is investigated \citep[see, e.g.,][]{oeberg2011,ali-dib2014,helling2014,marboeuf2014,thiabaud2014}.
These kind of studies are interesting, as they may help to predict the spectral appearance of atmospheres of planets formed via different pathways in the circumstellar disks.
In a further example, \citet{madhusudhan2014} studies the range of possible C/O ratios for 2 different disk models, depending on the formation and migration
mechanism invoked to form hot jupiters.
The result of these studies is that large planetary C/O ratios, close to unity,
are possible even when considering disks of solar composition \citep[the solar value is C/O$_\odot$ $\sim$ 0.55, see][]{asplund2009}.
For disks with supersolar C/O ratios the planetary C/O ratios should be even higher, although stars with C/O ratios close to and bigger
than 1 may be quite rare \citep{fortney2012}.
The C/O ratio is particularly interesting for the spectral appearance of exoplanets because for high enough temperatures
($T$ $\gtrsim$ 1000 K) a C/O $<1$ giant planet will have appreciable amounts of H$_2$O in its atmosphere
and almost no CH$_4$, whereas for C/O $>1$ the situation is the opposite and CH$_4$ is much more abundant than H$_2$O.
This transition happens quite sharply \citep[see, e.g.,][]{kopparapu2012,madhusudhan2012}.
Furthermore, condensation processes can potentially lead to local C/O ratios of $\sim$ 1-2 in the gas phase,
even if the global atmospheric C/O ratio is smaller than 1 \citep[see][]{helling2014}.
The reason this is for the locking up of oxygen in silicates, as has already been suggested by \citet{fortney2006}.

Both H$_2$O and CH$_4$ have strong absorption features and their main absorption bands between $\sim$ 1.3 and 5 $\mu$m 
are alternately located in wavelength space. Thus hot gaseous planets with C/O $<$ 1 and C/O $\gtrsim$ 1 in the spectrally 
active regions should be quite easily distinguishable and might give hints on the planet's formation history such as the location 
of formation in the protoplanetary disk and its migration through it \citep{madhusudhan2014}.
For even higher temperatures ($T$ $\gtrsim$ 1750 K), and C/O $>$ 1, HCN takes over as the most important carbon-carrying 
infrared absorber as it becomes more abundant than CH$_4$ in the spectrally active parts of the atmospheres 
\citep[see, e.g.,][]{moses2013}. ``Spectrally active'' denotes the regions where the radiation seen in the planet's emergent spectrum
originates. The respective atmospheres are then not dominated by CH$_4$ anymore, but by HCN.
Distinguishing H$_2$O and HCN absorption features should be possible, due to the different spectral
signatures of HCN and H$_2$O in the NIR and IR. Therefore a distinction between O and C dominated atmospheres
is possible also at high temperatures.

Motivated by the fundamentally different spectral appearances of the two C/O cases, \citet{madhusudhan2012} proposed a 2-d 
classification scheme for characterizing giant extrasolar planets, using the C/O ratio and the incident stellar flux as dimensions.
In his work, the importance of CH$_4$ for the C/O $>$ 1 cases is most strongly emphasized,
but the possible importance of HCN and C$_2$H$_2$ is mentioned as well. 
A 1-d  classification scheme for hot giant planets, based only on the stellar flux, had already been proposed by \citet{fortney2008} 
before, featuring ``cold'' hot jupiters without a temperature inversion and ``hot'' hot jupiters with a temperature inversion caused by TiO 
and VO absorption.
\begin{table*}[t]
\centering
\begin{tabular}{c|cccc}
Opacity source & Spectral range [$\mu$m] & Line list & Partition function & Pressure broadening \\ \hline \hline
CH$_4$ & 0.83 $< \lambda$ & \citet{yurchenko2014} & (a) & (a) \\
CH$_4$ & 0.86 $< \lambda$ & \citet{rothman2013} & \citet{fischer2003} & $\gamma_{\rm air}$, \citet{rothman2013} \\
C$_2$H$_2$ &  1 $< \lambda < $ 16.5 & \citet{rothman2013} & \citet{fischer2003} & $\gamma_{\rm air}$, \citet{rothman2013} \\
CO  &  1.18 $< \lambda$ & \citet{rothman2010} & \citet{fischer2003} & $\gamma_{\rm air}$, \citet{rothman2010}\\
CO & 0.112 $< \lambda < $ 0.43 & \citet{kurucz1993} & \citet{fischer2003} & Eq. (15), \citet{sharp_burrows2007} \\
CO$_2$ & 1 $< \lambda < $ 38.76 & \citet{rothman2010} & \citet{fischer2003} & $\gamma_{\rm air}$, \citet{rothman2010} \\
H$_2$S & 0.88 $< \lambda$ & \citet{rothman2013} & \citet{fischer2003} & $\gamma_{\rm air}$, \citet{rothman2013} \\
H$_2$ & 0.28 $< \lambda$ & \citet{rothman2013} & \citet{fischer2003} & $\gamma_{\rm air}$, \citet{rothman2013} \\
H$_2$ & 0.08 $< \lambda < $ 0.18 & \citet{kurucz1993} & \citet{fischer2003} & Eq. (15), \citet{sharp_burrows2007} \\
HCN & 2.92 $< \lambda$ & \citet{harris2006}, & \citet{fischer2003} & Eq. (15), \citet{sharp_burrows2007}  \\
& & \citet{barber2014} & & \\
H$_2$O & 0.33 $< \lambda$ & \citet{rothman2010} & \citet{fischer2003} & $\gamma_{\rm air}$, \citet{rothman2010} \\
K & 0.05 $< \lambda$ & \citet{piskunov1995} & \citet{sauval_tatum1984} & N. Allard, \citet{schweitzer1996} \\
Na & 0.1 $< \lambda$ & \citet{piskunov1995} & \citet{sauval_tatum1984} & N. Allard, \citet{schweitzer1996} \\
NH$_3$ & 1.43 $< \lambda$ & \citet{rothman2013} & \citet{fischer2003} & $\gamma_{\rm air}$, \citet{rothman2013} \\
OH & 0.52 $< \lambda$ & \citet{rothman2010} & \citet{fischer2003} & $\gamma_{\rm air}$, \citet{rothman2010} \\
PH$_3$ & 2.78 $< \lambda$ & \citet{rothman2013} & \citet{fischer2003} & $\gamma_{\rm air}$, \citet{rothman2013} \\ \hline
\end{tabular}
\caption{References for the atomic and molecular opacities used in the \emph{PETIT} code. (a): We use precalculated
cross-sections from the ExoMol website. For these only Doppler broadening has been taken into account.}
\label{tab:opa_sources}
\end{table*}
However, some ``hot'' hot jupiters are not thought to have a inversion, contradicting the 1-d classification system.
\citet{madhusudhan2012} argued that this could possibly be explained using the 2-d classification scheme, as TiO and VO 
should not be very abundant in planets with a high C/O ratio. In addition, there could be further reasons why TiO and VO should 
not be in the upper part of the atmosphere, e.g. due to settling and inefficient vertical mixing,
cold-trap depletion or photodissociation \citep{spiegel2009,showman2009,parmentier2013,knutson2010}.

Observational evidence for planets with C/O $\gtrsim$ 1 is scarce and the most prominent case,
WASP-12b \citep{madhusudhan2011}, is controversial \citep{crossfield2012,swain2013,stevenson2014}.
Current analyses of the photometric data indicate a C/O ratio $<$ 1:
\citet{line2014} estimated C/O ratios for 9 hot jupiters and found that while in 7 out of 9 cases
(HD 209458b, GJ436b, HD 149026b, WASP-12b, WASP-19b, WASP-43b, TrES-2b) a C/O value of 1 was within
their 1 $\sigma$ confidence interval, in 6 out of 9 cases the solar value was within the 1 $\sigma$ interval as well.
\citet{benneke2015} analyzed 8 hot jupiters (HD 209458b, WASP-19b, WASP-12b, HAT-P-1b, XO-1b, HD189733b,
WASP-17b and WASP-43b) using a self-consistent retrieval analysis, ruling out C/O $>$ 1 for all of them.
Clearly the quality and quantity of the photometric and spectroscopic observations needs to improve before more
conclusive results can be obtained for many of these planets \citep{line2013}.

A further example for a planet with a C/O ratio close to 1 is  HR 8799b,
for which C/O = $0.96\pm 0.01$ or $0.97^{+0.00}_{-0.01}$ has been estimated \citep{lee2013}, depending
on whether clouds are included in the model or not.

Although all the C/O ratios obtained by the above studies are depending on the assumptions made in the various
retrieval models, the current analysis of data does not indicate any planet with C/O $>$ 1.
Further, while the current quality of data is still too low for obtaining reliable retrieval results in many cases,
upcoming observing facilities such as JWST should greatly help to decipher the composition of hot jupiters.

In conclusion, the C/O ratio, together with the effective temperature, should be a key parameter constraining a hot jupiter's spectral
appearance and thus we want to study how the interplay between the C/O ratio and other parameters affect the atmospheres.
Systematic studies of exoplanet atmospheres have been published in the literature before \citep[see, e.g.,][]{sudarsky2003},
and although the C/O has been suggested to be of importance already a decade ago \citep{seager2005}, no systematic study
of the atmospheric characteristics as a function of the C/O ratio has been carried out so far.
Therefore, we publish a grid of emission spectra and pressure temperature ($PT$) structures for self-consistent hot jupiters atmospheres
for varying C/O ratio, [Fe/H], distance to the star, stellar host spectral type and planetary ${\rm log}(g)$.

The results were calculated with our new \textbf{P}r\textbf{e}ssure-\textbf{T}emperature \textbf{It}erator and Spectral Emission
Calculator for Planetary Atmospheres (\emph{PETIT}) code. \emph{PETIT} solves the 1-d plane parallel
structure of the atmosphere assuming local thermal equilibrium (LTE), radiative equilibrium or convection
and equilibrium chemistry.
Our goal is to investigate the behavior of planetary atmospheres in the parameter space
covered by our grid.
Furthermore we make the atmospheric $PT$-structures, abundance profiles, and resulting spectra publicly available for use in, e.g.,
the evaluation of observational data of planetary emission spectra.

In Section \ref{sect:code_struc} we introduce our code and explain its individual modules. We also show some of the tests
we carried out to check the results of our code for consistency. In Section \ref{sect:grid_valid} we discuss how the assumptions
in our code constrain the parameter range of the atmospheric grid. In Section \ref{sect:grid_setup_calc} we report on how the grid was set up
and how the calculations were carried out. The results can be found in Section \ref{sect:results},
the discussion and conclusions are in Section \ref{sect:conclusion}.

\section{Description of the code}
\label{sect:code_struc}
\subsection{Opacity database}
The current version of our opacity database comprises atomic and molecular line and continuum opacities from ultra-violet to
infrared wavelengths. So far only absorption processes are treated.
\rch{Scattering of incoming stellar radiation 
by molecules in the planetary atmosphere causes the stellar photons to traverse, on average, a somewhat longer distance through the atmosphere before reaching a certain pressure level. Hence, the photons will on average be absorbed at slightly lower pressures (higher altitudes) than if absorption only is considered. Because the reported optical albedos of hot jupiters are very low, in the low single digit percentage range, as summarized by \citet{madhusudhan2014b}, absorption appears much more important than scattering in these objects. Therefore, include only absorption in the radiative transfer calculation One exception is mentioned,
however, with Kepler 7-b having a geometric albedo of 0.32 $\pm$ 0.03 \citep{demory2011}.}
\subsubsection{Molecular and atomic line opacities}
\label{sect:mol_atom_line_opa}
A list of all line opacity sources, together with a reference to the corresponding line lists, pressure broadening parameters and partition
functions can be found in Table \ref{tab:opa_sources}. Our method to speed up molecular opacity calculations is explained in Appendix \ref{sect:line_calc}.

All molecular and atomic lines are considered to have a Voigt profile (except for the Na and K doublet)
and no truncation of the lines at large distances from the
line cores has been applied. This choice of the far-wing treatment of the line shape is arbitrary.
It is well known that the molecular lines should become sub-Lorentzian at large distances
from the line core \citep[see e.g.,][and the references therein]{freedman2008}.
However, the position of the cut-off, and the line wing shape itself, depend on
the pressure and temperature and the perturber gases which broaden the molecular and atomic transitions.
The choice to use Voigt profiles and not truncating the lines
is thus only made because of the lack of knowledge regarding the actual line profiles.
\citet{grimm2015} show that the differences when applying
no line cut-off, when compared to an arbitrary cut-off, are at least of the order of 10 \% when
considering layer transmissions. In order to calculate the Voigt profiles we use the code provided by \citet{humlicek1982}.

The calculations are performed on a pressure-temperature grid with 10 grid points in pressure going from 10$^{-6}$ to 10$^3$ bars
(equidistantly spaced in log-space).
\rch{Because the line wing strength due to pressure broadening is well
behaved (the strength is simply linear in $P$), we found this grid spacing to be sufficient when interpolating to the actual
pressures of interest.}
The temperature grid consists of 10 points going from 200 to 3000 K, equidistantly
spaced in log-space as well. Opacities with temperatures $\le$ 270 K are only calculated up to 1 bar, temperatures up to 670 K only
up to 10 bar,  and temperatures up to 900 K only up to 100 bar.
\rch{This choice was made because it was found, using the simple \citet{guillot2010} atmospheric model, that even cold planets such as Jupiter
and Uranus should not be cooler than 270, 670 and 900 K at the pressures cited above. As we concentrate on hot jupiters in this paper
we therefore did not extend the grid to cool temperatures at high pressures. We plan to extend the grid in the future, however.}
In total the above considerations yield 87 pressure--temperature grid-points.

Our fiducial wavelength range goes from 110 nm to 250$\mu$m. We calculate the opacities in this range on a grid with a spacing
of $\lambda/\Delta \lambda=10^6$. \rch{This resolution is sufficient to resolve the line cores at all pressures and temperatures.}
From these calculations we construct opacity distribution tables (k-tables) for later use (see Section \ref{sect:rad_trans_mod}).
These tables are then interpolated to the pressure-temperature values of interest.

Most of the line lists are obtained from the \emph{HITRAN}/\emph{HITEMP} databases \citep{rothman2013,rothman2010},
together with additional data from the VALD, Kurucz and ExoMol line lists \citep{piskunov1995,kurucz1993,harris2006,barber2014}.
For methane we use \emph{HITRAN} for temperatures below 300 K. For temperatures above 300 K the ExoMol cross-sections
are used \citep{yurchenko2014}, as this line list is much more complete at higher temperatures.
The ExoMol cross-sections
(in units of cm$^2$ molecule$^{-1}$) can be obtained, tabulated as a function of wavelength,  from the ExoMol website.
No pressure broadening has been applied when calculating these cross-sections, as pressure broadening information
is not readily available. However, due to the sheer number of methane lines the cross-sections should be dominated by the Gaussian line cores.
{
In general, for all molecular and atomic line opacities, pressure broadening information is often not available, especially when taking into account
arbitrary mixtures of various molecular and atomic gaseous species. We therefore estimate the pressure broadening by using
the air broadening coefficients $\gamma_{\rm air}$ of the \emph{HITRAN}/\emph{HITEMP} database when these are available for a given molecule of interest.
In cases where this information is missing as well we resort to the use of the pressure broadening approximation provided
by Eq. (15) in \citet{sharp_burrows2007}.

A special line shape treatment is needed when considering the Na (589.16 \& 589.76 nm) and K (766.7 \& 770.11 nm) doublet lines.
Na and K are very important to correctly describe the atmospheric absorption in the optical, as these two species are one of the main
absorbers in this spectral range and their line wings act as a pseudo-continuum contribution to the total opacity \citep[see, e.g.,][]{sharp_burrows2007,freedman2008}.
Different groups have tried to estimate the line shapes for Na and K taking into account collisions with H$_2$ and He \citep{burrows2003,allard2003,zhu2006},
and the efforts are ongoing \citep{allard2012}. In particular \citet{allard2003} showed that for brown dwarfs the use of correct Na and K wing
profiles improves the agreement between synthetic spectra and observations.
The line profiles we use for Na were obtained from Nicole Allard (private communication)
using \citet{rossi1985} pseudo potentials.
For K we use the profiles available on the website of Nicole Allard\footnote{\url{http://mygepi.obspm.fr/~allard/alkalitables.html}},
which include C$_{2{\rm v}}$ and C$_{\infty{\rm v}}$ interaction symmetries.
As H$_2$ should be the main perturber for alkali atoms in the atmosphere of giant planets, only H$_2$-broadening is currently considered.
The other lines of Na and K, which are much weaker than the doublet transitions, are modeled using van der Waals (vdW) broadening as
described in \citet{schweitzer1996}:
\beq
\gamma_{\rm vdW} = \alpha C_0^{2/5}v^{3/5}N_{\rm H_2},
\eeq
where $C_6$ is the van der Waals interaction constant, $v$ is the mean relative velocity between H$_2$ and the alkali atom,
$N_{\rm H_2}$ is the H$_2$ number density and $\alpha$ is a dimensionless number. \citet{schweitzer1996} report that $\alpha=17$, but it is a factor of 10 smaller in \citet{sharp_burrows2007}. We found that if we want to reproduce the vdW line widths given in \citet{allard2007}, then
we need to use the smaller $\alpha$ value. The required ionization energies were taken from the NIST database\footnote{\url{http://www.nist.gov/pml/data/asd.cfm}}.

\begin{figure}[t!]
\centering
\includegraphics[width=0.485\textwidth]{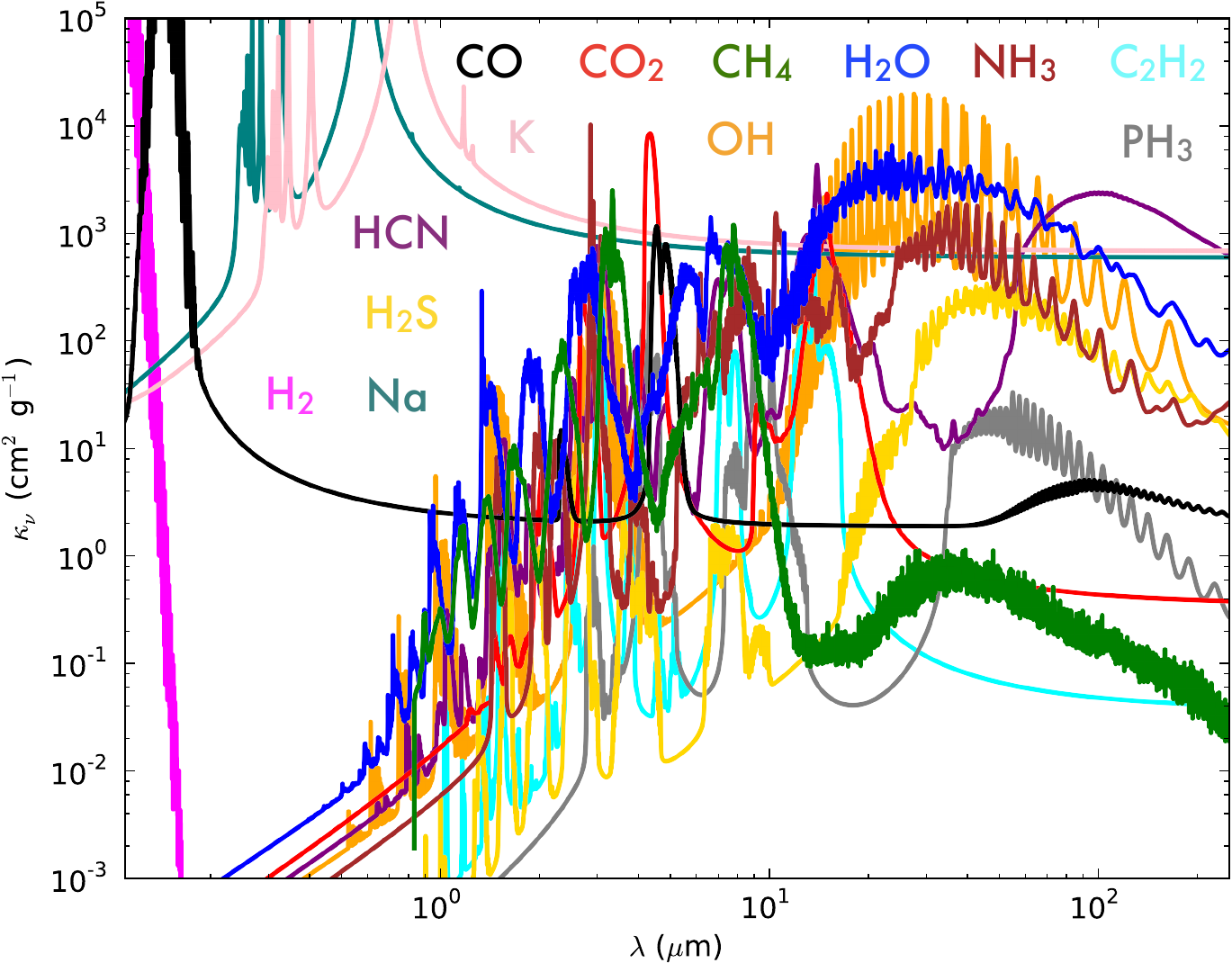}
\caption{Molecular and atomic line opacities of our database at a temperature of 1650 K and a pressure of 100 bar in our
fiducial wavelength range going from 110 nm to 250 $\mu$m. Every color stands for a different species, with the names of the
species indicated in the plot.}
\label{fig:opas}
\end{figure}

Because the non-Lorentzian line profile calculations for the Na and K wings by Nicole Allard are only valid up to a certain H$_2$ density (10$^{20}$ or 10$^{19}$
cm$^{-3}$ for Na or K, respectively) we revert to the use of Voigt profiles for higher densities. This occurs in the range of $\sim$ 3-30 bars in
hot jupiters, where the atmosphere becomes optically thick in the IR and the stellar light has been absorbed.

In Figure \ref{fig:opas} we show all molecular and atomic line opacities of our database at a temperature of 1650 K and a pressure of 100 bar in our
fiducial wavelength space going from 110 nm to 250 $\mu$m. The pressure of 100 bars is far higher than where the radiation in the planetary emission 
spectra usually stems from.
However, as the pressure broadening smoothes out individual lines, the large scale opacity features can more easily be seen at higher
pressures. This figure has been generated from our opacity distribution database and shows
the mean value $\kappa_{\bar{\nu}} = \sum_i \kappa_i \Delta g_i$ of every wavelength subgrid which are spaced at a resolution of $\lambda/\Delta \lambda = 1000$. Here, $g$ is the cumulative opacity distribution function, see Section \ref{sect:rad_trans_mod} for more information.

VO and TiO opacities have not been added yet. We explained in Section \ref{sect:intro} that the role of these two absorbers is quite controversial, as they might not be present in the atmospheres due to a potential rain-out, cold-trap capture or photodissociation.
Nonetheless, we plan to add VO and TiO opacities in the next version of the code.

\subsubsection{Continuum opacities}
As a continuum opacity source we currently consider collision induced absorption (CIA) arising from H$_2$-H$_2$ and H$_2$-He collisions.
Tabulated data and programs from \citet{borysow1988,borysow1989a,borysow1989b,borysow2001,borysow2002} were used to obtain the
cross-sections\footnote{Tables and code were obtained from \url{http://www.astro.ku.dk/~aborysow/programs/index.html}}.

\subsection{Stellar spectra}
For the host star spectral templates we use \emph{PHOENIX} models of main-sequence
stars which have evolved to 1/3 of their main sequence lifetime.\footnote{The results for the stellar spectra depend
only very mildly on this choice, the main effect being that the stars slowly increase their luminosity
with time. Because the transiting hot jupiters that can be best studied orbit K-type stars, which typically
have ages less than half of their main sequence lifetime, we chose a value of 1/3.}
For the stellar evolution
we use Yonsei-Yale tracks \citep{yi2001,kim2002,yi2003,demarque2004}
as well as the evolutionary calculations of
\citet{baraffe1998}. More details can be found in \citet{vanboekel2012}.
\subsection{Code structure and modules}
The basic principle for solving for the atmospheric structure is based on \citet{dullemond2002},
which we adapted to the case of 1-d plane parallel planetary atmospheres.
The code starts with an initial guess for the temperature structure, computes the corresponding
molecular and atomic abundances and the resulting opacities and then calculates the temperature
assuming radiative-convective equilibrium. The code then starts again with the newly found $PT$-
structure until the solution converges.
Because the $PT$-structure, the abundances and opacities mutually depend on each other,
we solve for atmospheric structure in an iterative fashion.

From a given atmospheric temperature structure we obtain the
molecular abundances using the \emph{CEA} equilibrium chemistry code \citep{gordon1994,mcbride1996}.
When the current opacities are calculated, we solve the full angle and frequency dependent radiative
transfer problem of the planetary radiation field.
From this we obtain the intensity-mean, flux-mean and Planck mean opacities as well as the
variable Eddington factors. These opacities and Eddington factors are then used in the
Variable Eddington Factor (VEF) module to find
the temperature structure using the moments of the radiation field \citep[see, e.g.,][]{hubeny2014}.
The temperature is found using a two-stream
approximation for the planetary and stellar radiation field. Furthermore we check if a given atmospheric layer is
convective by applying the Schwarzschild criterion. We switch to an adiabatic temperature gradient in the layers
that are found to be convective.

The iteration is stopped once the maximum
change in temperature between the current iteration and the temperature found 60 iterations ago is smaller than
0.01 K and if the planetary emerging flux obtained from the full angle and frequency dependent radiative transfer solutions is equal to the imposed
total flux with a relative maximum deviation of 0.001.
In rare situations the iteration will slowly oscillate within a a given temperature range and not find a solution
and therefore not converge to a solution with a maximum flux deviation of 0.001.
In this case we flag the files of the atmospheric structures with {\verb|_nconvergence_YYY|} where {\verb|YYY|} is the relative
deviation to the imposed flux in percent.

In sections \ref{sect:rad_trans_mod}, \ref{sect:VEF} and \ref{sect:CEA} we give further detailed information
on the code modules (full radiative transfer module, VEF temperature iteration module and the equilibrium
chemistry module, respectively).
\subsubsection{Radiative transfer module}
\label{sect:rad_trans_mod}
In protoplanetary disks, for which the VEF based approach by \citet{dullemond2002} has been developed,
the main contribution to the total opacity is coming from dust and ice grains \citep[see, e.g.,][]{semenov2003}.
An important feature of dust grain opacities is that they vary only slowly with wavelength, making it possible
to use a small number of wavelength grid points.

In the case of planetary atmospheres molecules contribute strongly to the total opacity.
The molecules which are important for the spectrum can have hundreds of millions to tens of billions of very sharply
peaked lines \citep[see, e.g.,][for H$_2$O and CH$_4$, respectively]{rothman2010,yurchenko2014}.
Therefore, especially at low pressures, radiative transfer calculations need to be carried out at high spectral resolution.
We thus have adopted the opacity distribution method and the correlated-k (c-k) assumption \citep{goody1989,fu_liou1992,lacis_oinas1991} 
to carry out our radiative transfer calculations. Opacity distribution tables (k-tables)
should yield a good description of the detailed high resolution opacities while keeping the numerical costs of
the radiative transfer calculations minimal.

We combine the k-tables of all molecular species contributing to the total opacity using a fast combination method
which has a computational cost linear in the number of species $N_{\rm sp}$, see Section \ref{sect:corr-kapp}.

The calculations to obtain the mean opacities and Eddington factors are carried out on our fiducial grid
(going from 110 nm to 250 $\mu$m) with a grid spacing of $\lambda / \Delta \lambda = 10$, which results
in 78 spectral bins. In order to test the accuracy of these results we have carried out calculations at a
grid spacing of $\lambda / \Delta \lambda = 50$ as well, but the differences in the results are negligible.

To calculate the emission spectrum of an atmosphere after the $PT$-structure has converged we carry
out a c-k radiative transfer calculation at a grid spacing of $\lambda / \Delta \lambda = 1000$, which results
in 7729 spectral bins.

In every spectral bin of the $\lambda / \Delta \lambda$ = 10, 50 and 1000 cases we employ a $g$-grid. $g$ replaces
the spectral coordinate $\lambda$ or $\nu$ in c-k, it is equal to the value of the cumulative opacity distribution function, i.e.
\beq
{\rm d}g = f(\kappa){\rm d}\kappa \ ,
\eeq
where $f(\kappa){\rm d}\kappa$ is the fraction of the opacity values between $\kappa$ and $\kappa+{\rm d}\kappa$ within a
given frequency interval. 
We carry out the radiative transfer calculation using $g$, instead of the frequency or wavelength.
The frequency averaged mean $Q_{\bar{\nu}}$ of any radiative quantity $Q_\nu$ within the spectral bin can then be calculated as
\begin{align}
\nonumber Q_{\bar{\nu}} &= \frac{1}{\Delta \nu}\int_{\nu}^{\nu+\Delta\nu}Q_{\nu'}{\rm d}\nu' \\
&= \int_{0}^{1} Q_g {\rm d} g \ ,
\end{align}
where $Q_g$ is the quantity corresponding to $Q_\nu$ in $g$-space within the spectral bin of interest.

For the $\lambda / \Delta \lambda = 1000$ case we approximate $g$ on a grid of 20 Gaussian quadrature points, consisting of
a 10 point Gaussian quadrature grid going from $g=0$ to $g=0.9$ and a 10 point Gaussian quadrature grid going
from $g=0.9$ to $g=1$.

For the $\lambda / \Delta \lambda$ = 10 and 50 cases we take a finer grid in $g$. The g-grid has 36 points, consisting
of a 6 point Gaussian quadrature grid ranging from $g=0$ to $g=0.95$, an 8 point Gaussian grid ranging from $g=0.95$ to $g=0.99$,
a 20 point Gaussian grid ranging from $g=0.99$ to $g=0.99999$ and a two point trapezoidal quadrature grid ranging from
$g=0.99999$ to 1.

The different methods for obtaining the combined c-k opacity of all species at the resolutions of $\lambda / \Delta \lambda = 1000$, $\lambda / \Delta \lambda$ = 10 and $\lambda / \Delta \lambda$ =  50 are explained in Section \ref{sect:corr-kapp}.

The radiative transfer calculations are made using a 2nd order Feautrier method, considering 20 $\mu$ (i.e. cos\,$\theta$) angles
on a 20-point Gaussian quadrature grid. 
\subsubsection{Variable Eddington factor method}
\label{sect:VEF}
A description of the variable Eddington factor method, and how we use it to find the temperature in the atmosphere,
can be found in Section \ref{sect:VEF-method-app}.

\begin{figure}[t!]
\centering
\includegraphics[width=0.485\textwidth]{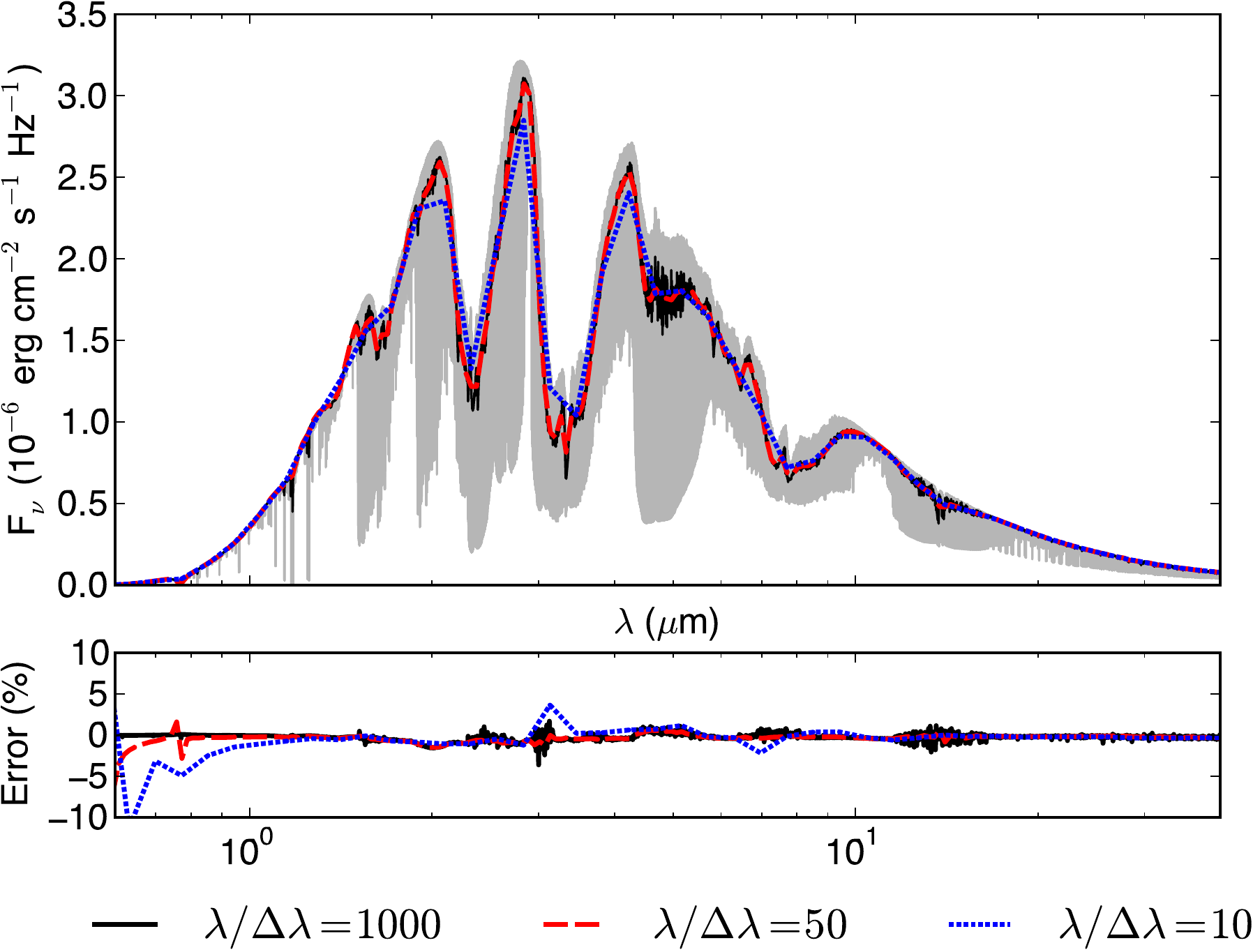}
\caption{{\it Upper panel:} Emission flux of a hot jupiter calculated with the code introduced in this paper.
The gray solid line shows the full line-by-line radiative transfer calculation at a resolution of $\lambda/\Delta\lambda=10^6$.
Overplotted one can see the correlated-k calculations at $\lambda/\Delta\lambda=10^3$ (black dashed line),
$\lambda/\Delta\lambda=50$ (red long dashed line) and at $\lambda/\Delta\lambda=10$ (blue short dashed line).
{\it Lower panel:} Relative error of the $\lambda/\Delta\lambda=10^3, \ 50, \ 10$ calculations when comparing to
the rebinned $\lambda/\Delta\lambda=10^6$ calculation.}
\label{fig:spec_test}
\end{figure}

In our code the radiation from the star can be received in 3 different ways: (i) The angle between the atmospheric vertical and the
stellar irradiation is $\mu_* = {\rm cos}(\theta_*)$; (ii) The stellar flux is absorbed by the planet with a cross-section of
$\pi R_{\rm Pl}^2$ but distributed over the dayside hemisphere (dayside average): The incident vertical irradiation is reduced by
a factor of $1/2$; (iii) The stellar flux is absorbed by the planet with a cross-section of $\pi R_{\rm Pl}^2$ but distributed
over the full $4\pi R_{\rm Pl}^2$ area (global average):
The incident vertical irradiation is reduced by a factor of $1/4$.
In the dayside or global average cases the stellar irradiation field is treated to be shining at the atmosphere isotropically.
For our atmospheric grid we chose option (ii). We therefore assume that there is no efficient redistribution of the insolation energy
to the night side. We will revisit the validity of this assumption in Section \ref{sect:grid_valid}.
\subsubsection{Equilibrium Chemistry}
\label{sect:CEA}
We use the NASA Chemical Equilibrium with Applications (\emph{CEA}) code by \citet{gordon1994,mcbride1996}. The code
minimizes the total Gibbs free energy of all possible species while conserving the number of atoms of every atomic species.
Given a pressure and temperature together with the atomic composition of the gas the output of the code is
the mass and number fraction of all possible outcome species (atoms, ions, molecules), the resulting density, as well as
the adiabatic temperature gradient $\nabla_{\rm ad}$ of the gas mixture.

\subsection{Testing the code}
\label{sect:code_test}
To characterize the quality of the results produced by the \emph{PETIT} code a series of tests were carried out.
\subsubsection{Correlated-k radiative transfer}
\label{sect:ck-rt_test}
First it was tested whether the correlated-k opacity combination methods introduced in Section \ref{sect:corr-kapp}
yield results of sufficient accuracy. To this end we calculated the emission spectrum of a hot jupiter at our three
different resolutions $\lambda/\Delta\lambda=10^3, \ 50, \ 10$, using correlated-k and compared to the
results of a line-by-line calculation at a resolution $\lambda/\Delta\lambda=10^6$.
As an example $PT$-structure we took a self-consistent result from our code for a 1 $\mj$, 1 $\rj$ planet\footnote{Where $\mj$ and $\rj$ are Jupiter's
mass and radius, respectively.} around a sun-like star with an effective temperature $T_*$ = 5730 K with radius $R_*=\rsun$. The
planet was assumed to be in a circular orbit at a distance of $d$ = 0.04 AU, have an internal temperature $T_{\rm int}$ = 200 K and
a C/O ratio of 1.17.
We calculated the $PT$-structure for a day-side averaged hemisphere.

\begin{figure}[t!]
\centering
\includegraphics[width=0.485\textwidth]{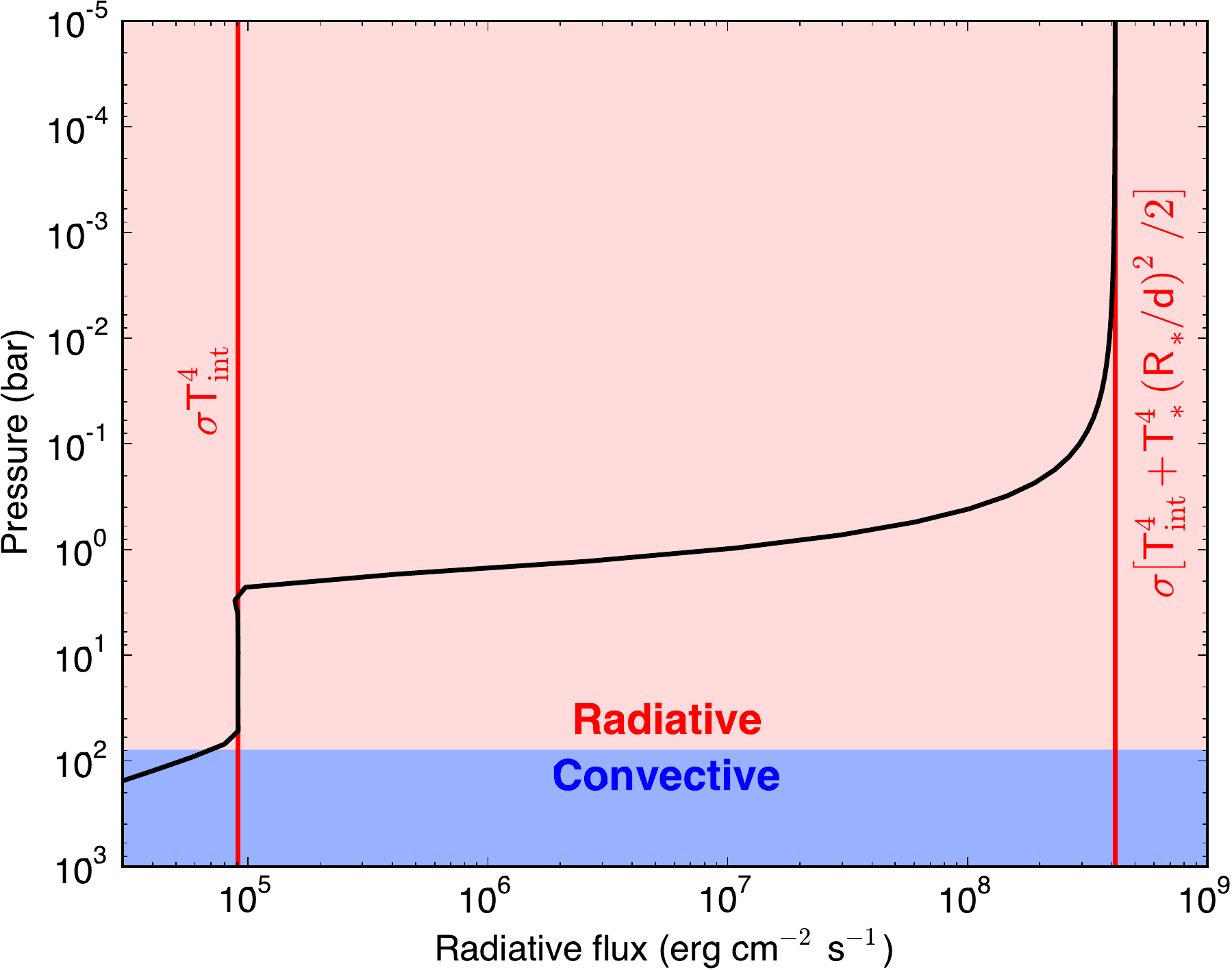}
\caption{Bolometric flux of the converged atmospheric structure calculated from the $\lambda/\Delta\lambda=10$
correlated-k radiation field integrated over angle and frequency space.
The bolometric flux is shown as a black solid line. The two red solid vertical lines denote the imposed total and internal fluxes of the planet.
The red shaded area denotes the radiative region of the atmosphere, whereas the blue shaded region shows the convective region.}
\label{fig:flux_test}
\end{figure}

The resulting emission spectra of the planet can be seen in \rch{the upper panel of} Figure \ref{fig:spec_test}.
\rch{In the lower panel we calculate the relative errors of the correlated-k calculations when compared to the frequency
averaged line-by-line calculation. If the c-k assumption was perfectly valid the error would be zero, as the flux values of
a c-k calculation at resolution (e.g.) 10 should be identical to the flux of a higher resolution line-by-line calculation,
after having been frequency averaged to the same resolution}. 
One sees that in regions of appreciable flux the relative deviation between the rebinned $\lambda/\Delta\lambda=10^6$ line-by-line calculation
and the correlated-k calculations is always smaller than 5 $\%$ and usually much less. Thus our results are within
the accuracy limits commonly found for correlated-k \citep[see, e.g.,][]{fu_liou1992,lacis_oinas1991}.

\subsubsection{Energy balance}
As a next step we tested whether the converged solution is consistent with the input parameters.
This was done by checking whether the final $PT$-structure, together with the molecular abundances
and their corresponding opacities gives rise to the correct total emergent flux.
For a day-side averaged $PT$-spectrum the total emergent flux should be
\beq
F_{\rm imposed} = \sigma\left[T_{\rm int}^4+\frac{T_*^4}{2}\left(\frac{R_*}{d}\right)^2\right] \ .
\eeq
Furthermore, deep within the atmosphere, but at lower pressure than the radiative-convective boundary
$P_{\rm conv}$, the radiation field only needs to carry the internal flux of the planet.
The reason for this is that all the stellar flux has been absorbed.
One thus finds that
\beq
F_{\rm deep}(P<P_{\rm conv}) = \sigma T_{\rm int}^4 \ .
\eeq

Even further down the $PT$-structure will eventually become convective such that
the radiative flux becomes negligible when compared to the convective flux.
In Figure \ref{fig:flux_test} one can see the result obtained from integrating the angle and frequency dependent
radiation field of the $\lambda/\Delta\lambda=10$ correlated-k structure calculation. The radiation field was integrated
to yield the bolometric flux in the atmosphere as a function of pressure. It can be seen that the surface flux converges
to $F_{\rm imposed}$. Furthermore, at approximately 3 bar, the stellar flux has been absorbed and the radiative flux
is equal to $\sigma T_{\rm int}^4$. At even higher pressures ($P\sim70$ bar) the atmosphere becomes convective and
the flux transported by radiation starts to dwindle. The radiation field thus behaves as expected and the converged
solution indeed fulfils the input parameters of the problem.
The relative difference between the converged solution of the total emergent flux and the imposed flux was 0.08 \%.

\begin{figure}[t!]
\centering
\includegraphics[width=0.485\textwidth]{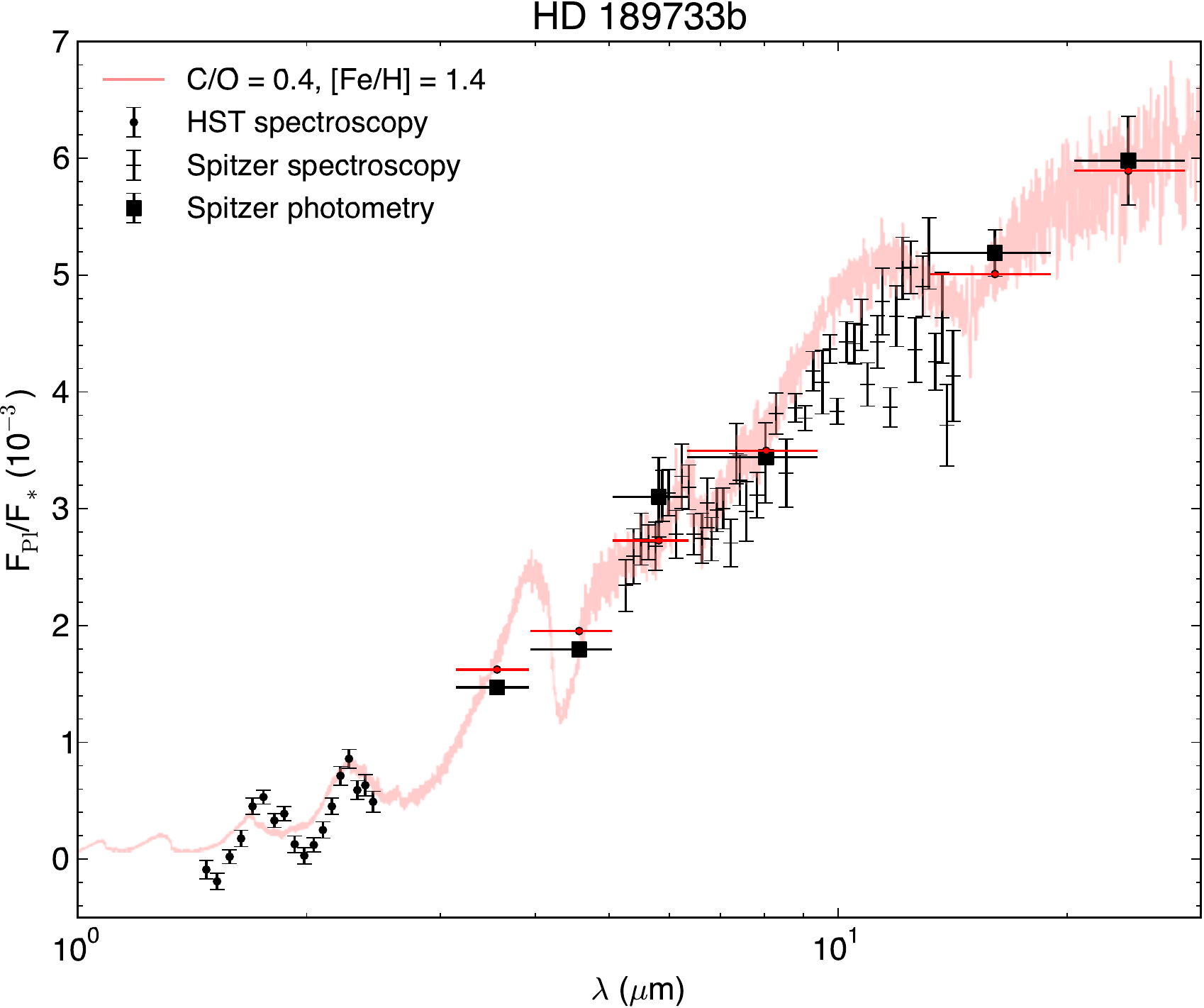}
\caption{Secondary eclipse measurement of HD 189733b using data from \citet{knutson2012,charbonneau2008,agol2010,swain2010,grillmair2008}. \emph{Spitzer} photometric and spectroscopic points are shown using black squares and crosses, respectively. The \emph{HST} spectra is shown as black dots.
Our model for HD 189733b at C/O = 0.4 and [Fe/H] = 1.4 is shown as a red line. For the photometric points we overplot boxcar-averaged red points obtained
from our spectrum.}
\label{fig:hd189}
\end{figure}

\subsubsection{Comparison to data: HD 189733b}
In order to get to get a qualitative impression of the comparability of our calculations with actual data we chose to look at HD 189733b, as it has quite a lot 
of available measurements.
We used the following data: \emph{Spitzer IRAC} photometry: 3.6 and 4.5 $\mu$m \citep{knutson2012}, 
5.8 $\mu$m \citep{charbonneau2008} and 8 $\mu$m \citep{agol2010}, \emph{Spitzer IRS} broadband at 16 $\mu$m 
, \emph{Spitzer MIPS} at 24 $\mu$m \citep[both][]{charbonneau2008}, \emph{HST NICMOS} spectroscopy \citep{swain2010} and \emph{Spitzer IRS} spectroscopy \citep{grillmair2008}.
For the stellar, planetary and orbital parameters we adopted $T_*$ = 5040 K, $R_*$ = 0.756 $\rsun$, $R_{\rm pl}$ = 1.138 $\rj$ \citep{torres2008},
$M_{\rm pl}$ = 1.137 $\mj$ \citep{butler2006,agol2010} and $d$ = 0.031 AU \citep{butler2006}.
The comparison between the data and our model for HD 189733b assuming C/O = 0.4 and [Fe/H] = 1.4 can be seen in Figure \ref{fig:hd189}.
We chose these values as they are within the Bayesan regions with the highest credibility identified by \citet{benneke2015} for this planet.
One can see that between 5 and 8 $\mu$m our model fits the IRS spectroscopy quite well, including the water feature at 6.6 $\mu$m, while it is somewhat too high at larger wavelengths.
While the \emph{IRAC} points for wavelengths below 8 $\mu$m are not fitted very well, the \emph{IRAC} photometry point at 8 $\mu$m, the \emph{IRS} broadband and \emph{MIPS} photometry points are all well fitted by our model. Further, the water absorption features between 1.5 and 2.5 $\mu$m in our model seem to correlate with the location of maxima and minima in the \emph{HST} data. The depth of the absorption features is much bigger in the
\emph{HST} data, however, although some of the values in the \emph{HST} spectra are negative, which is unphysical and related to the observational process.
We conclude that the comparison between observations and our model seems to already work quite well in certain parts of the spectra.
Dedicated fitting studies might improve the results further.
\section{Atmospheric processes in hot jupiters}
\label{sect:grid_valid}
As outlined in Section \ref{sect:intro} our goal is to set up an atmospheric grid for hot jupiters.
The range of effective temperatures we study for these objects is extending from 1000 K to 2500 K.
We discuss important physical effects which govern the atmospheres of this class of planets below
and assess how well the \emph{PETIT} code is able to describe them.
\begin{itemize}
\item {\bf Chemistry} \\
We are using a chemical equilibrium model for obtaining molecular and atomic abundances in our code
\rch{and we assess the viability of this assumption below}.
As outlined before, the knowledge of these abundances is crucial to construct the atmospheric opacities.

There are different regions in hot jupiter atmospheres, in which different chemical assumptions are fulfilled.

In the deep regions of the atmosphere temperatures and densities are high.
Therefore the chemical reaction timescales are short.
Here the chemistry is in equilibrium, i.e. the system is in a state of minimal Gibbs free energy.
By definition an equilibrium chemistry code will then suffice to obtain the molecular abundances.

Further, there are two more important effects, which are often summarized in the term ``non-equilibrium chemistry'':

In higher portions of the atmosphere the density is lower and the gas is often at lower temperatures:
under these conditions vertical eddy diffusion can quench the abundances if the timescale for attaining chemical
equilibrium is longer than the vertical mixing timescale.

In even higher regions the density is very low. Here photodissociation, i.e. photochemistry can
become the governing process if the insolation of the atmosphere is strong enough. In these
regions the photodissociation timescale will be shorter than the relevant chemical timescales.

It is obvious that if the effective temperature (which translates into a distance to the star) is high,
photodissociation acts on ever shorter timescales. However, this is compensated by the fact that a hotter
atmosphere of a planet closer to its star will have shorter chemical timescales, such that planets at
smaller semi-major axes are actually less affected by photochemistry than planets further outside.

We study how strongly quenching and photochemistry are expected to affect hot jupiters below.
Our emphasis is on the regions which will shape the spectral appearance of the planet, i.e. the spectrally active
regions.

For emission spectra the spectrally active region of a planetary atmosphere usually lies in the pressure range from $10^{-3}$ to 10 bars
\citep[see, e.g., supplementary material in][]{madhusudhan2011}. We obtain a reasonable assessment of the
importance of non-equilibrium chemistry by considering the work by \citet{miguel2014}, who analyzed the chemical properties
of planetary atmospheres around FGKM-stars. They used stellar model spectra compiled by \citet{rugheimer2013}
for the FGK stars and a spectral model for an inactive M dwarf by \citet{allard2001}.
Furthermore they consider vertical mixing. \\
By comparing figures 6 and 7 in \citet{miguel2014}, we identify the effective temperature region where the spectrally
active region is not affected by non-equilibrium effects to be $T_{\rm eff}\in [1500 \ {\rm K}, 2600 \ {\rm K}]$.
Given that only little stellar light is absorbed in the regions above $10^{-3}$ bar
we do not expect the $PT$-structure for $P>10^{-3}$ bar to be compromised within this
$T_{\rm eff}$-range.\footnote{We found that even in cases where the atmospheres are enriched in metals by up to 25 \% (in mass)
less than 10 \% of the incident stellar radiation has been absorbed above \rch{the} $10^{-3}$ bar \rch{altitude}.}
However, we want to remind the reader that our above choice of the $T_{\rm eff}$-range is subjected to the assumptions
made in \citet{miguel2014}, in particular concerning the stellar model spectra, the eddy diffusion parameter
(they took $10^9$ cm$^2$ s$^{-1}$) and the analytical model used for the atmospheric $PT$-structure.
It is reassuring, however, that also \citet{venot2015} did not find any significant differences in the emission spectra
of hot C-rich planetary spectra when comparing different chemical schemes for the treatment of photochemistry.
In their paper they compare a more sophisticated chemical network to the results of a less complete
carbon-chemistry network. Although the abundances of methane obtained by using these two different networks can
vary by roughly an order of magnitude it leaves the emission spectra in their calculations unchanged (although methane
is one of the strongest absorbers in these atmospheres).
The reason for this is that the region where photochemistry becomes important lies above the spectrally active
region of the atmosphere.
We thus conclude the atmospheres of the lowest temperatures in our grid could be affected by non-equilibrium
chemistry. We thus flag the file names of all results with $T_{\rm eff}$ $<$ 1500 K  with the ``{\verb|_neqc|}'' flag to make the user aware of this.

\item {\bf Clouds} \\
Clouds appear to be widespread in all planetary atmospheres.
The most commonly stated evidence for clouds or hazes in hot jupiter atmospheres is the fact that
the transmission spectra of many of these objects show no or only weak features at optical wavelengths.
This is striking as in general one would expect strong features from Na and K absorption
in the case of cloud free atmospheres.
HD 189733b represents a very prominent example, \rch{featuring a nearly flat transmission spectrum at
optical wavelengths, except for the alkali line cores \citep[e.g.][]{sing2011}}. Further (potential) examples for clouds or hazes weakening
absorption features in hot jupiter transmission spectra are HD 209458b \citep{charbonneau2002},
 XO-2b \citep{sing2012}, WASP-29b \citep{gibson2013a}, HAT-P-32b \citep{gibson2013b} and WASP-6b \citep{jordan2013}.

Clouds in hot jupiters may consist of silicates such as MgSiO$_3$ or Mg$_2$SiO$_4$,
liquid iron droplets, corundum (Al$_2$O$_3$) and others. A further possibility is the photochemical creation of
hydrocarbon hazes, arising from the photodissociation of CH$_4$ in the upper layers of the atmosphere.
For a more detailed discussion of possible cloud and haze forming species see, e.g., \citet{marley2013}.

Assessing the influence of clouds on the $PT$-structure and emission spectrum
of hot jupiters is not an easy task.
In the case of HD 189733b, \rch{which shows a featureless optical transmission spectrum
\citep[except for the alkali line cores, see][]{sing2011}},
\citet{barstow2014} find that the $PT$-structure they can retrieve using the planet's emission spectrum
is more or less insensitive to whether or not a cloud model is included (they use various MgSiO$_3$ models).
At the same time many of their cloud models are able to reproduce HD 189733b's transmission spectrum.
This indicates that for hot jupiters, at least for HD 189733b, the treatment of clouds is important for
the appearance of the planet's transmission spectrum, but not so much for the actual absorption of the bulk of
the stellar light in the deeper layers of the dayside atmosphere. In this case the influence of clouds on the $PT$-structure
and the emission spectrum would be minor.
This is in agreement with the earlier work by \citet{fortney2008}, who also find that clouds have a minor effect on their
self-consistently calculated PT-profiles and emission spectra of hot jupiters and therefore neglect clouds.
The obvious importance of clouds in the case of transmission spectroscopy
is due to the slant optical depths of possible cloud species being $\sim$35-90 bigger than the
vertical optical depth \citep{fortney2005}.

We do not currently consider the formation of clouds and the associated effect on the planet's opacity.
However, from the previous discussion we conclude that it might be permissible to neglect clouds
in our calculations. Nonetheless we want to note, following \citet{fortney2005}, that in cases of high 
metallicity planets the effects of clouds may become important, especially if appreciable amounts of
silicate, iron or corundum condensates can form.
This has to be stressed in light of the fact that hot jupiters seem to be most prevalent
in stellar systems of high metallicity \citep{fischer2005}.

\item {\bf Winds} \\
Based on GCM simulations and theoretical considerations, winds are expected to be present on hot jupiters,
driven by the temperature contrasts between the day and nightside, and the polar and equatorial regions
\citep[see, e.g.,][]{heng2014}.
The question of whether these winds will have an effect on the thermal structure of the planetary atmosphere
depends on whether the advection timescale of the winds $\tau_{\rm adv}$ is shorter than the radiative cooling
timescale $\tau_{\rm rad}$ and/or chemical timescale $\tau_{\rm chem}$ of the atmosphere.
If $\tau_{\rm adv}$ is indeed shorter than one of those two timescales, then energy or molecules will
be transported, and the assumptions of local radiative or chemical equilibrium breaks down.
To properly carry out this timescale comparison one would have to couple GCM simulations with radiative
transport and chemical non-equilibrium calculations, which is beyond the scope of this work.
The fact that one sees a day-night temperature variation when looking at the thermal phase curve of, e.g.,
HD 189733b \citep{knutson2012}, shows that winds are not able to perfectly redistribute the energy from
the incident stellar radiation across the whole planetary surface. However, the results in \citet{knutson2012}
also show that the hottest and coldest points in the atmosphere are offset from the substellar and antistellar
point, respectively. This indicates that winds play a role in distributing energy across the planet.
In general, it is found that the higher the effective temperature of a hot jupiter, the less efficient the
transport of energy by wind becomes \citep{perez-becker2013}. For ``cool'' planets with effective
temperatures of $\sim$ 1000 K redistribution of energy may be quite efficient unless the planet has
a mass of a few Jupiter masses or more \citep{kammer2015}.

In order to at least partially accommodate the effect of heat redistribution by winds,
our code has 3 possible ways to treat the distribution of the incident stellar
light across the atmosphere: (i) no wind transport of energy, (ii) day-side averaging or (iii)
global averaging, the latter approximating the case where winds highly efficiently distribute the energy
received by the star across the planetary surface (see Section \ref{sect:VEF}).
Our treatment of the stellar energy input in the cases (ii) and (iii) are only approximative ways to inject
the stellar energy into the planetary atmosphere.
A fourth way would be to use a redistribution parameter for the incident stellar irradiation which adds
a fraction of the absorbed stellar energy to the night side internal temperature and decreases the amount
of light to be absorbed on the dayside \citep{burrows2006}. Other possibilities include the mimicking of 
planetary winds by assuming that the atmosphere carries out a rigid body rotation, as it was done in \citet{iro2005}.

\end{itemize}

\section{Setup and calculation of the grid}
\label{sect:grid_setup_calc}
\subsection{Grid setup}

We set up a grid of 10,640 models which is defined by the following parameters:
\begin{enumerate}
\item {\bf $T_{\rm eff}=$ 1000, 1250, 1500, 1750, 2000, 2250, 2500 K} \\
We chose to go to
temperatures somewhat lower than where we are unaffected by non-equilibrium chemistry
effects (1500 K, see Section \ref{sect:grid_valid}). The files of models with $T_{\rm eff}<1500$ K will be
flagged with ``{\verb|_neqc|}'' to make the user aware of potential differences when including non-equilibrium chemistry.
Furthermore, high metallicity models with low ${\rm log}(g)$ and high $T_{\rm eff}$ will have temperatures
larger than 3000 K in the higher pressure parts of the atmosphere. If this happens before
the atmosphere becomes convective we flag these models with ``{\verb|_t3000k|}'', as our opacity grid only extends
to 3000 K (see Section \ref{sect:mol_atom_line_opa}).
At atmospheric layers where $T>3000$ K we use the opacities at 3000 K.
\item {\bf [Fe/H] = -0.5, 0.0, 0.5, 1.0, 2.0} \\
The metallicity is chosen to range from slightly subsolar to strongly enriched and we
use scaled solar compositions according to \citet{asplund2009}. It is not
generally expected that enriched exoplanets have a scaled solar composition.
Nonetheless, we use this approximation as a proxy for various degrees of enrichment.
A further degree of freedom regarding the composition is introduced to our grid by
varying the C/O ratio. In this work we focus on metallicities
higher than the solar value. The reason for this is that giant exoplanets are expected to be
enriched in metals, with objects of several hundred Earth masses having metallicities
of up to several tens of the solar metallicity \citep{forntey2013}.
\item {\bf C/O = 0.35, 0.55, 0.7, 0.71, 0.72, 0.73, 0.74, 0.75, 0.85, 0.9, 0.91, 0.92, 0.93, 0.94, 0.95, 1.0, 1.05, 1.12, 1.4} \\
We investigate C/O values which are subsolar or supersolar but $<$ 1 (C/O$_\odot$ $\sim$ 0.55),
as well as values around and above 1. We use a finer sampling around C/O $\sim$ 0.73 and C/O $\sim$ 0.92,
because we want to resolve the transition from oxygen to carbon-dominated spectra and atmospheres
at low and high temperatures. Commonly, the transition is expected to happen quite sharply
at C/O values around 1 \citep[see, e.g.,][]{kopparapu2012,madhusudhan2012}. We find C/O = 0.92 for the high
temperature atmospheres.
Furthermore, the infrared opacity of the atmospheres is minimal when C/O is close to 1,
because most of the C and O atoms are locked up in CO and neither H$_2$O nor CH$_4$ of HCN
are very abundant. This gives rise to inversions for the hottest atmospheres
($T_{\rm eff} \gtrsim$ 1500) K, where the alkali atoms absorb the stellar irradiation quite effectively
but the cooling is inefficient due to the IR opacity minimum (see Section \ref{sect:results}).
The C/O ratio at a given metallicity was obtained from varying the O abundance.
This means that for supersolar C/O ratios the O abundance was decreased, corresponding
to the accretion of water depleted gas or planetesimals during the planet's formation.
\item {\bf Spectral type of host star: F5, G5, K5, M5} \\
In order to assess the dependence of the atmospheric structure on the spectral shape of the
stellar radiation field we calculated our grid using 4 different spectral types for the host star.
For the earlier spectral types the energy received by the planet is absorbed predominantly
by the alkalis in the optical wavelengths, whereas for the later spectral types the wavelength
range of the absorption shifts more and more to the IR, leading to increasingly isothermal
planetary atmospheres.
\item {\bf ${\rm log}(g)$ = 2.3, 3.0, 4.0, 5.0} \\
Our ${\rm log}(g)$ grid was chosen such that it encompasses hot jupiters of every conceivable
mass--radius combination, including bloated hot jupiters as well as compact ($R_{\rm Pl} \sim \rj$)
planets of varying masses (all planets listed on \url{http://exoplanets.org} with a mass and
radius measurement fall within our adopted ${\rm log}(g)$ range).
\end{enumerate}

\subsection{Chemical model}
\label{sect:chem_model}
The following atomic species were considered in the equilibrium chemistry network:
H, He, C, N, O, Na, Mg, Al, Si, P, S, Cl, K, Ca, Ti, V, Fe and Ni.
Based on \citet{seager2000} we consider the following reaction products:
e$^-$, H, He, C, N, O, Na, Mg, Al, Si, P, S, K, Ca, Ti, Fe, Ni,
H$_2$, CO, OH, SH, N$_2$, O$_2$, SiO, TiO, SiS, H$_2$O, C$_2$, CH, CN, CS, SiC, NH, SiH,
NO, SN, SiN, SO, S$_2$, C$_2$H, HCN, C$_2$H$_2$, CH$_4$, AlH, AlOH, Al$_2$O, CaOH, MgH,
MgOH, VO, VO$_2$, PH$_3$, CO$_2$, TiO$_2$, Si$_2$C, SiO$_2$, FeO, NH$_2$, NH$_3$, CH$_2$, CH$_3$,
H$_2$S, KOH, NaOH, NaCl, KCl, H$^+$, H$^-$, Na$^+$, K$^+$, Fe (condensed), Al$_2$O$_3$ (condensed),
MgSiO$_3$ (condensed), SiC (condensed).

The choice of condensed species is motivated by \citet{seager2000,sudarsky2003}.
Additionally, we also added SiC as a condensable species, to account for condensation of
C in atmospheres with a high C/O ratio, as has also been suggested by \citet{seager2005}.

A reaction pathway that is of prime interest is the one connecting
H$_2$O, CH$_4$ and CO. In Section \ref{sect:intro} we already introduced the C/O ratio as a useful quantity for characterizing planetary atmospheres,
as it allows to interpret the relative abundances of CH$_4$ and H$_2$O for temperatures $T$ $<$ 1750 K.
CH$_4$ and H$_2$O are important molecules because they are abundant, have a high infrared opacity and therefore shape
the overall appearance of the atmosphere's emission spectrum. 

The net chemical equation of interest for this case is
\beq
{\rm CH}_4 + {\rm H}_2{\rm O} \xrightleftharpoons[T \ \lesssim \ 1000 \ {\rm K}]{T \ \gtrsim \ 1000 \ {\rm K}} {\rm CO} + 3{\rm H}_2 \ ,
\label{equ:COchem}
\eeq
leading to a quite sharp transition of CH$_4$ vs. H$_2$O
rich atmospheres at C/O $\sim$ 1 \citep[see, e.g.,][]{kopparapu2012,madhusudhan2012}:

In chemical equilibrium CO is the most common C and O bearing molecule in planetary atmospheres,
where the temperatures are high enough ($T \gtrsim 1000$ K). In an oxygen-rich atmosphere (C/O $<$ 1)
the remaining oxygen is then partly found in
the form of H$_2$O and almost no CH$_4$ is present, as most C is locked up in CO.
In a carbon-rich atmosphere (C/O $>$ 1) the excess C is put partly into CH$_4$,
with no O left to form water. For $T \lesssim$ 1000 K the low temperature direction in Eq. (\ref{equ:COchem}) is dominant, leading to
appreciable amounts of both CH$_4$ and H$_2$O and negligible amounts of CO.

The main effect of including condensation is the removal of oxygen from the gas phase through the condensation of MgSiO$_3$ for
temperatures smaller than $\sim$ 1500 K, leading to a spectrally noticeable decrease of H$_2$O and CO at C/O values as low as
0.7. This effect is observed for the $T_{\rm eff}$ = 1000, 1250 and 1500 K cases. Effectively it shifts
the spectrally visible transition from H$_2$O dominated atmospheres to CH$_4$ dominated atmospheres away
from C/O $\sim$ 0.9 to somewhat smaller values of C/O $\sim$ 0.7, as the formation of MgSiO$_3$ acts as a sink for
the O atoms available to form H$_2$O.

The depletion of O-bearing gas phase species due to condensable O-bearing species has been found in much more complete cloud models as well \citep{helling2014}. We describe some of the incompletenesses of our cloud model below:
One of the effects our condensation model does not treat is the the problem of homogeneous or heterogeneous
nucleation, which could potentially shift the formation of condensates in the atmospheres
toward layers of higher supersaturation if initial condensation seeds are not present in the atmospheres \citep[see, e.g.,][]{marley2013}.

Further we want to stress that the condensed species in each layer remain in chemical contact with the gas phase in our model
and do not rain out to deeper layers of the atmosphere.

\rch{The consequences of a potential rainout for a planetary atmosphere can be manyfold.
First of all the rainout removes metals from the atmosphere, relocating them to deeper layers.
Hence the corresponding grain or droplet opacity will be missing from higher atmospheric layers.
Because we do not include cloud opacities in our calculations we make the implicit
assumption of a rainout of the condensed particles, although we do not model it,
the net effect being the removal of metals from the higher layers.
It has to be kept in mind, however, that the chemical equilibrium solution of the gas abundances
in chemical contact with the condensed species is not necessarily the same as it would be when assuming
a rainout.
Our implicit assumption of a rainout is also applicable when considering the gaseous Na and K alkali abundances.
In our models MgSiO$_3$ condenses at temperatures below $\sim$ 1600 K. In principle this silicate
material can further react with the alkali atoms to form alkali feldspars (such as albite and orthoclase),
removing the gaseous alkalis from the gas for $T$~$\lesssim$~1600~K \citep[see, e.g.,][]{lodders2010}.
We do not consider these feldspars in our condensation model, such that the alkali atoms stay in the gas,
as they would in a silicate rainout scenario.
It has been found that alkali atoms are present in cool brown dwarf atmospheres,
indicating that silicate rainout may occur in these objects \citep{marley2002,morley2012}.
Another consequence of condensed material can be the formation of a cloud deck, close to
and above the layers of the atmosphere hot enough the evaporate the in-falling
cloud particles again. Such cloud decks can heat the atmosphere locally and in the layers below,
by making the atmosphere more opaque to the planet's intrinsic flux, effectively
acting like a blanket covering the lower layers of the atmosphere \citep[see, e.g.,][]{morley2014,helling2014b}.
If the cloud layer is optically thick close to the planet's photosphere it will leave an
imprint on the planet's spectral appearance and and may reduce the contrast of absorption features.
The height of the cloud deck depends critically on the planets effective temperature and also on its surface
gravity since the condensation temperature is pressure-dependent. The cooler an object is, the deeper in its interior
the clouds will reside.
Therefore the spectral imprint of clouds will vary with temperature, similar to the behavior in brown
dwarf atmospheres. Silicate clouds with a high optical depth are thought to reside in the photospheres
L4-L6 type brown dwarfs ($T_{\rm eff}$~$\sim$~1500-1700 K) where they affect the spectra.
For cooler objects the cloud deck lies below the photosphere and the clouds are no longer seen
\citep[see, e.g.][]{lodders2006}.
In our atmospheres we checked the possible locations of the cloud decks
(i.e. the layers below which the condensates
evaporate). We found that the silicate evaporation layer of planets with
$T_{\rm eff}$~=~1000 K and $T_{\rm eff}$~=1250 K is always located at
pressures far higher than that of the photosphere,
such that we do not expect any spectral impact of a cloud layer.
For effective temperatures between 1500 K and 1750 K
the evaporation layer lies close to and above the photosphere (in altitude),
such that a cloud deck could potentially affect the spectrum.
For increasing ${\rm log(g)}$ the photosphere shifts to layers of deeper pressure, but so does the evaporation layer,
as condensation is pressure dependent.
Note that this temperature range is close to the effective temperature where L4-L6 dwarfs are thought to be most
strongly affected by silicate clouds.
For higher temperatures the evaporation layer is far above the photosphere such that we do not expect clouds to be of importance.}

For C/O ratios $>$ 1 and temperatures $>$ 1750 K we find, in agreement with previous studies, that the spectrally most 
important carbon bearing molecule is no longer CH$_4$, but HCN \citep[see, e.g.,][]{kopparapu2012,venot2012,moses2013}.
In general the lower the pressure and the higher the temperature the more important HCN
becomes.
Therefore we see that the spectra at the highest effective temperatures are dominated by HCN absorption.

\subsection{Calculation of the grid}
\label{sect:grid_calc_descr}
The calculations were carried out using 150 atmospheric layers spaced equidistantly in log$(P)$ between
$10^{-14}$ and 9$\times 10^{4}$ bar. Note that our opacity grid is only calculated between 10$^{-6}$ and 10$^3$ bar. For pressures outside
this range we use the opacities at the pressures at the boundaries of our opacity grid.
The grid calculations were extended to smaller pressures
to not introduce any kinks at the 10$^{-6}$ boundary:
the alkali line cores are already optically thick at these low pressures,
and a cut-off of the atmospheric structure at 10$^{-6}$ bar would result in no alkali core flux coming from above at the highest point in the atmosphere, making the temperature there to cool.
We provide the $PT$-structures only
between 10$^{-6}$ and 10$^3$ bar.
\rch{However, at altitudes above the 10$^{-6}$ bar level the contribution of the pressure-broadened line wings is to the total opacity is negligible and the opacity is dominated by the line cores, whose shape is given by thermal broadening and is independent of pressure. 
As only little mass is above any given pressure lower than 10$^{-6}$ bar, the line wings are not able to significantly alter the radiation field.
Therefore, adapting the 10$^{-6}$ opacity curves at all lower pressures should not affect the resulting PT structures; in all this range the line cores are
of significant optical thickness, whereas the line wings are highly optically thin.
Hence, the line cores govern both the absorption and the re-emission of energy, and thus the PT-structure.}

The calculations were extended to pressures larger than 10$^{3}$ bar as we consider quite large surface gravities, which
essentially rescale the temperature structures to higher pressures. We wanted to make sure that we do not cut-off the atmospheric structures at 10$^{3}$ bar for high ${\rm log}(g)$ cases when the atmosphere is not yet optically thick at all
wavelengths. We found, however, no differences in the $PT$ structures nor the emission spectra when comparing cases
extending down to either 10$^{3}$ or 9$\times 10^{4}$ bar.

\begin{figure*}[t!]
\centering
\includegraphics[width=0.97\textwidth]{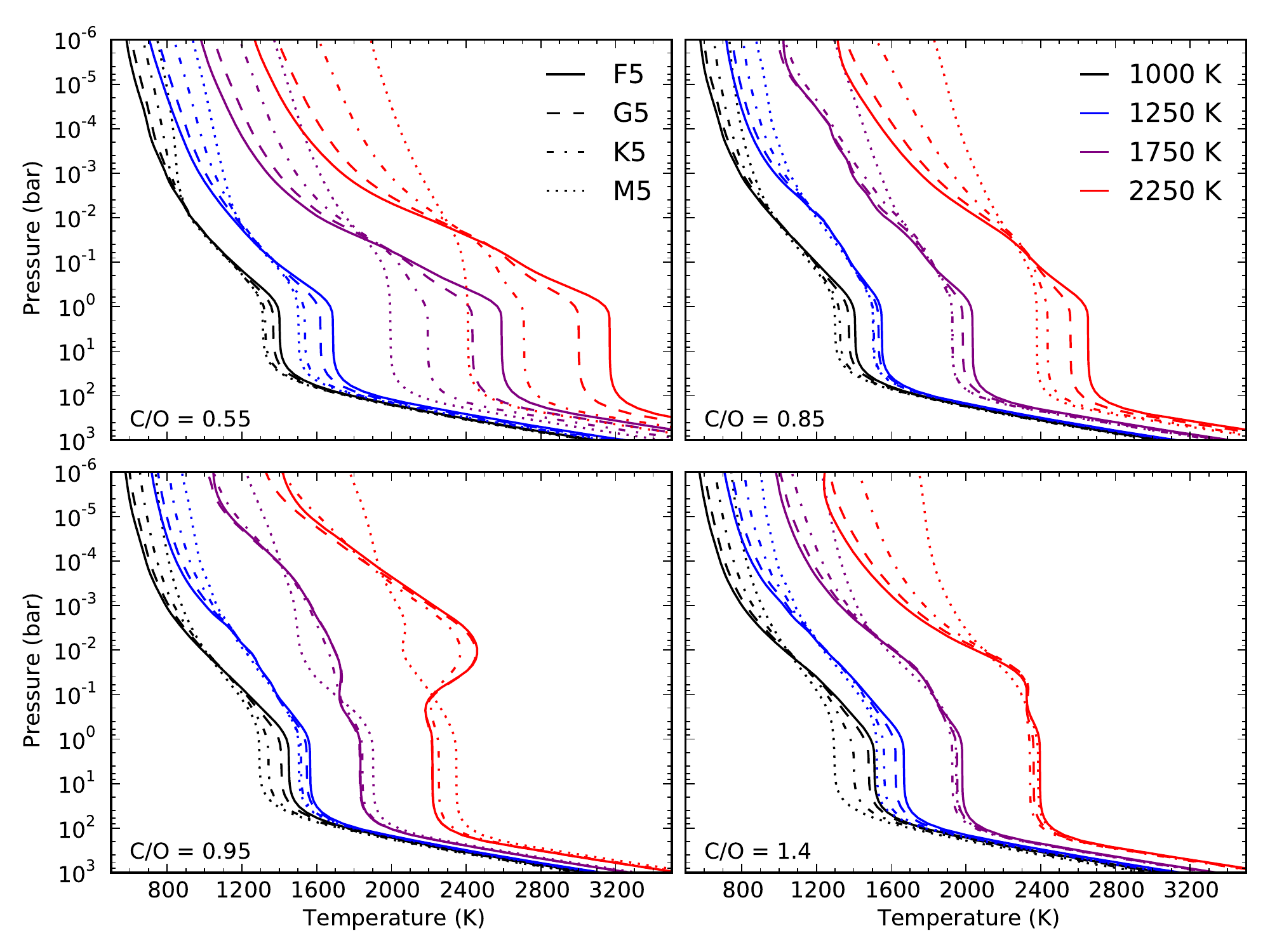}
\caption{Atmospheric $PT$-structures for planets of varying host star spectral types, effective temperatures and C/O ratios with
${\rm log}(g) = 3$ and [Fe/H] = 1. 
The line style varies with host star spectral type as follows: F5 (solid), G5 (dashed), K5 (dot-dashed), M5 (dotted). The line color
indicates the following planetary effective temperatures: 1000 K (black), 1250 K (blue), 1750 K (purple), 2250 K (red). The four
different panels correspond to 4 different C/O ratios: C/O = 0.55 ({\it upper left panel}), C/O = 0.85 ({\it upper right panel}),
C/O = 0.95 ({\it lower left panel}), C/O = 1.4 ({\it lower right panel}).}
\label{fig:COtypeTEMPpts}
\end{figure*}

\rch{For the temperature iteration the pressure-, temperature- and abundance-dependent combination of the individual species' opacity tables
is the computationally most demanding part of the atmospheric structure calculation.}
Thus, for numerical convenience, we precalculated the opacity tables for every atmospheric structure on 40 $\times$ 40 pressure and
temperature grid points (taking about 2 minutes) before the iterations were run. We then interpolated in this table during the iterations
\rch{and verified that the results were consistent with those obtained when re-calculating the opacity tables for every individual iteration.}

\subsubsection{Convection and convergence}
As described in Section (\ref{subsect:treatment_of_convection}), we use the Schwarzschild criterion to
assess whether a given layer in the atmosphere should be convective, and if so we switch to an adiabatic temperature
gradient. We find that the lowest layers of the atmospheres (at the highest pressure) become convective, with a radiative gradient
much bigger than the adiabatic temperature gradient.
For hot atmospheres ($T_{\rm eff}\geq 2000$ K) with high metallicities [Fe/H]  $\gtrsim 1$ we find that regions with a
steep temperature gradient high in the atmosphere (10$^{-2}$ bar $>$ P $>$ 10$^{-6}$ bar) can become convective.
In these situations the solutions can become unstable, as the layers switch back and forth between being either
radiative or convective, introducing jumps and kinks in the $PT$-spectra. This suggests that these layers are in the continous
transition region between being fully radiative or convective, which cannot be resolved by the binary Schwarzschild criterion.
A better treatment would be to implement convection via the mixing length theory (MLT), as it allows for a continuous transition from
a fully radiative to a fully convective solution.
For now, we decided to rerun the $PT$-structures affected by this convergence problem and to forbid the occurrence of convection
in the uppermost layers (10$^{-2}$ bar $>$ P $>$ 10$^{-6}$ bar) of the atmosphere.
The corresponding atmospheric structure files have been
flagged with ``{\verb|_conv|}''. We plan to implement MLT in a future version of the code.

\vspace{5mm}
\section{Results}
\label{sect:results}
We first discuss some general characteristics of our results in Section \ref{sect:first_glance}.
We will study the atmospheric properties systematically as a function of effective temperature for all atmospheric
parameters in sections \ref{sect:LT_atmos} -- \ref{sect:HT_atmos}.
\vspace{5mm}

\subsection{A first glance}
\label{sect:first_glance}

To give a first overview of our of results we show atmospheric $PT$-structures of ${\rm log}(g) = 3$ and [Fe/H] = 1
planets for varying host star spectral types (F5, G5, K5, M5) and effective temperature (1000 K, 1250 K, 1750 K, 2250 K) at four
different C/O ratios (0.55, 0.85, 0.95, 1.4) in Figure \ref{fig:COtypeTEMPpts}.

\begin{figure*}[t!]
\centering
\begin{minipage}{0.48\textwidth}
\includegraphics[width=1.0\textwidth]{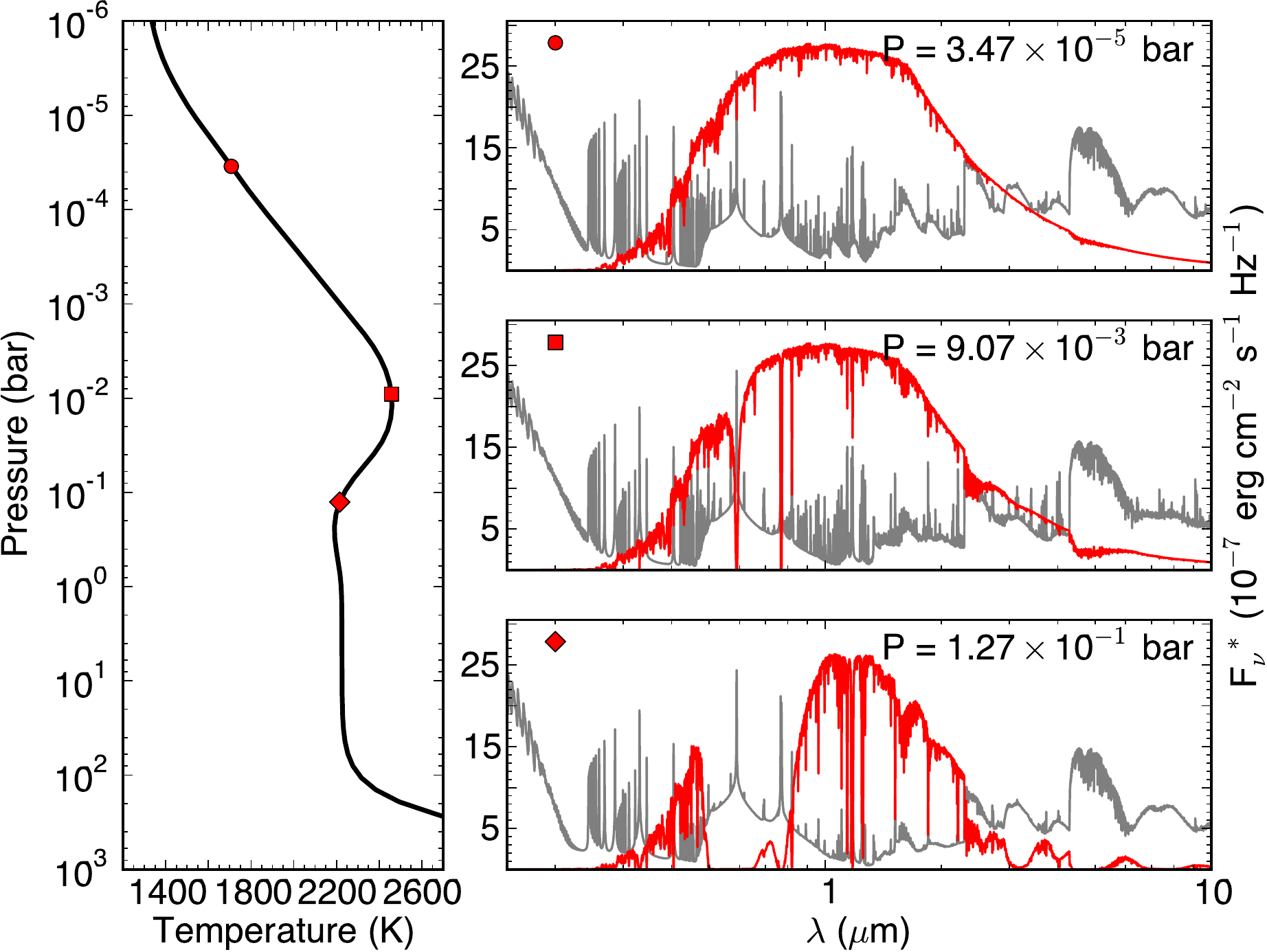}
\caption{{\it Left panel:} $PT$-structure of a ${\rm log}(g)$ = 3, [Fe/H] = 1, $T_{\rm eff}$ = 2250 K, C/O = 0.95 atmosphere of a planet in orbit around
a G5 star. {\it Right panels:} Local stellar flux (red solid line) at the three pressure levels at 3.47 $\times$ 10$^{-5}$ {\it (top panel)}, 9.07 $\times$ 10$^{-3}$ {\it (middle panel)} and 1.27 $\times$ 10$^{-1}$ bar {\it (bottom panel)} in the atmosphere. The local opacity ${\rm log}(\kappa)$ for each layer is shown as a grey solid line (rescaled). The respective pressure levels are indicated by a red circle, square and diamond in the $PT$-structure.}
\label{fig:stellar_abs_inv}
\end{minipage}
\begin{minipage}{0.03\textwidth}
\end{minipage}
\begin{minipage}{0.48\textwidth}
\includegraphics[width=1.0\textwidth]{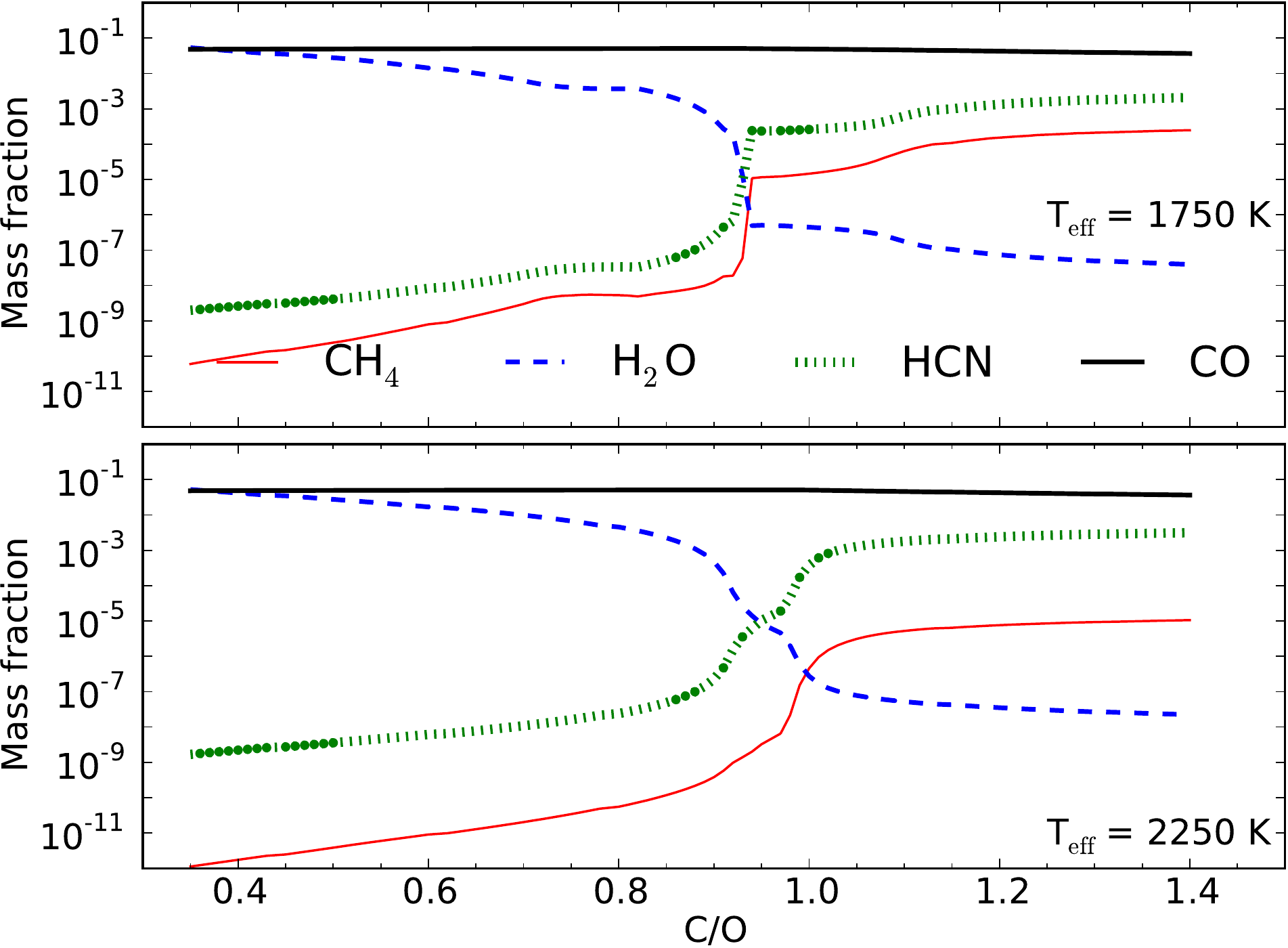}
\caption{Mass fractions of CH$_4$ (thin solid red line), H$_2$O (dashed blue line), HCN (dotted green line) and CO (thick solid black line) as a function of the C/O ratio for a ${\rm log}(g)$ = 3, [Fe/H] = 1 atmosphere of a planet in orbit around a G5 star at a pressure level of 9.07 $\times$ 10$^{-3}$ bar. The {\it top panel} shows the mass fractions for a planet with $T_{\rm eff}$ = 1750 K while
the {\it bottom panel} shows the mass fractions for a planet with $T_{\rm eff}$ = 2250 K.}
\label{fig:highCOinv}
\end{minipage}
\end{figure*}

Some general, expected trends can quite easily be made out from looking at this plot:
\begin{itemize}
\item The later the host star spectral type, the more isothermal the atmospheric structure
becomes. This is expected because the wavelength range of the received stellar irradiation becomes
more and more similar to the wavelength range of the internal planetary radiation field, such that the radiation
field absorbed by the gas at the top of the atmosphere is similar to the radiation field absorbed by the gas
at the bottom of the atmosphere, hence leading to similar temperatures.
\item The $PT$-structures with C/O = 0.55 are hotter than the $PT$-structures with C/O = 0.85.
The main reason for this is that the atmosphere with the lower C/O ratio has, everything else being equal,
more oxygen and thus a higher opacity due to a higher H$_2$O abundance. This results in a stronger green house
effect, as the excess H$_2$O leads to a less efficient escape of radiation from the atmosphere. 
In order to radiate away the required amount of energy (set by $T_{\rm eff}$) the atmospheres need to be hotter.
\end{itemize}

Another very striking result is that for C/O ratios close to 1 temperature inversions form in the atmospheres for
effective temperatures above 2000 K. 
In general, they can even occur at effective temperatures as low as 1500 K, see Section \ref{sect:MT_atmos}.
This is interesting, as no extra optical opacity sources such as TiO and VO except for the ones given in Table \ref{tab:opa_sources} are being considered.
For host stars later than K5 there are no inversions in the planetary atmospheres.
This phenomenon will be further studied in Section \ref{sect:highCOinv}.

\subsubsection{Inversions at high C/O ratios}
\label{sect:highCOinv}

As outlined above, C/O ratios of $\sim$ 1 can lead to inversions in atmospheres with high enough effective temperature
if the stellar host is of K spectral type or earlier.
The reason for the inversions to occur for these spectral types is that an appreciable amount of stellar flux is received
from the star in the optical wavelength regime.
This means that the alkali lines,  and the pseudo-continuum contribution of the alkali line wings, will become very effective
in absorbing the stellar irradiation.

At the same time, close to C/O = 1, most of the oxygen and carbon is locked up in CO, leading to low H$_2$O, CH$_4$ and HCN abundances and opacities.

The combined effect of the effective absorption of the strong 
irradiation and a decreased ability of the atmospheric gas to cool,
because of too little CH$_4$, H$_2$O and HCN leads to the inversion in the atmospheres.

The absorption of the stellar light as a function of depth can be seen in Figure \ref{fig:stellar_abs_inv}, where we plot the
$PT$-structure of a ${\rm log}(g)$ = 3, [Fe/H] = 1, $T_{\rm eff}$ = 2250 K, C/O = 0.95 atmosphere of a planet in orbit around
a G5 star, as well as the local stellar flux at the pressure levels 3.47 $\times$ 10$^{-5}$, 9.07 $\times$ 10$^{-3}$
and 1.27 $\times$ 10$^{-1}$ bar in the atmosphere.
Also a plot of the logarithm of the (rescaled)
opacity ${\rm log}(\kappa)$ is shown in the Figure for each pressure level.
The respective pressure levels are indicated by red points in the $PT$-structure.

Figure \ref{fig:stellar_abs_inv} nicely shows how the alkali pseudo-continuum absorbs the full stellar flux in its wavelength
domain at the position of the inversion: At the highest pressure shown in the spectral plots (3.47 $\times$ 10$^{-5}$ bar) the 
stellar flux is still completely unaffected by any absorption effects as the atmosphere is still optically thin at all wavelengths 
(except for right at the core of the alkali lines). At the hottest point in the temperature inversion (at 9.07 $\times$ 10$^{-3}$ bar) 
one can see that the alkali wings have already started to absorb non-negligible amounts of energy, and just after the inversion 
(at 1.27 $\times$ 10$^{-1}$ bar) the stellar flux in the alkali wings has been completely absorbed.
\rch{Interestingly}, the inversions obtained in our calculations \rch{due to alkali heating seem to} abide by the rule that the tropopause, i.e. the atmospheric
layer at minimum temperature just after the inversion, should commonly be found at $\sim$ 0.1 bar
for a wide variety of possible atmospheres \citep{robinson2014}.

As can be seen in the stellar flux spectrum at the highest pressure the absorption of the stellar light outside of the alkali wings
is rather sluggish, showing the importance of the alkali wings in the formation of the inversion.

As mentioned above, in a small region of C/O  around 1, the atmosphere's ability to efficiently radiate away the 
absorbed stellar light decreases due to the involved chemistry.
This can be understood by looking at Figure \ref{fig:highCOinv}, which shows the CH$_4$, H$_2$O, HCN and CO mass 
fractions in a ${\rm log}(g)$ = 3, [Fe/H] = 1 atmosphere of a planet in orbit around a G5 star as a function of C/O
at a pressure level of 9.07 $\times$ 10$^{-3}$ bar, i.e. close to the pressure where the inversion temperature, if an inversion occurs, 
is maximal.
Two cases for planets with $T_{\rm eff}$ = 1750 K and $T_{\rm eff}$ = 2250 K are shown and we carried out 100
self-consistent atmospheric calculations for both cases with C/O going from 0.35 to 1.4 in equidistant steps.

One sees that for the $T_{\rm eff}$ = 2250 K case, at C/O = 0.95, the H$_2$O abundance has already decreased by 4 orders 
of magnitude when compared to the lowest C/O values, while the CH$_4$ abundance is still more than 2 magnitudes smaller 
than its highest abundance at the highest C/O values. Further, HCN has not yet risen to a high enough abundance to take over
the cooling. The C/O = 0.95 point at $T_{\rm eff}$ = 2250 K thus is very close
to the aforementioned point of minimum IR opacity, leading to the inversions seen in our results for all host 
spectral types except M5. For higher C/O ratios the IR opacity and the atmosphere's ability to cool increases, such that no 
inversions are observed anymore, mainly because HCN takes over the cooling.

For the particular case of $T_{\rm eff}$ = 1750 K in Figure \ref{fig:COtypeTEMPpts} the situation
must be different, as there is no inversion present in the atmosphere.
The reason for this can be seen in the panel for $T_{\rm eff}$ = 1750 K in Figure \ref{fig:highCOinv}:
for this atmosphere the transition from water-rich to methane-rich atmospheres occurs
much quicker as a function of C/O than it does for the $T_{\rm eff}$ = 2250 case.
The methane mass fraction jumps from 10$^{-8}$ to 10$^{-5}$ at C/O = 0.93 and the HCN mass fraction
jumps from 10$^{-6}$ to 10$^{-4}$ and no extended region of
low water, methane and HCN abundance is seen. Further, as this atmosphere is cooler, the overall
CH$_4$ content is higher than in the hotter case.
This is expected to occur and has been studied before both in equilibrium and disequilibrium chemical networks \citep[see, e.g.,][]{moses2013},
showing that CH$_4$ becomes less abundant as the temperature increases in carbon-rich atmospheres.
In conclusion, this atmosphere can cool more efficiently.

\subsubsection{Inversions and line list completeness for HCN and C$_2$H$_2$}
\label{sect:hcn_importance}
We want to issue a word of caution regarding the cooling efficiency of atmospheres.
At high temperatures for C/O $>$ 1 and $T_{\rm eff}$ $\gtrsim$ 1750 K
we find that HCN is more abundant than CH$_4$.
It is therefore very important to use HCN line lists which are as complete as possible.
In fact we found that if we use HCN from the \emph{HITRAN} database, which is made for low
atmospheric temperatures, we got strong inversions occurring even for C/O $>$ 1
if the effective temperatures were high.
\begin{figure}[t!]
\centering
\includegraphics[width=0.485\textwidth]{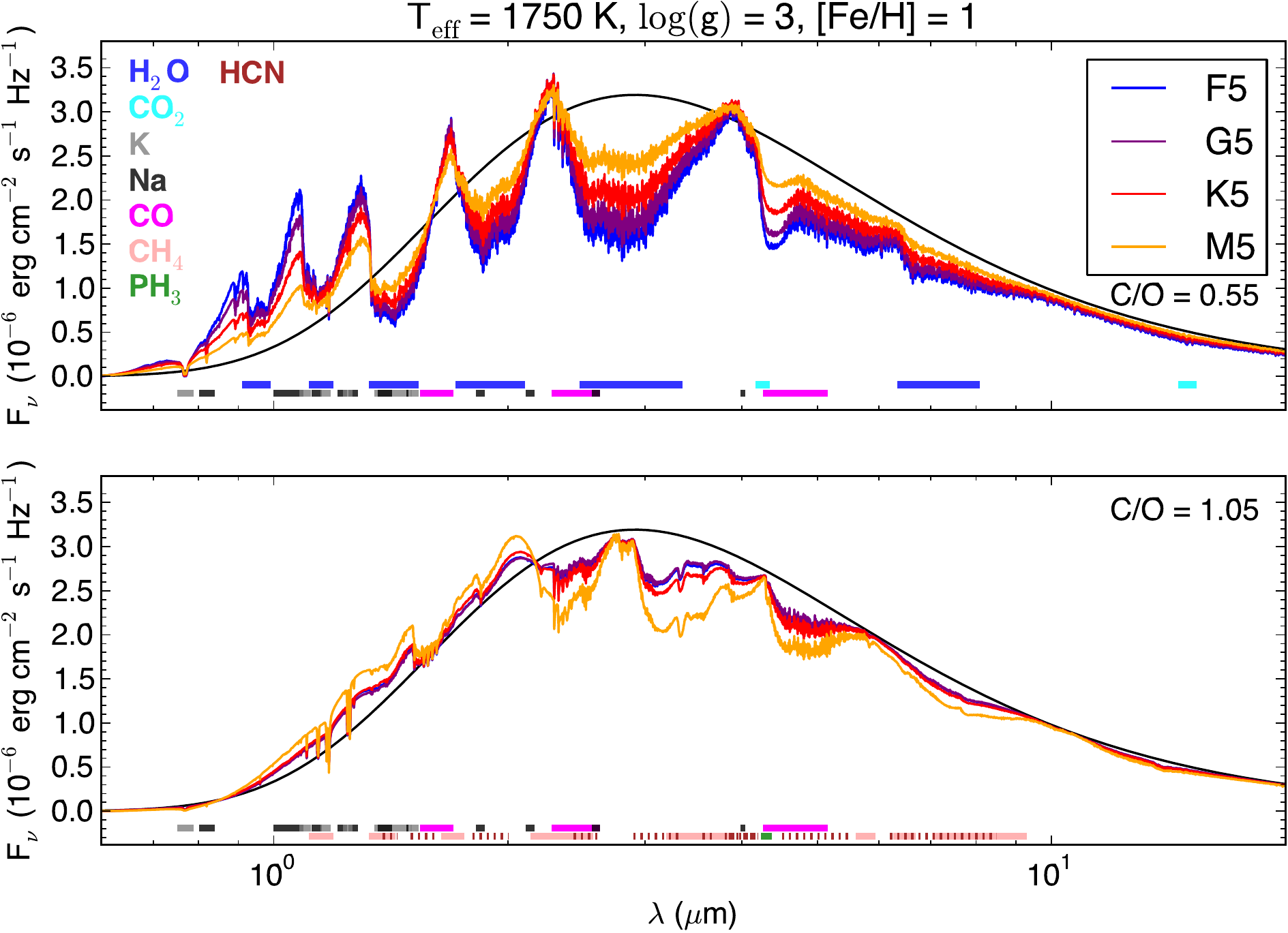}
\caption{Emission spectra as a function of host star spectral type for a $T_{\rm eff}$ = 1750 K, ${\rm log}(g)$ = 3, [Fe/H] = 1 planet
with C/O = 0.55 ({\it upper panel}) and C/O = 1.05 ({\it lower panel}). The spectra are shown for a F5 (blue lines), G5 (purple lines),
K5 (red lines) and M5 (orange lines) host star. The colored bars indicate the position of the absorption maxima of various species.
The black line shows the blackbody flux at the atmosphere's effective temperature.}
\label{fig:spec_spT}
\end{figure}
Only once we switched to the ExoMol line list
for HCN we got the results presented in this paper, where inversions only occur for
C/O $\sim$ 1.
The ExoMol line list is much more complete for HCN, containing many more lines.
\rch{The line list is made specifically for high temperatures, optimized for temperatures up to 3000 K
and compares well to a high temperature laboratory measurement made at $T$ = 1370 K \citep{barber2014}.\footnote{More
comparisons are not possible as there are not many high temperature measurements for this molecule.}}
This allows the atmospheres to cool more efficiently, making the inversions go away
in many cases.

\begin{figure*}[htb!]
\centering
\begin{minipage}{0.48\textwidth}
\includegraphics[width=1.\textwidth]{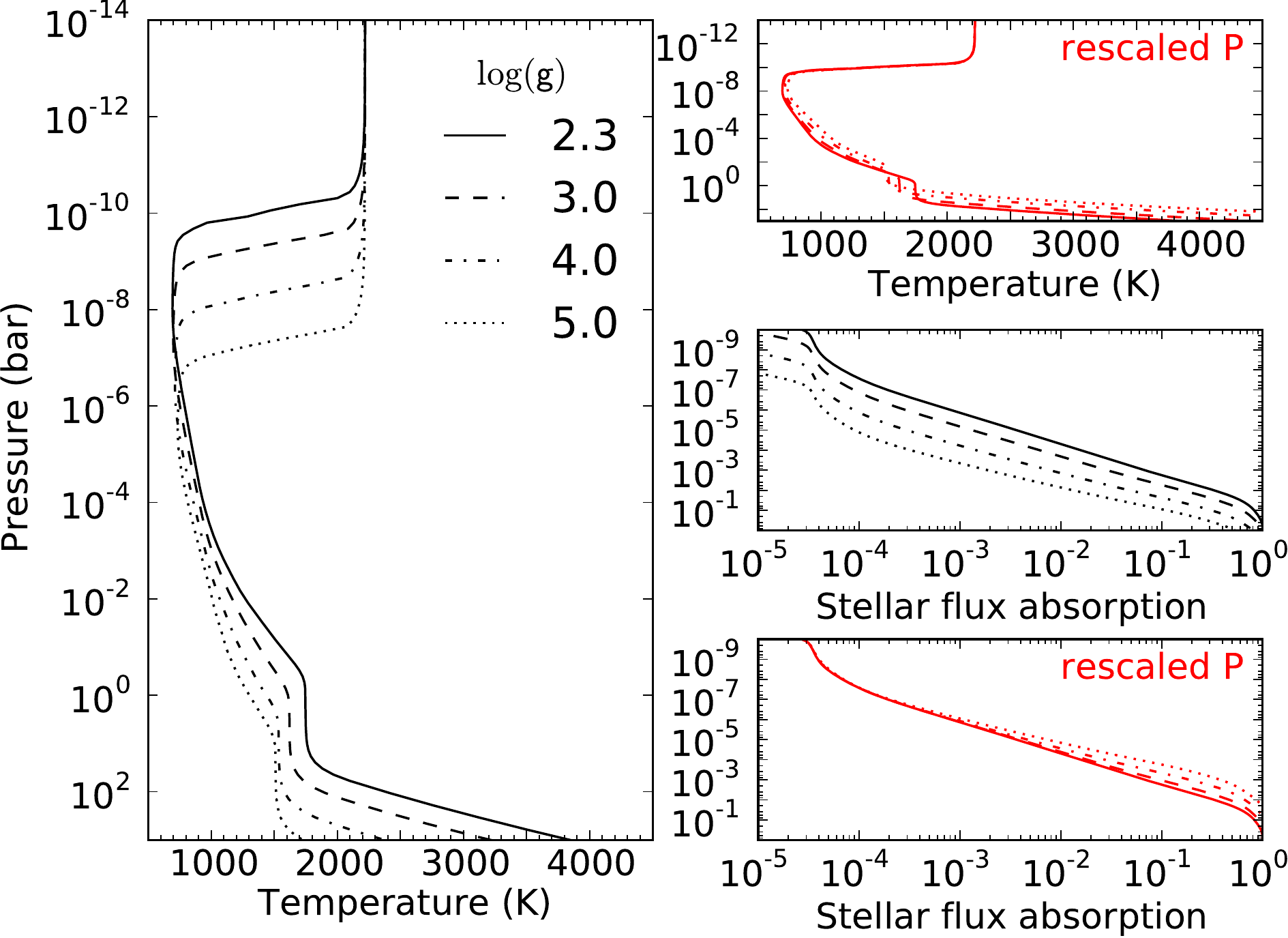}
\caption{Atmospheric $PT$-structures and stellar light absorption as a function of ${\rm log}(g)$ for planets with
$T_{\rm eff}$ = 1250 K, [Fe/H] = 1 and C/O = 0.55 in orbit around a G5 star. The linestyles correspond to ${\rm log}(g)$ =
2.3 (solid line), 3.0 (dashed line), 4.0 (dot-dashed line), 5.0 (dotted line). {\it Left panel}: $PT$-structures. {\it Top right panel}:
$PT$-structures with pressure rescaled by $10^{2.3-{\rm log}(g)}$. {\it Middle right panel}: Fraction of absorbed stellar flux
as a function of pressure. {\it Bottom right panel}: Fraction of absorbed stellar flux as a function of rescaled pressure.}
\label{fig:loggPT}
\end{minipage}
\begin{minipage}{0.03\textwidth}
\end{minipage}
\begin{minipage}{0.48\textwidth}
\includegraphics[width=1.\textwidth]{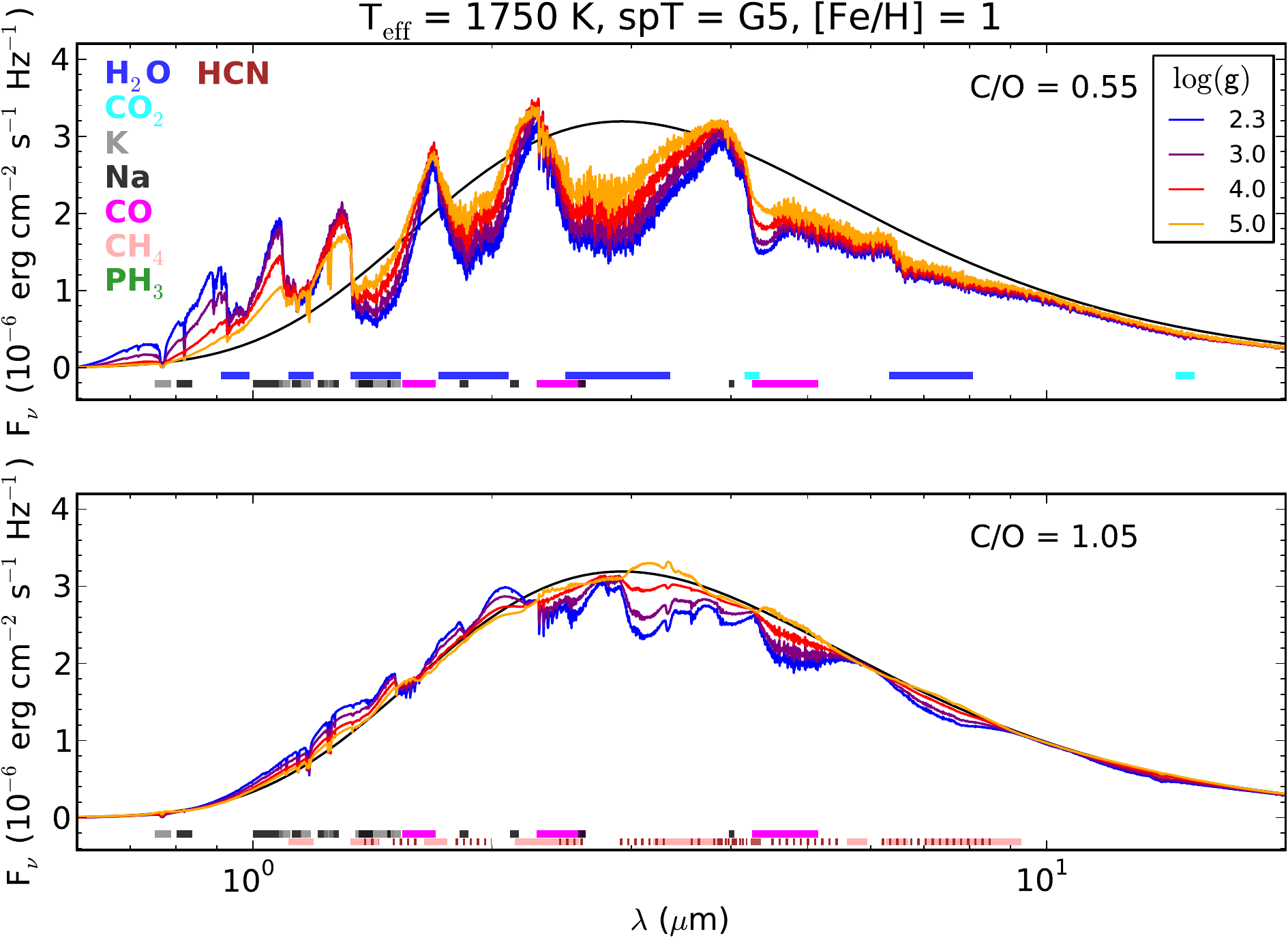}
\caption{Emission spectra as a function of surface gravity for a $T_{\rm eff}$ = 1750 K, [Fe/H] = 1 planet
with C/O = 0.55 ({\it upper panel}) and C/O = 1.05 ({\it lower panel})  in orbit around a G5 star.
The spectra are shown for ${\rm log}(g)$ = 2.3 (blue lines), ${\rm log}(g)$ = 3 (purple lines),
${\rm log}(g)$ = 4 (red lines) and ${\rm log}(g)$ = 5 (orange lines).
The colored bars indicate the position of the absorption maxima of various species.
The black line shows the blackbody flux at the atmosphere's effective temperature.}
\label{fig:spec_logg}
\end{minipage}
\end{figure*}

Likewise, we want to stress that we use the \emph{HITRAN} line list for the C$_2$H$_2$ molecule,
as an ExoMol version is not available.
C$_2$H$_2$ is quite common in our results for C/O $\gtrsim$ 1 in the cases where HCN is common as well. 
This suggests that the atmospheres ability to cool might be further enhanced
if high temperature line lists for C$_2$H$_2$ were to be considered.

\subsection{Host star dependance of the atmospheres}
\label{sect:host_dep}
\subsubsection{Spectra}
As described in Section \ref{sect:first_glance} planets orbiting increasingly cooler host stars will approach an
increasingly isothermal atmospheric structure, because the spectral energy distribution of the insolation becomes
more and more comparable to the SED of the planetary radiation field.

We show the emission spectra of atmospheres with varying host star spectral type for a planet with $T_{\rm eff}$ = 1750 K,
${\rm log}(g)$ = 3, [Fe/H] = 1 for two different C/O ratios (0.55, 1.05) in Figure \ref{fig:spec_spT}.
We indicate the positions of absorption features of H$_2$O, CO$_2$, K, Na, CO, CH$_4$, PH $_3$ and HCN in the plots.
For the atmospheres with C/O = 0.55 the emission spectra
clearly become more blackbody-like as the host star gets cooler:
the excess emission (with respect to the blackbody curve at 1750 K) of the atmospheres for $\lambda$ $<$ 1.3 $\mu$m
decreases for cooler host stars. Furthermore the molecular absorption bands in the emission spectra start to get shallower.
As expected for a C/O ratio $<$ 1, the spectra are clearly water-dominated.

For the atmospheres with C/O = 1.05 the situation is somewhat different.
First, the atmospheres are clearly carbon-dominated, showing strong HCN features.
Moreover, the latest type host star (M5) causes the least isothermal planetary spectrum,
while all earlier type host stars result in a much
more isothermal atmospheric structure and, therefore, spectra.
This is the contrary of what we saw for the C/O = 0.55 atmosphere,
now host stars of an earlier type are making the planetary spectra more isothermal.
This is merely the spectral consequence of early type host stars creating inversions or isothermal atmospheres
for planets with C/O $\sim$ 1, which we explained in Section \ref{sect:highCOinv}.
As the M5 star is not able to heat the atmosphere enough due to a lack of energy in the optical wavelengths
the corresponding $PT$-structure and spectra are less isothermal.
The $PT$-structures producing the spectra shown here for C/O = 1.05 do not have inversions,
they are just more isothermal due to the heating.
As we will see in Section \ref{sect:HT_atmos}, atmospheres at $T_{\rm eff}$ = 1750 K can, in general, exhibit inversions.

\subsection{${\rm log}(g)$ dependence of the atmospheres}
\label{sect:logg_PT_dep}

\subsubsection{$PT$-structures}

The behavior of the $PT$-structures as a function of ${\rm log}(g)$ is studied in Figure {\ref{fig:loggPT}}.
If one considers gray opacities which are constant as a function of $P$ and $T$ and assumes hydrostatic
equilibrium one obtains the following simple relation between the optical depth $\tau$ and the pressure $P$
\beq
\tau = \frac{\kappa}{g}P \ ,
\eeq
where $\kappa$ is the gray opacity and g is the gravitational acceleration (taken to be constant).
In this case, changing the gravitational acceleration will conserve the
temperature structure as a function of $\tau$, as $\tau$ is the effective spatial coordinate for the radiation field.
The mapping from $\tau$ to $P$, however, will change, resulting in locations of a given optical depth
and temperature to move to larger pressure values when $g$ is increased.
This is equivalent to saying that the location of the planetary atmospheric photosphere moves
in terms of pressure if the surface gravity is changed.

Thus, when plotting the $PT$-structures as a function of planetary gravitational acceleration,
as can be seen in Figure {\ref{fig:loggPT}},
one notices that at higher ${\rm log}(g)$ the temperature structure appears to be shifted to larger pressures
when comparing to cases with lower ${\rm log}(g)$.
For demonstration purposes we show the $PT$-structures up to 10$^{-14}$ bar.
Note, however, that we only calculate the opacities down to pressures of 10$^{-6}$ bar
and adopt the 10$^{-6}$ bar values at all smaller pressures,
i.e. $\kappa(P<10^{-6} \ {\rm bar}) = \kappa(P=10^{-6} \ {\rm bar})$.
\rch{The $PT$-structures for pressures below 10$^{-6}$ bar are not necessarily unphysical, however (see Section \ref{sect:grid_calc_descr} for a discussion).
The ``highest altitude inversion'' visible in this plot for pressures much smaller than 10$^{-6}$ bar
is due to the heating by the alkali line cores.}

In the top right panel of Figure {\ref{fig:loggPT}} we show the $PT$-structures once more.
In this case we have re-scaled the pressures in $PT$-structures with ${\rm log}(g)$ higher than 2.3
(which is the lowest ${\rm log}(g)$ value we consider)
with $10^{2.3-{\rm log}(g)}$. To first order, his should counterbalance the pressure shift of the temperature structure
induced by gravity when compared to the ${\rm log}(g)$ = 2.3 case.
However, as the opacities are non-gray and varying vertically we expect differences.
Nonetheless, the resulting $PT$-structures lie on top of each other quite well.

When comparing in greater detail one notices that the deep isothermal regions (at $\sim$ 1-100 bars) are at higher temperatures for lower
${\rm log}(g)$. Here the pressure dependence of the opacity comes into play:
for lower ${\rm log}(g)$ values the stellar light is absorbed at lower pressures, where the atomic and
molecular lines are less broadened.
This results in the stellar light being able to penetrate deeper in terms of rescaled pressure when comparing
to high ${\rm log}(g)$ atmospheres.
This means that more stellar light reaches regions of the atmosphere which are optically thick in the near-infrared,
which does, in turn, heat up the atmosphere deep in these IR optically thick regions.

In the middle and bottom panel on the right side of Figure {\ref{fig:loggPT}} we show the fraction of the absorbed
stellar flux with respect to the stellar flux at the top of the atmosphere. The middle panel shows this fraction
as a function of pressure, the bottom panel shows this fraction as a function of rescaled pressure.
One sees that the stellar light is able to penetrate deeper in terms of rescaled pressure
in the case of low ${\rm log}(g)$.

In Figure {\ref{fig:loggPT}} we have shown an oxygen-dominated atmosphere, where the abundance
of the main coolant and absorber, H$_2$O, is
roughly independent of pressure.
For carbon rich atmospheres the pressure dependent abundances of H$_2$O, CH$_4$ and HCN
might play a role in addition to the pressure shift of the temperature structures.

In order to test that our above observations for the oxygen rich atmosphere are not caused by pressure and temperature
dependent chemistry effects, we calculated self-consistent structures with vertically constant abundances
of molecules and varied the surface gravity.
\begin{figure*}[htb!]
\begin{minipage}{0.48\textwidth}
\centering
\includegraphics[width=1.\textwidth]{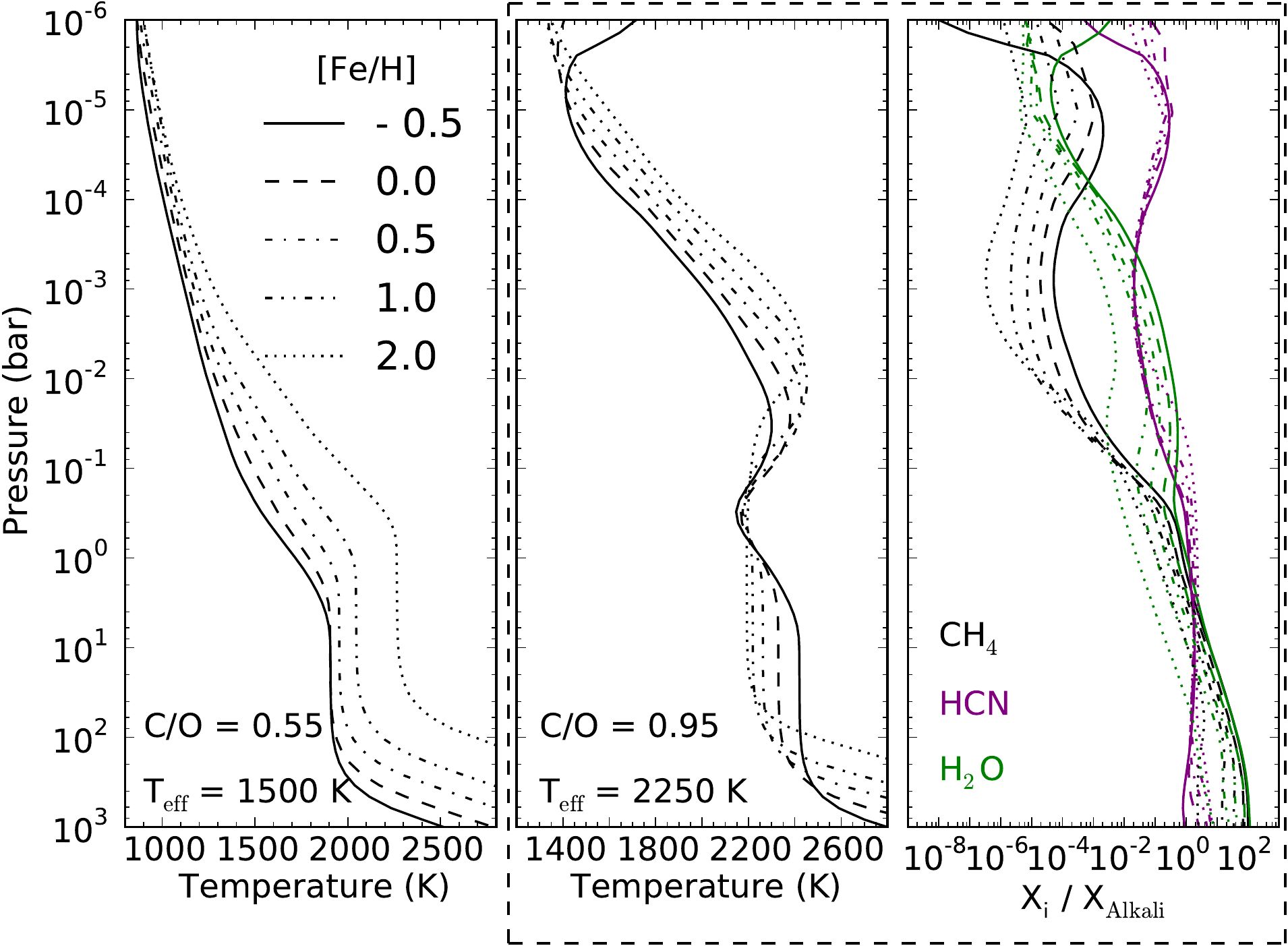}
\caption{Atmospheric $PT$-structures and mass fractions as a function of metallicity for ${\rm log}(g)=3$ planets
around a G5 star. The {\it left panel} shows the $PT$-structures of the cases with C/O = 0.55, $T_{\rm eff}$ = 1500 K planets,
the {\it middle panel} shows the cases with C/O = 0.95, $T_{\rm eff}$ = 2250 K. The {\it right panel} shows the mass
fractions of CH$_4$ (black lines), HCN (purple lines) and H$_2$O (green lines) divided by the alkali mass fraction for the
$PT$-structures shown in the middle panel. The different line styles in all panels stand for different metallicities:
[Fe/H] = -0.5 (solid line), 0.0 (dashed line), 0.5 (dot-dashed line),
1.0 (double dotted dashed line), 2.0 (dotted line).}
\label{fig:met_PT}
\end{minipage}
\begin{minipage}{0.03\textwidth}
\end{minipage}
\begin{minipage}{0.48\textwidth}
\centering
\includegraphics[width=1.\textwidth]{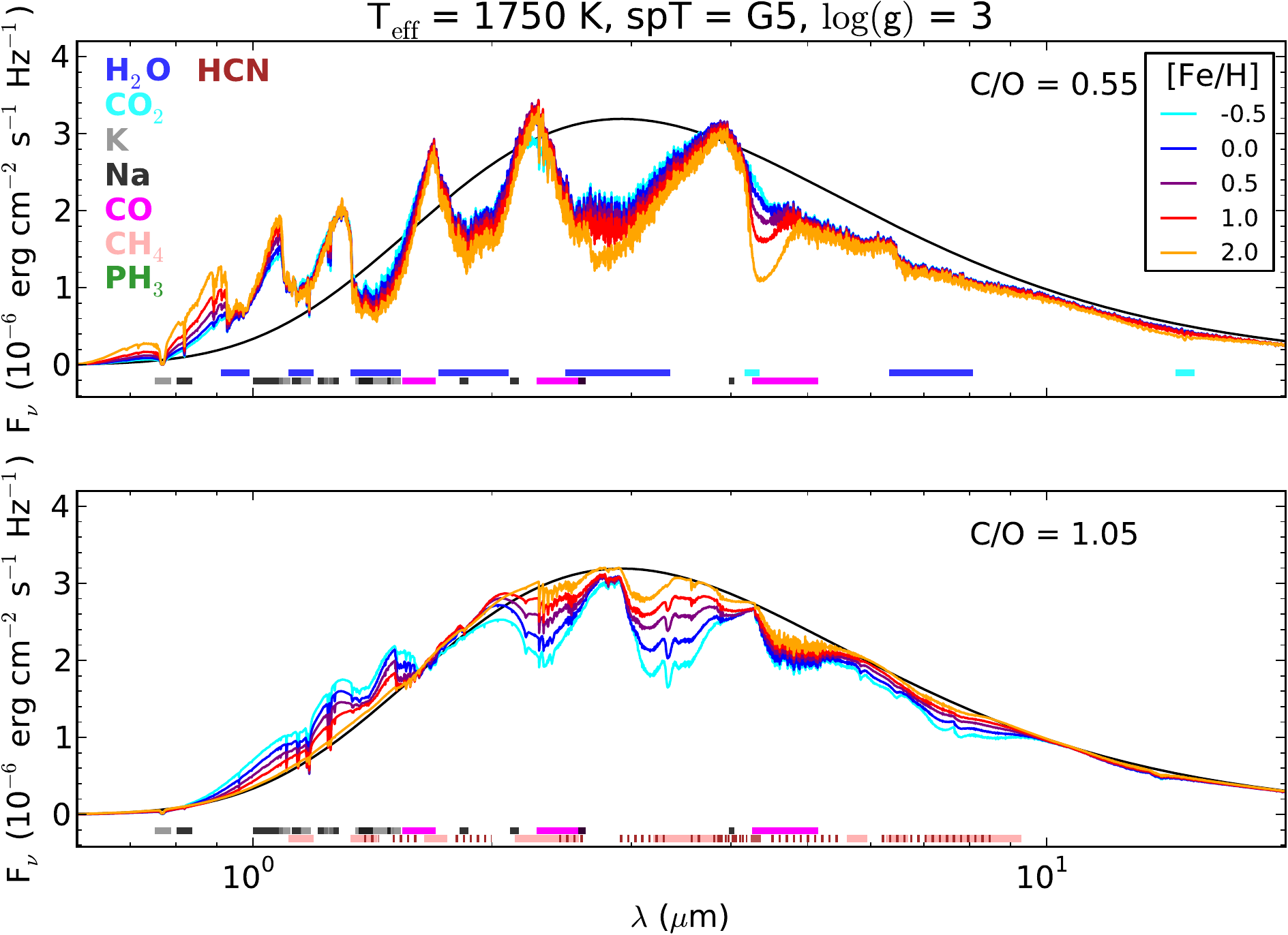}
\caption{Emission spectra as a function of metallicity for a $T_{\rm eff}$ = 1750 K, [Fe/H] = 1 planet
with C/O = 0.55 ({\it upper panel}) and C/O = 1.05 ({\it lower panel})  in orbit around a G5 star.
The spectra are shown for [Fe/H] = -0.5 (cyan lines), [Fe/H] = 0.0 (blue lines), [Fe/H] = 0.5 (purple lines),
[Fe/H] = 1 (red lines) and [Fe/H] = 2 (orange lines).
The colored bars indicate the position of the absorption maxima of various species.
The black line shows the blackbody flux at the atmosphere's effective temperature.}
\label{fig:spec_feh}
\end{minipage}
\end{figure*}
We found the same behavior of the structures as described above, verifying that the pressure
dependent line wing strengths are responsible.

\subsubsection{Spectra}
In Figure \ref{fig:spec_logg} we show the emission spectra of atmospheres with varying surface gravity for a planet with $T_{\rm eff}$ = 1750 K,
and [Fe/H] = 1 in orbit around a G5 host star, again for two different C/O ratios (0.55, 1.05). As mentioned above, a
variation in the surface gravity rescales the temperatures profiles in terms of pressure. We also found that the deep isothermal regions are hotter
for the lower surface gravity cases, because the insolation can probe deeper into the atmosphere. In the pressure rescaled $PT$-structures
(see upper right panel of Figure \ref{fig:loggPT}) one can see that above the isothermal region the atmospheres of planets with higher surface gravity
are hotter for a given rescaled pressure: The photosphere is located at higher pressures for a higher surface gravity. It is therefore less
transparent, due to the line wing pressure broadening. In order to radiate away the required amount of flux the temperature therefore
needs to be higher.
The flux in the absorption features then originates in hotter regions, making the absorption troughs shallower in the C/O = 0.55 case.
This behavior was verified by the atmospheric structures with vertically constant molecular abundances as well.

In the C/O = 1.05 case the same behavior can be seen, except for the atmospheres with the highest ${\rm log}(g)$, which shows
emission features. Here the stellar light is absorbed over narrower and higher rescaled pressure ranges because the alkali line wings are much broader (the light is absorbed at higher actual pressure). The atmospheric cooling ability, however, is largely independent of pressure, because the emission of light depends on the Planck opacity $\kappa_{\rm P}$ and $\partial \kappa_{\rm P} / \partial P = 0$,
if the pressure dependence of the chemistry is omitted.ÊThis causes the atmosphere at highest ${\rm log}(g)$ to develop an
inversion.

\subsection{Metallicity dependence of the atmospheres}
\subsubsection{$PT$-structures}

The influence of the metallicity on the $PT$-structures at low C/O ratios is as one would expect:
An increased [Fe/H] value in atmospheres leads to higher temperatures in the deep isothermal part of the atmosphere
in the cases where no inversions are observed: the temperature structure is scaled to lower pressures as the metallicity
increases, as a higher optical depth is reached earlier in the atmosphere. The stellar light can penetrate deeper
than suggested by a simple pressure scaling, however: the pressure dependent line wings are weaker
(as the atmospheric structures shift to smaller pressures for higher metallicities).
This increases the temperature of the atmospheres in the deep isothermal regions at 1-100 bars
(see left panel of Figure \ref{fig:met_PT}), just like it did for low surface gravities studied in Section \ref{sect:logg_PT_dep}.
Similar to the test carried out for varying surface gravities in
Section \ref{sect:logg_PT_dep} we calculated test atmospheres with vertically constant molecular abundances,
scaling the abundances by different factors for different structures, mimicking variations in metallicity
without having to deal with effects introduced by chemistry. These calculations showed the same behavior as the nominal
calculations when varying the metallicity.

\begin{figure*}[t!]
\centering
\begin{minipage}{0.48\textwidth}
\includegraphics[width=1.0\textwidth]{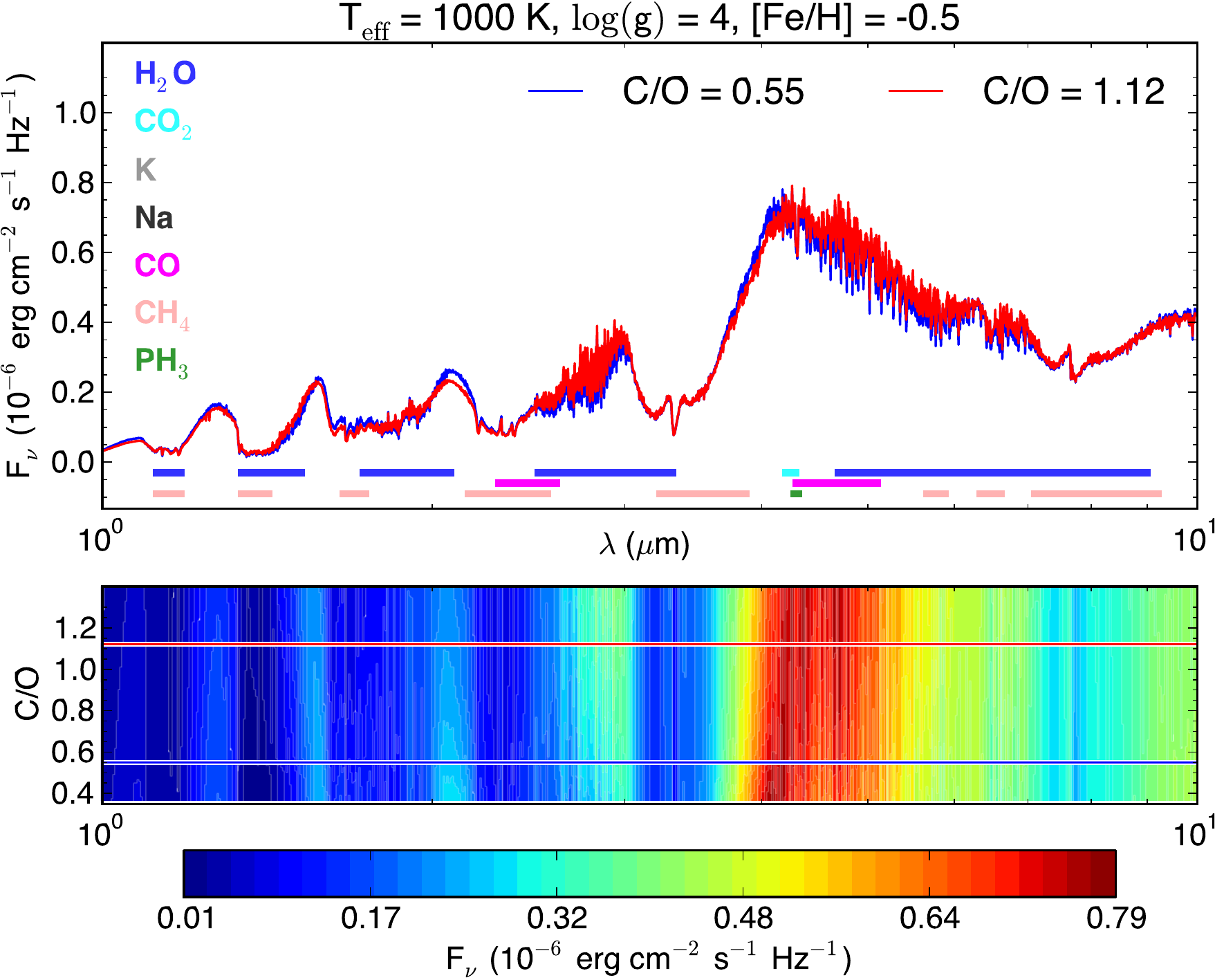}
\end{minipage}
\begin{minipage}{0.03\textwidth}
\end{minipage}
\begin{minipage}{0.48\textwidth}
\includegraphics[width=1.0\textwidth]{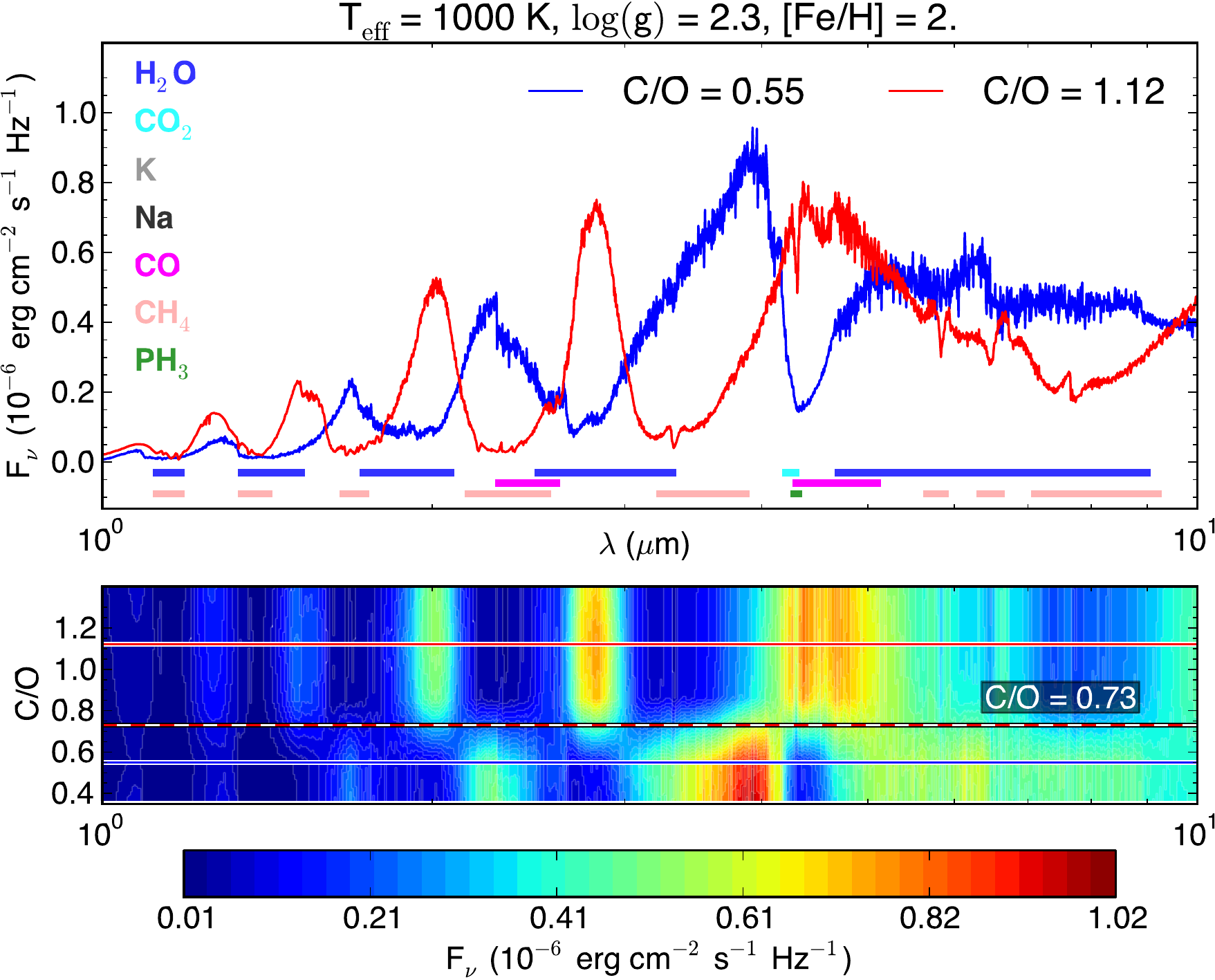}
\end{minipage}
\caption{Atmospheric emission spectra of planets in orbit around a G5 star with $T_{\rm eff}$ = 1000 K.
{\it Left panel:} Planets with ${\rm log}(g)$ = 4, [Fe/H] = -0.5. {\it Right panel:} Planets with
${\rm log}(g)$ = 2.3, [Fe/H] = 2. {\it Upper subpanels:} Emission spectra as a function of wavelength
for planets with C/O = 0.55 (blue solid line) and C/O = 1.12 (red solid line). The absorption bands of
dominant absorbers are indicated by the colored bars below the spectra. {\it Lower subpanels:}
Emission spectra as a function of wavelength ($x$-axis) and C/O ratio ($y$-axis).
The flux values are indicated as a color map. The red-white dashed horizontal line in the right
panel indicates the C/O value where the atmosphere switches from being rich in water to being
methane-rich. The corresponding C/O value of this transition is indicated in the plots.
The red and blue horizontal lines indicate the C/O values of the wavelength
dependent spectra shown in the upper subpanels.}
\label{fig:low_temp_co_diff_sim}
\end{figure*}

In the case of $PT$-structures with C/O $\sim 1$, which have inversions, the inversion temperature increases
and the region directly beneath (i.e. at higher pressure) the inversions has a lower temperature if the
metallicity increases (see middle panel of Figure \ref{fig:met_PT}).
It is, at first, not evident why this should happen, because if all the metal atomic abundances scale with 10$^{\rm [Fe/H]}$ one would expect the same for the resulting molecular abundances and opacities,
and therefore the heating vs. cooling ability of the atmosphere should stay the same.
This interpretation is consistent with the analytical double-gray atmospheric models as published, e.g., by
\citet{guillot2010,hansen2008,thomas2002}, where the inversion temperature should stay constant unless the ratio
\beq
\gamma = \frac{\kappa_{\rm vis}}{\kappa_{\rm IR}}
\eeq
changes, where $\kappa_{\rm vis}$ and $\kappa_{\rm IR}$ are the mean opacities in the
visual and IR wavelengths in the atmosphere. The behavior we see in the atmospheres
suggests that
\beq
\frac{{\rm d} \gamma}{{\rm d [Fe/H]}} > 0 \ ,
\eeq
which should only be possible if $\kappa_{\rm vis}$ and $\kappa_{\rm IR}$ (and the molecular
abundances giving rise to these opacitites) do not just
simply scale linearly with metallicity.
In order to test this we checked the abundances of the major absorbers and emitters
as a function of metallicity throughout the atmospheres for the $PT$-structures shown
in the middle panel of Figure \ref{fig:met_PT}.
Indeed we found that the ratios of mass fractions $X_{\rm H_2O}/X_{\rm Alkali}$ and
$X_{\rm CH_4}/X_{\rm Alkali}$ decreased when the metallicity was increased
(see right panel of Figure \ref{fig:met_PT}). $X_{\rm HCN}/X_{\rm Alkali}$ increases,
at the relevant temperatures already being the dominant carbon opacity carrier.
However, the increase in $X_{\rm HCN}/X_{\rm Alkali}$ is apparently not enough to
act as a counterweight compensating the loss of infrared opacity due to the lower
$X_{\rm H_2O}/X_{\rm Alkali}$. This leads to less efficient cooling as [Fe/H] increases.
This abundance change is likely caused by the pressure dependence of the chemistry,
as higher metallicities shift the temperature structure to smaller pressures, where,
for carbon-dominated atmospheres, CH$_4$ and H$_2$O are less abundant,
while the HCN abundance increases.
\subsubsection{Spectra}
Analogous to the ${\rm log}(g)$ case an increase in metallicity (and thus opacity) can be regarded as a similar pressure rescaling, as we found
for a gray atmosphere with vertically constant opacity $\kappa$ that $\tau=\kappa/g P$.
As $\kappa$ is in the numerator, atmospheric structures with increased metallicity should behave similarly to structures with {\it de}creased
surface gravity, featuring a higher temperature in their isothermal regions, but a lower temperature (as a function of rescaled pressure) in the higher
atmosphere: Because the photosphere will be located at smaller pressures (in actual, non-rescaled pressure) for an increased metallicity,
the line wings will be less strong (less pressure broadening).
The atmosphere is therefore more transparent and cools better. In order to radiate
away the imposed flux, the temperature in this more transparent photosphere needs thus to be decreased.
The minima in the spectrum, stemming from the opacity maxima, i.e. the line's Gauss-cores, will originate from the same region in terms of rescaled pressure. As these pressures are now at a lower temperature, this leads to deeper absorption troughs in the spectra.
This can be seen in the upper panel of Figure \ref{fig:spec_feh} and was confirmed by the vertically constant molecular abundance calculations as well,
when rescaling the abundances as described above.
In summary, more pronounced absorption troughs can mean either a lower surface gravity or a higher metallicity (see figures \ref{fig:spec_feh} and \ref{fig:spec_logg}).

In the C/O = 1.05 case we can again draw on our studies of the $PT$-structures:
we saw that for atmospheres with inversions, due to the chemistry involved,
the cooling ability of the atmospheres relative to the heating by the alkalis decreases if the metallicity is increased (see middle 
and right panel of Figure \ref{fig:met_PT}).
The spectra shown in the lower panel of Figure \ref{fig:spec_feh}, although they do not exhibit inversions,
are consistent with these observations, showing absorption spectra which become more and more isothermal 
as the metallicity is increased.

\subsection{Low temperature atmospheres ($T_{\rm eff} \lesssim 1250$ K)}
\label{sect:LT_atmos}
At low enough temperatures ($T_{\rm eff} \lesssim 1250$ K) HCN does not yet play a
significant role for the atmospheric spectra.
Additionally the left pointing arrow of the chemical
reaction in equation (\ref{equ:COchem}) can still be of importance, meaning that H$_2$O and
CH$_4$ are significant carriers of C and O atoms.
It is important to note, however, that the chemical equilibrium abundances are not only temperature,
but also pressure dependent:
\begin{figure*}[htb!]
\centering
\begin{minipage}{0.48\textwidth}
\includegraphics[width=1.0\textwidth]{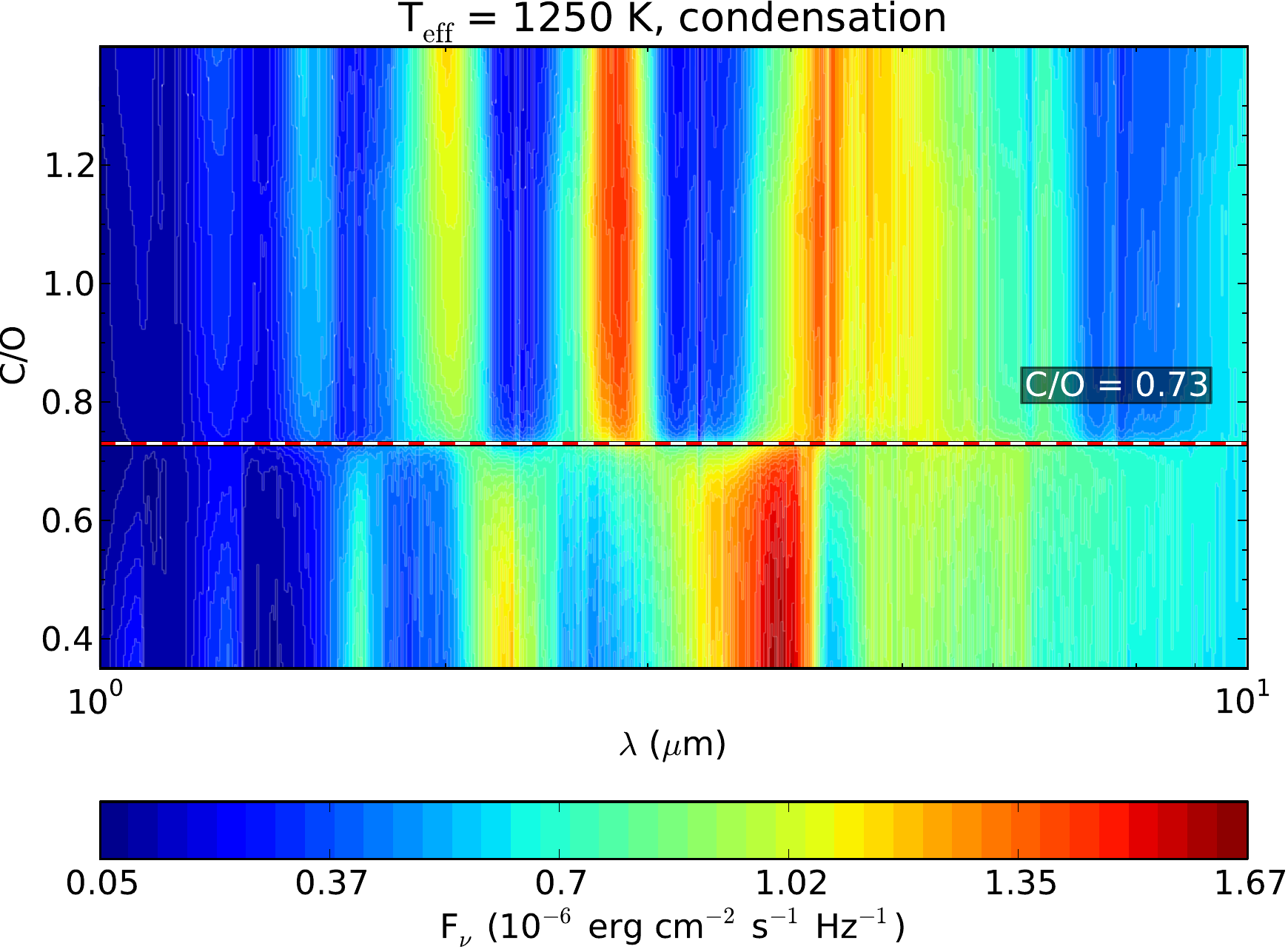}
\end{minipage}
\begin{minipage}{0.03\textwidth}
\end{minipage}
\begin{minipage}{0.48\textwidth}
\includegraphics[width=1.0\textwidth]{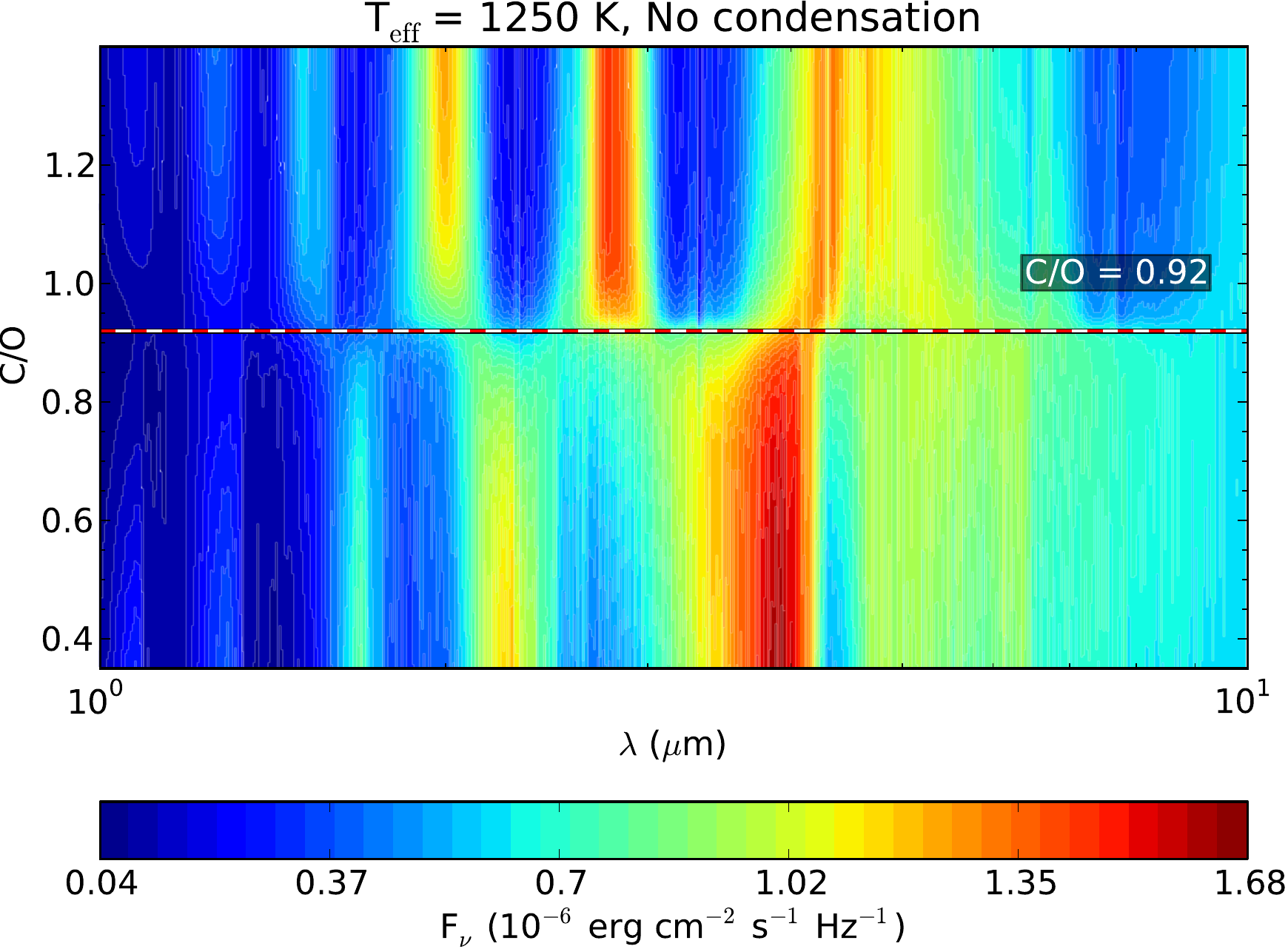}
\end{minipage}
\caption{Emission spectra as a function of wavelength ($x$-axis) and C/O ratio ($y$-axis) of planets with
$T_{\rm eff}$ = 1250 K, ${\rm log}(g)$ = 3, [Fe/H] = 1 in orbit around a G5 star.
The flux values are indicated as a color map.
The red-white dashed horizontal lines indicate the C/O values where the atmospheres switch from being rich in
water to being methane-rich. The corresponding C/O value of this transition is indicated in the plots.
{\it Left panel}: Nominal chemical model (including condensation), {\it right panel}:
Chemical model without condensation.}
\label{fig:cond_change}
\end{figure*}
at low temperatures and high pressures CH$_4$ will be abundant also in an oxygen-rich
atmosphere, while H$_2$O will be abundant in carbon-rich atmospheres.
At low temperatures and low pressures CO will become increasingly important, such that the oxygen-rich
atmospheres do not contain a lot of methane and the carbon rich ones do not contain a lot of
water.

As seen in the above discussions, [Fe/H] and ${\rm log}(g)$ can strongly influence to which 
pressure levels the optical depth-dependent temperature structure will be be scaled, as for a gray
atmosphere it would hold that $\tau = \kappa /g P$. Therefore, low metallicity atmospheres (causing
a small $\kappa$) at high surface gravities cause the temperature structure to be scaled to high pressures.
In Figure \ref{fig:low_temp_co_diff_sim} we show emission spectra of planets with $T_{\rm eff}$ = 1000 K
in orbit around a G5 star. The spectra are shown for C/O = 0.55 and 1.12 in the upper subpanels. 
Furthermore we indicate the positions of absorption features of H$_2$O, CO$_2$, K, Na, CO, CH$_4$ and PH$_3$ in the plots.
The left panel shows the emission spectra for planets with ${\rm log}(g)$ = 4, [Fe/H] = -0.5. This means
that here the surface gravity is high and the metallicity is low, causing the temperature structures to be
scaled to high pressures.
The right panel shows planets with ${\rm log}(g)$ = 2.3, [Fe/H] = 2, i.e. with
low surface gravities and high metallicities, leading to temperature structures to be scaled to low pressures.
In the lower subpanels we show color maps of emission spectra as a function of wavelength ($x$-axis) and C/O ratio ($y$-axis).

In the right upper subpanel, one sees that the two spectra are very different, showing either water
or methane features for the atmospheres with C/O =0.5 or 1.12, respectively.
As described above, this is expected, corresponding to a low pressure scaling of the temperature
structure \rch{and due to the pressure dependence of the CO--CH$_4$--H$_2$O chemistry}.
In the lower right subpanel there is an overall shift from H$_2$O to CH$_4$ dominated spectra at C/O
$\sim$ 0.73.

As expected, in the left panel there is only little difference between the oxygen-rich and \rch{carbon}-rich case.
Further, the lower left subpanel does not show any transition between a water- and methane-dominated
atmosphere, as both molecules are present in the atmospheres at all C/O ratios.
\rch{Once more, this is expected, as in this case, i.e. for low metallicity and high ${\rm log}(g)$ the photosphere of the atmosphere is scaled to high
pressures, where the chemistry dictates that CO is not the major carbon and oxygen carrier, but instead CH$_4$ and H$_2$O dominate,
at least at the low atmospheric temperatures considered here. Therefore, although the CH$_4$/H$_2$O number ratio
may change as a function of C/O, this change is not sufficient to affect the spectrum significantly.}

Therefore, the spectral appearance of a planet is not only given by the C/O ratio and the effective
temperature but also by a factor
\beq
\beta = {\rm[Fe/H]} - {\rm log}(g) \ ,
\label{equ:beta_factor}
\eeq
which is a measure for the optical depth -- pressure mapping in the atmospheres and gives
insight to which pressure levels a given atmospheric temperature profile $T(\tau)$ is scaled.
We found that transitions between water- and methane rich atmospheres occur at
$\beta \gtrsim -4$ or $-3.5$ for $T_{\rm eff}$ = 1000 K.
For $T_{\rm eff}$ = 1250 K we found that $\beta \gtrsim -5.0$, indicating that a transition between
water and methane dominated spectra should always be expected at these temperatures.
However, values of $\beta$ close to this threshold should always exhibit some methane or water features,
even if the atmosphere is water or methane dominated, respectively. 

\subsubsection{C/O dependence with and without condensation}
\label{sect:co_dep_cond}

For atmospheres with effective temperatures $\lesssim$ 1750 K, the spectrally active parts of the atmosphere
have temperatures low enough for the condensation of MgSiO$_3$ \citep[for the temperature
dependent saturation vapor pressure of MgSiO$_3$ see, e.g.,][]{ackerman2001}.

Condensation of O in MgSiO$_3$ does not have a too strong effect on the spectra in the sense that they are either water or methane dominated at high enough temperatures, i.e. $T_{\rm eff} >$ 1000 K. It does shift the C/O ratio-dependent transition between the 2 cases, however, as we detail below.
Note that we do not include cloud opacities yet,
so the presence of MgSiO$_3$ grains will not affect the radiation field.

For atmospheres with C/O values in the vicinity to, but less than 1, the condensation of MgSiO$_3$ decreases the amount
of oxygen available to form CO and H$_2$O considerably.
\begin{figure*}[htb!]
\centering
\begin{minipage}{0.48\textwidth}
\includegraphics[width=1\textwidth]{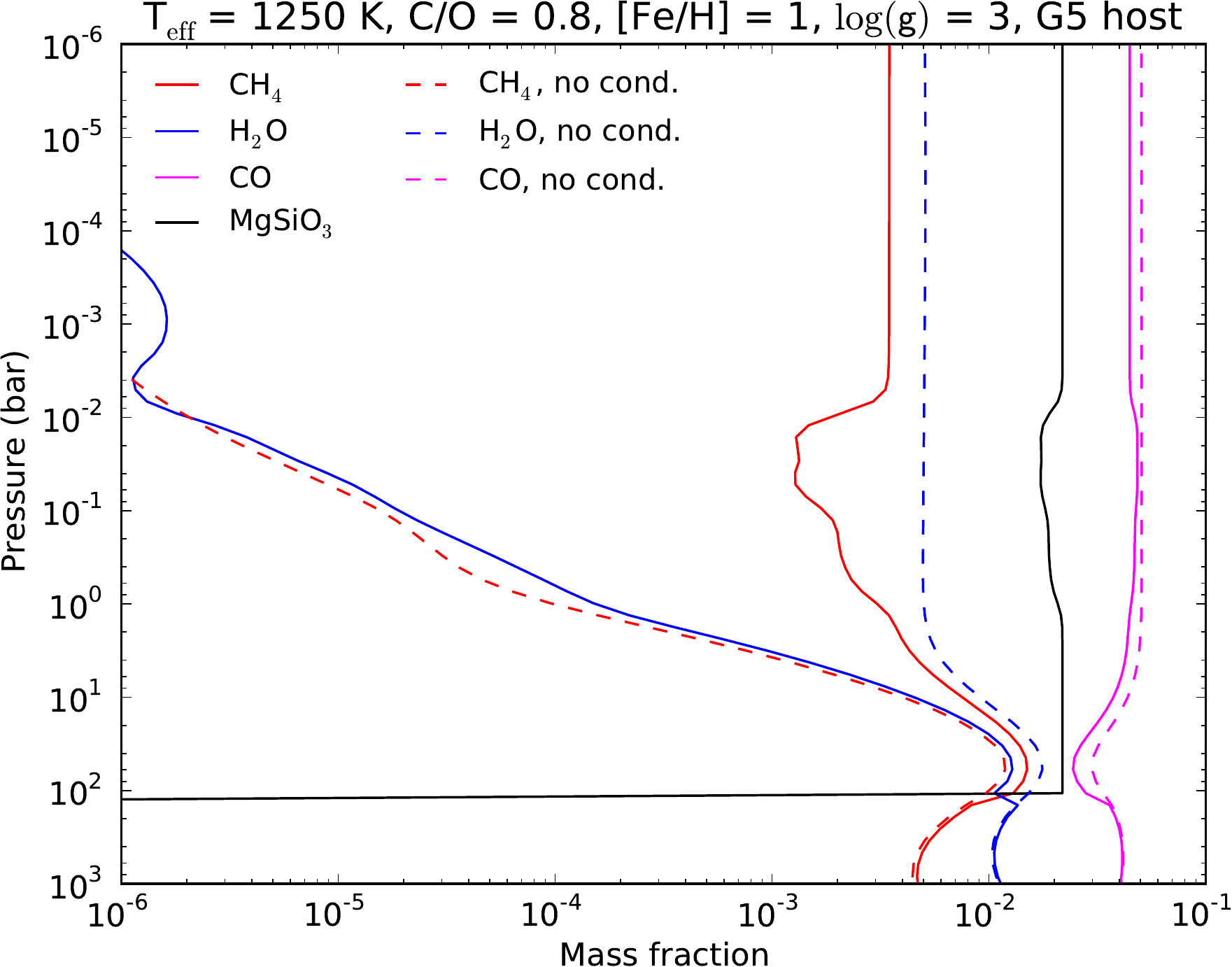}
\caption{Mass fractions of components in the atmosphere of a planet with C/O = 0.8,
$T_{\rm eff}$ = 1250 K, ${\rm log}(g)$ = 3, [Fe/H] = 1
in orbit around a G5 star. The solid lines show the mass fractions of H$_2$O (blue), CH$_4$ (red), CO (magenta)
and MgSiO$_3$ (black) for our nominal model, including condensation, while the dashed lines show the results
for an atmosphere without condensation.}
\label{fig:cond_abund}
\end{minipage}
\begin{minipage}{0.03\textwidth}
\end{minipage}
\begin{minipage}{0.48\textwidth}
\includegraphics[width=1\textwidth]{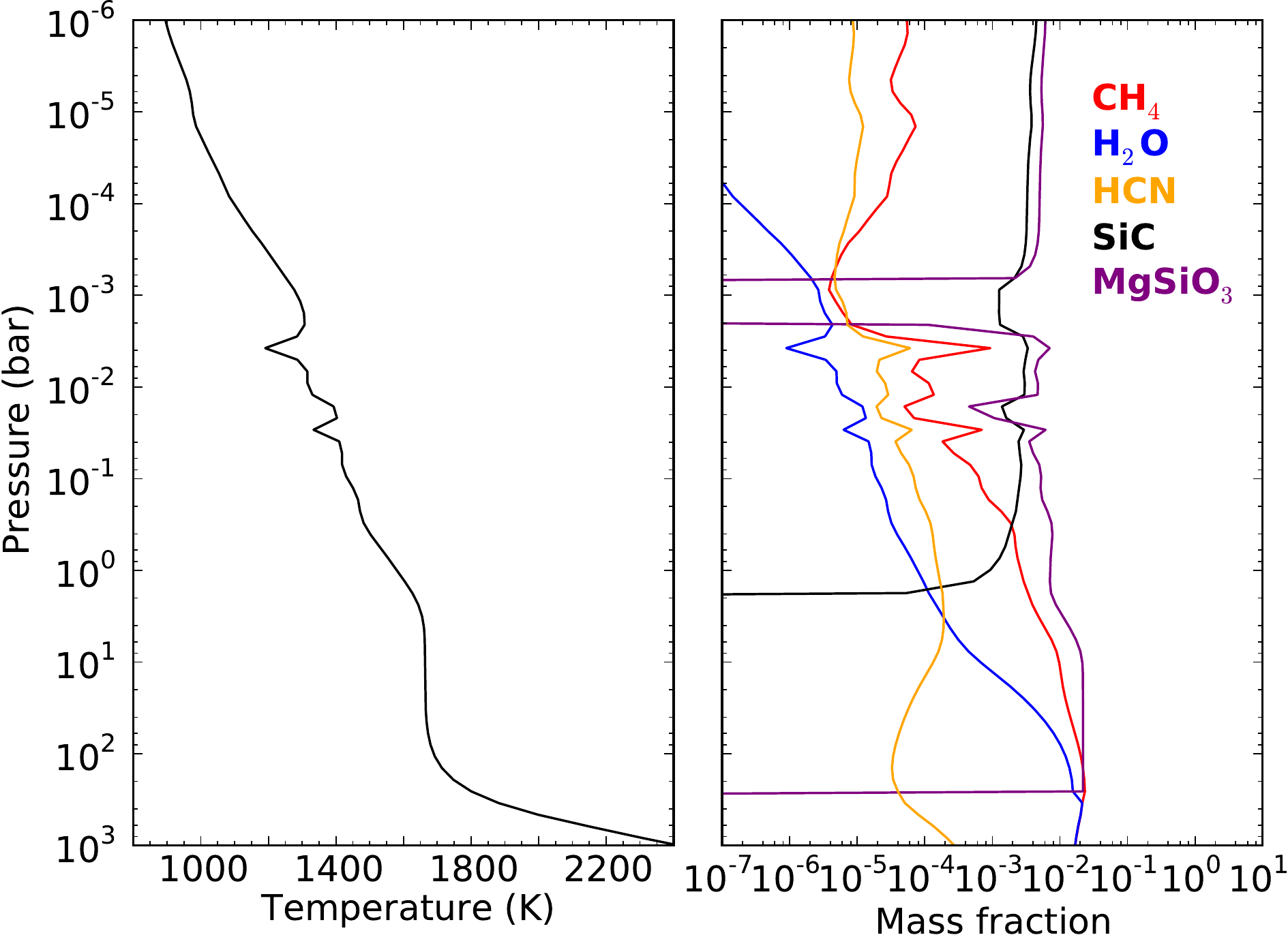}
\caption{{\it Left panel:} $PT$-structure of the atmosphere of a planet with $T_{\rm eff}$ = 1500 K, ${\rm log}(g)$ = 4, [Fe/H] = 1,
C/O = 0.95 in orbit around a G5 star. {\it Right panel:} Mass fractions of CH$_4$ (red solid line), H$_2$O (blue solid line),
HCN (orange solid line), SiC (black solid line) and MgSiO$_3$ (purple solid line) as a function of pressure for the $PT$-structure shown in the left panel.}
\label{fig:t_1500_kinks}
\end{minipage}
\end{figure*}
In turn the H$_2$O features in the spectra will weaken and
CH$_4$ can form in noticeably higher abundances as C atoms are more available due to the lower amount of CO
being formed.

This results in shifting the transition from H$_2$O to CH$_4$/HCN dominated spectra from C/O = 0.92, which we obtain for atmospheres with $T_{\rm eff}$ $\gtrsim$ 1750 K, to C/O = 0.73 which we obtain for $T_{\rm eff}$ $\lesssim$ 1750 K, as we described in the previous section.

In order to test this condensation dependance further we carried out atmospheric calculations at $T_{\rm eff}$ = 1250 K,
neglecting the effect of condensation.

A comparison of the resulting emission spectra as a function of C/O for both cases ($T_{\rm eff}$ = 1250 K, with and without considering condensation) can be seen in Figure \ref{fig:cond_change}. We calculated atmospheres with C/O ratios spaced equidistantly between 0.35 and 1.4 using 100 grid points for both cases.
The difference in location for the shift from water to methane dominated
spectra, moving from C/O = 0.73 (condensation) to C/O = 0.92 (no condensation), is very prominent in these plots.

To further verify this finding we plot the mass fractions of H$_2$O, CH$_4$, CO and MgSiO$_3$ in Figure \ref{fig:cond_abund}
for a planetary atmosphere with $T_{\rm eff}$ = 1250 K, C/O = 0.8, ${\rm log}(g)$ = 3, [Fe/H] = 1 in orbit around a G5 star.
The C/O value is chosen such that the atmosphere is water dominated in the model neglecting condensation, but it is methane
dominated in our nominal atmospheric model, which includes condensation.

One clearly sees that for high pressures, where the temperatures are too high for MgSiO$_3$ to condense, the abundances of
H$_2$O, CH$_4$ and CO for both models are nearly identical. The small differences are due to differences in the $PT$-structures
found for the 2 chemical models.
For pressures smaller than 10$^{-2}$ bar, however, MgSiO$_3$ starts to condense, noticeably decreasing the CO and H$_2$O
abundances. CH$_4$ becomes much more abundant than H$_2$O, which is in contrast to the behavior of the model
without condensation, where H$_2$O stays more abundant than CH$_4$ throughout the atmosphere.

We therefore conclude that the transition from water- to methane-rich spectra may happen at C/O ratios
considerably smaller than 1 if the planetary effective temperature is not too high.
Especially for retrieval analyses of planetary spectra, which measure the local {\it gas} C/O ratio in the
spectrally active regions of the atmosphere, the above findings are relevant.
If condensation is expected to occur, a result such as
``C/O~$<$1'', due to the absence of methane features, could actually indicate an even lower total (gas + condensates) C/O~ratio$\lesssim$0.7.
\rch{If a given atmosphere were enriched in Mg and Si one would expect this effect to be even stronger,
shifting the transition between carbon and oxygen rich spectra to even lower C/O ratios.}

Finally, we want to remind the reader of the simplifications of our chemistry model, which does neither include settling
nor properly accounts for the effects of homogeneous or heterogeneous nucleation (see Section \ref{sect:chem_model}). Furthermore the absence of quenching in our models might be problematic if the timescales
for condensation and chemistry in general are longer than the vertical eddy-diffusion timescales.
Nevertheless, similar results have been found with much more sophisticated condensation models:
\citet{helling2014} were able to produce local C/O $\sim$ 1-2 values in the gas phase
for an atmosphere with a global C/O = 0.99 due to the condensation of O in dust species.
Their model for condensation and cloud formation is much more complete and includes
homogeneous and heterogenous nucleation, settling, traces the growth and evaporation of grains,
and considers more condensable species than our model.

Given these differences in condensation modeling it will be very important to reinvestigate our findings presented
here with more sophisticated cloud models in the future.

\subsection{Intermediate temperature atmospheres ($T_{\rm eff} \sim 1500$ K)}
\label{sect:MT_atmos}
At $T_{\rm eff}$ = 1500 K the transition from oxygen to carbon dominated atmospheres is still at
C/O = 0.73 as silicate condensation still takes place.
Furthermore, the carbon dominated atmospheres show strong methane features, but HCN features start
to emerge as well.

\subsubsection{Inversions and kinks at $T_{\rm eff}$ = 1500 K and C/O $\sim$ 1}
At $T_{\rm eff}$ = 1500 K and C/O $\sim$ 1 condensation can lead to weak inversions and, occasionally,
to kinks in the $PT$-structures.
For C/O ratios $\sim$ 1 the atmospheres are already carbon dominated.
\begin{figure*}[htb!]
\centering
\begin{minipage}{0.48\textwidth}
\includegraphics[width=1\textwidth]{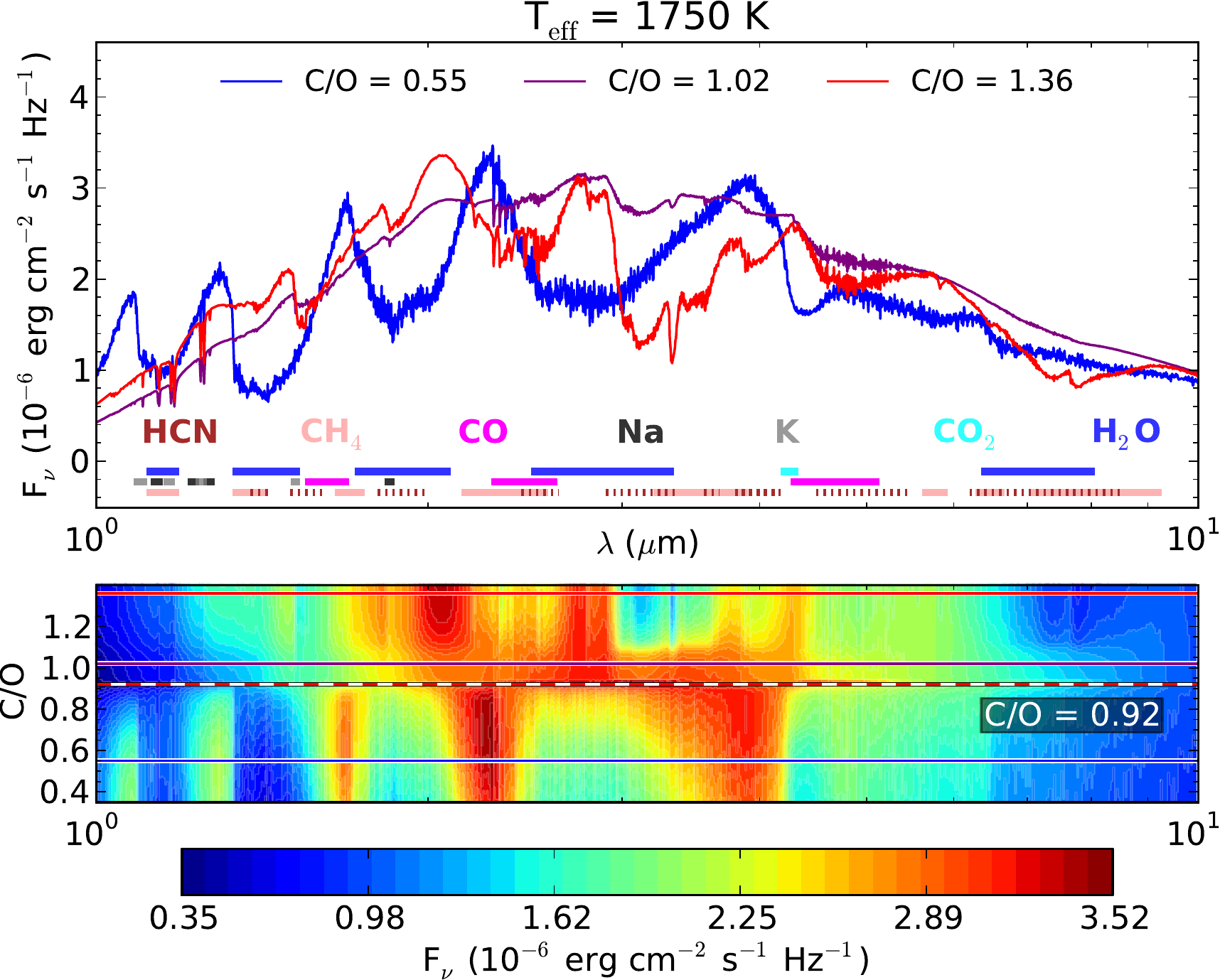}
\end{minipage}
\begin{minipage}{0.03\textwidth}
\end{minipage}
\begin{minipage}{0.48\textwidth}
\includegraphics[width=1\textwidth]{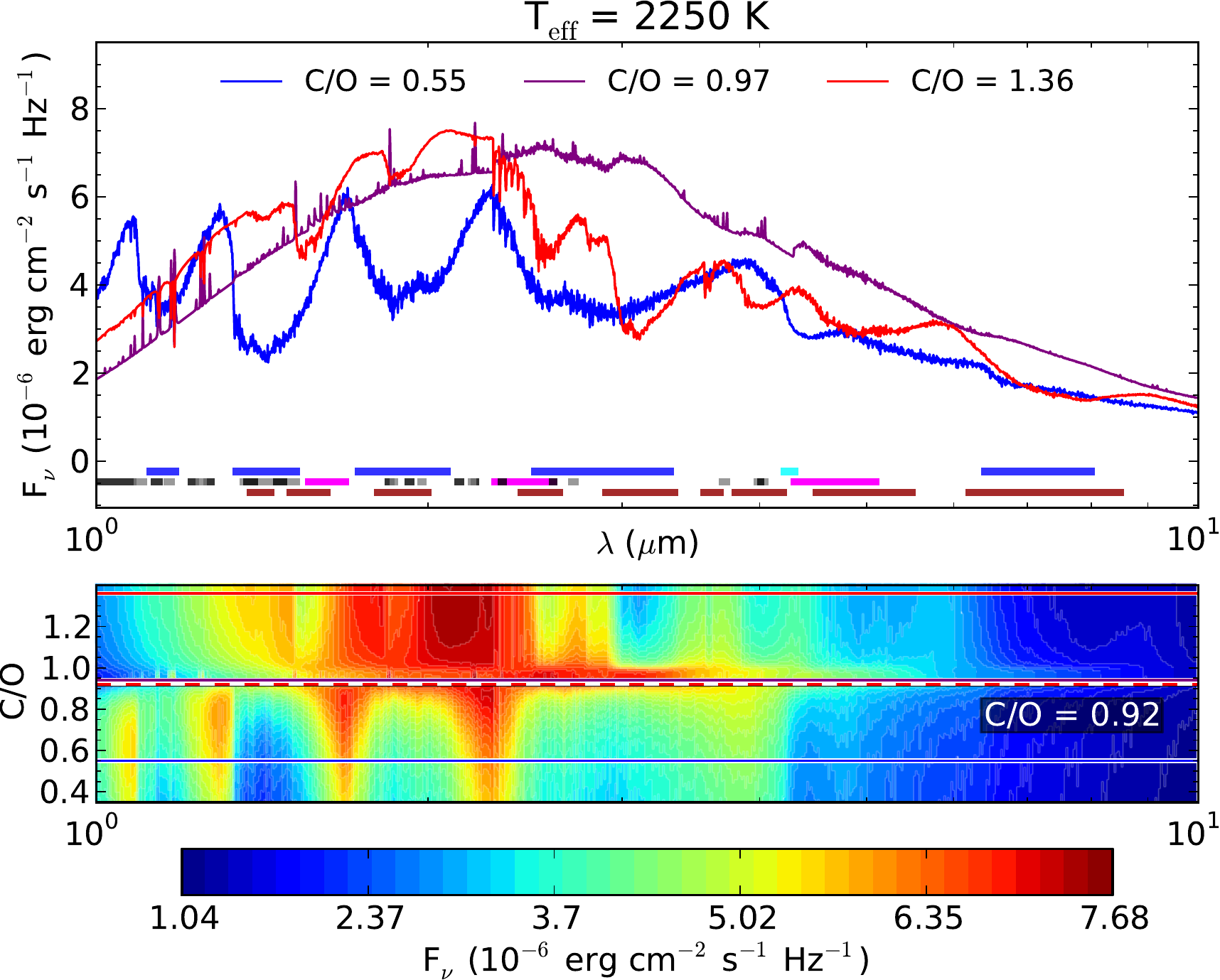}
\end{minipage}
\caption{Plots as shown in Figure \ref{fig:low_temp_co_diff_sim}, but for planets with
[Fe/H] = 1, ${\rm log}(g)$ = 3 in orbit around a G5 star. {\it Left panel:} $T_{\rm eff}$ = 1750 K,
{\it Right panel:} $T_{\rm eff}$ = 2250 K.}
\label{fig:hot_specs}
\end{figure*}
In general, the atmospheres are still too cool to contain enough HCN to efficiently radiate away
the absorbed stellar energy, such that H$_2$O and CH$_4$ are the main absorbers and the
H$_2$O--CH$_4$--CO chemistry is important.
At the intermediate atmospheric temperatures considered here,
inversions are likely to occur because of the condensation of SiC.
This results in a lower abundance of SiO, as less Si is available.
The O atoms which are not bound in SiO anymore form more CO and thus decrease
the C budget available to form CH$_4$,
therefore decreasing the atmosphere's ability to cool.
This effect can be further enhanced by the evaporation of MgSiO$_3$ in the inversion regions,
which frees additional O to be put into CO, subsequently locking up more C atoms.
As for the atmospheres which have inversions at C/O $\sim$ 1 at higher effective temperatures,
the inversions vanish for higher C/O ratios $>$ 1: less oxygen is present to form CO in the first place.
Therefore more CH$_4$ can be formed.

At $T_{\rm eff}$ = 1500 K kinks in the PT-structure can occur when condensation of MgSiO$_3$ and the associated locking up
of oxygen causes the cooling to become more strongly methane-dominated in a certain layer,
whereas there is less methane present to cool in an adjacent, hotter layer in which there is less MgSiO$_3$
and more oxygen is available in the gas phase to form CO and water, locking up carbon and decreasing the methane abundance.
The cooling ability of the methane-deprived layers is lower, leading to a strong temperature change from one layer to the next.
Such kinks depend on the choice of the grid spacing and
cell locations, such that they should not be treated as real physical phenomena but rather numerical artifacts.
The corresponding structure files have been flagged with ``{\verb|_kink|}''.
As inversions due to alkali heating and a low cooling ability are in general not seen in our results for M5 host stars, the kinks and inversions are not present for planets with M5 hosts.

In Figure \ref{fig:t_1500_kinks} we show an example for the kinks which are caused by the condensation.
One clearly sees that the kinks in the $PT$-structures going to hotter temperatures are caused by the (partial)
evaporation of MgSiO$_3$, reducing the amounts of coolants such as CH$_4$ and HCN.

\subsection{High temperature atmospheres ($T_{\rm eff} \gtrsim 1750$ K)}
\label{sect:HT_atmos}
At high temperatures condensation processes do not play an important role anymore.
Therefore, the transition between water and carbon-dominated spectra shifts from C/O = 0.73
to 0.92. Furthermore the carbon-rich atmospheres become more and more HCN dominated
and CH$_4$ becomes less and less important as the temperature increases.
As mentioned before, the chemistry is not only temperature but also pressure dependent,
favoring HCN over CH$_4$ at high temperatures and low pressures. \\ \\ \\
{\bf $T_{\rm eff}$ = 1750 K} \\
At $T_{\rm eff} = 1750$ K, we find that the higher the $\beta$-factor
(see Equation \ref{equ:beta_factor}) of an atmosphere is, the more HCN dominates the spectrum.
Methane features are visible for all $\beta$s, however.
Due to the chemistry, a low $\beta$-factor allows for some presence of water in the carbon-rich
atmospheres. Thus at low $\beta$s we find a weak water absorption signature imprinted
on the rather opacity free region extending from 2.4-3 $\mu$m, which is bracketed by
two CH$_4$ features.
Because of the strong stellar irradiation the atmospheric structures at C/O
$\sim$ 1 become either more isothermal or exhibit inversions.
We show spectra of atmospheres with $T_{\rm eff}$ = 1750 K and varying  C/O ratios in the left panel of
Figure \ref{fig:hot_specs}. \\Ê\\
{\bf $T_{\rm eff}$ = 2000 K} \\
At even higher temperatures HCN becomes more dominant.
Inversions at C/O $\sim$ 1 predominantly form for low $\beta$ $<$ -2.5 (or -2) in these atmospheres.
For the larger $\beta$ values the methane features fade away. \\Ê\\
{\bf $T_{\rm eff}$ = 2250 K} \\
For $T_{\rm eff}$ = 2250 K the atmospheres with C/O $>$ 1 are strongly HCN dominated.
Only for low $\beta$ values weak methane features are present.
Furthermore more or less all atmospheres with C/O $\sim$ 1 have inversions if the spectral type of the
host star is K or earlier.
We show spectra of atmospheres with $T_{\rm eff}$ = 2250 K and varying  C/O ratios in the right panel of
Figure \ref{fig:hot_specs}. \\ \\
{\bf $T_{\rm eff}$ = 2500 K} \\
For $T_{\rm eff}$ = 2500 K the atmospheres with C/O $>$ 1 are completely HCN dominated,
and the methane features have vanished.
All atmospheres with C/O $\sim$ 1 have inversions if the spectral type of the
host star is K or earlier.

\section{Summary and conclusion}
\label{sect:conclusion}
In this work we present a systematic parameter study of hot jupiter atmospheres.
In addition to ``classical'' grid parameters such as metallicity, effective temperature and surface gravity
we study the effects of the atmospheric C/O ratio as well as the host star spectral type. We summarize
the key findings of our study in Figure \ref{fig:CO_atmo_car} and in the text below.

\begin{itemize}
\item {\bf At low effective temperatures ($T_{\rm eff}$ $<$ 1500 K) the atmospheres can be either water or methane
dominated, but not always:} if $\beta = {\rm [Fe/H]} - {\rm log}(g)$ is small, the spectra at $T_{\rm eff}$
$\lesssim$ 1000 K are quite similar, showing both strong water and methane features.
The optical depth (and hence temperature) versus pressure profile scales approximately with $\beta$.
Hence, a given optical depth (temperature) is reached at high pressure when beta is low and vice versa.
We want to remind the reader, however, that we neglect quenching, which could potentially alter the methane
and water mixing ratios.
For high pressures and low temperatures CH$_4$ and H$_2$O co-exist as the dominant oxygen and carbon opacity carriers,
and are both visible in the spectrum.
At $\beta$ values above -4 to -3.5, the atmospheres are either water- or methane-dominated at $T_{\rm eff}$ = 1000 K.
For atmospheres with $T_{\rm eff}$ = 1250 K the spectra look similar only for the highest surface gravities (${\rm log}(g)$ = 5)
and lowest metallicities ([Fe/H] $\lesssim$ 0), such that these atmospheres should be either water or methane dominated
for most planets.
\item {\bf At $T_{\rm eff}$ $\lesssim$ 1500 K the condensation of MgSiO$_3$ is a relevant effect at the local atmospheric
temperatures.} The condensation effectively lowers the amount of oxygen which can be put into CO and H$_2$O,
such that more carbon atoms are available to form CH$_4$. {\bf As a result the atmospheres start to be methane
dominated at C/O = 0.73.} For higher temperatures MgSiO$_3$ can no longer condense, shifting the transition from
oxygen to carbon dominated spectral signatures to C/O = 0.92.
\item {\bf For planets with $T_{\rm eff}$ $\gtrsim$ 1500 K and C/O $\sim$ 1
host stars with spectral type earlier than M5 (we consider M5, K5, G5, F5) can lead to temperature inversions in the atmospheres.}
The reason for this is that under these circumstances all the main coolants of the atmosphere, H2O, HCN, and CH$_4$, are depleted,
whereas the absorption of optical radiation by the alkali metals remains highly effective.
For $T_{\rm eff}$ = 1500 K the condensation of SiC can sufficiently lower the cooling ability for inversions to develop.
At this effective temperature the condensation of MgSiO$_3$ can lead to kinks and numerical instabilities in the
solutions for the $PT$-structure. For $T_{\rm eff}$ = 2000 K all atmospheres with $\beta$ $<$ -2 to -2.5 will exhibit inversions.
For $T_{\rm eff}$ $\geq$ 2250 K all atmospheres with C/O $\sim$ 1 exhibit inversions.
\item {\bf The lower $\beta$ = [Fe/H] - ${\rm log}(g)$, the more methane-dominated the spectra are at C/O ratios $\gtrsim$1.}
At higher temperature and/or higher $\beta$ values, such planets have HCN-dominated spectra.
In general we show the dominant absorbers as a function of temperature and C/O ratio in Figure \ref{fig:CO_atmo_car}.
\item {\bf The host star spectral type is an important factor for the spectral appearance of the atmosphere}.
For planets with C/O $\sim$ 1 host stars of spectral type K or earlier can give rise to inversions
if they are at small enough distances, whereas for M-type host stars inversions do not occur.
Further, the later the host star spectral type, the more isothermal the planetary atmospheres become (if the C/O ratio is not $\sim$ 1).
This is because SED of the stellar irradiation becomes increasingly similar to the planetary radiation field.
\item{\bf Planetary metallicity and surface gravity determine the location of the planetary photosphere.} High surface gravities
or low metallicities will shift it to larger pressures, whereas low surface gravities or high metallicities shift it to low pressures.
As the molecular and atomic line wing strength scales approximately linearly with pressure, for photospheres at low pressures the
flux is originating at somewhat deeper layers (in terms of optical thickness), leading to somewhat cooler atmospheric temperatures
and hence deeper absorption troughs. The deep isothermal temperature increases in these cases, as the insolation can probe deeper
into the atmospheres. Similar results for the surface gravity dependence of the absorption troughs have also been reported in \citet{sudarsky2003}. 
\end{itemize}

\begin{figure}[t!]
\centering
\includegraphics[width=0.485\textwidth]{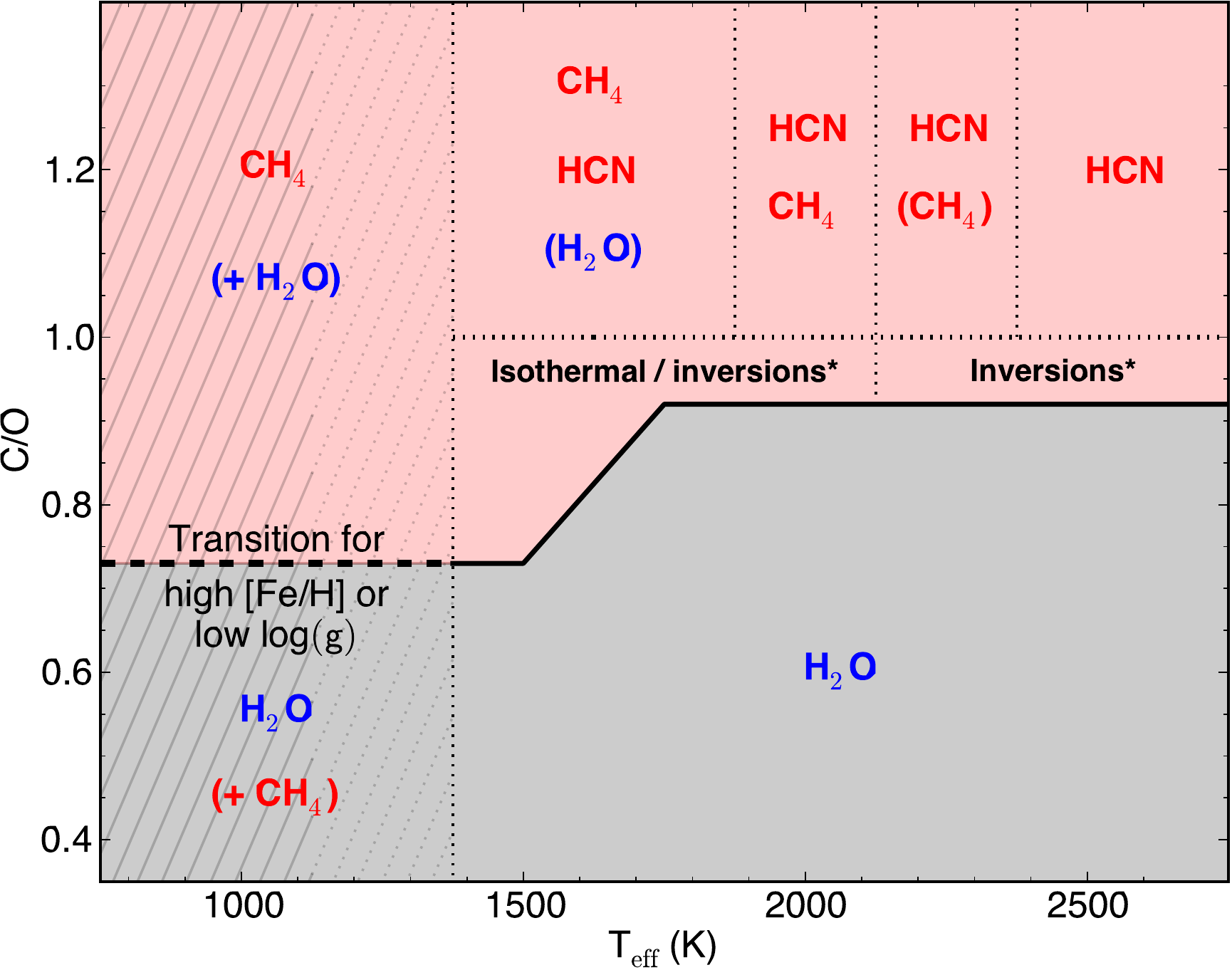}
\caption{Dominating IR absorbing/cooling species as a function of $T_{\rm eff}$ and C/O. The red shaded region denotes 
carbon-dominated atmospheres, whereas the grey shaded region denotes oxygen-dominated atmospheres.
The gray-hatched region denotes the temperature range where the atmospheric spectra can be dominated
by CH$_4$ and H$_2$O at the same time, independent of the C/O value. This occurs if [Fe/H] is low or ${\rm log}(g)$
is high. Within each region defined by the black solid lines the dominating IR absorbing/cooling species is indicated in the plot. The region in which inversions occur is shown in the plot as well. ${^*}$Only host stars of type K and earlier can cause inversions.}
\label{fig:CO_atmo_car}
\end{figure}

It is interesting to see that at low temperatures the strength of methane or water features does not only depend on the C/O ratio,
but also on the pressure level of the photosphere, which can be quantified using the the $\beta$ factor.
For higher temperatures the $\beta$ factor plays a role as well, as it determines whether CH$_4$ or HCN dominates the spectra of
carbon-rich atmospheres. Also the occurrence of an inversion at C/O $\sim$ 1 can be tied to the $\beta$ factor, at least for the atmospheres with
$T_{\rm eff}$ $\sim$ 2000 K. Therefore the $\beta$ factor can be used as a third dimension to characterize the spectral appearance of an
exoplanet, in addition to the effective temperature and the C/O ratio.

Moreover, the fact that the transition from water to methane rich spectra shifts due to the condensation of silicates, which lock up oxygen,
is important when carrying out retrieval analyses of planetary atmospheres.
The C/O ratio is often measured by taking into account the abundances of only the gaseous carbon and oxygen carrying molecules.
This can potentially overestimate the total (gas + condensates) C/O ratio.
It is important to note that our current condensation model is simplified, assuming instantaneous condensation once the
saturation vapor pressure is exceeded and no settling or mixing of the cloud particles.
It will therefore be very important to investigate this effect in the future in greater detail, using a more sophisticated
condensation model.

The fact that inversions can potentially occur at C/O $\sim$ 1 is interesting, as we did not require any additional
absorbers such as TiO and VO, the absorption of stellar light by the alkali atoms is sufficient.
To further study the inversions it is necessary to obtain molecular line lists as complete as possible
as their occurrence is very strongly dependent on the atmospheric cooling ability.

The grid of atmospheres presented in this work is made publicly available and can be found at the CDS.

\begin{acknowledgements}
We thank the anonymous referee for a very thorough assessment of the manuscript and many comments
that have improved the quality of this paper.
Further, P.M. thanks Robert L. Kurucz for answering questions related to his line lists, Alexandre Faure
for answering questions regarding line broadening parameters, Nicole Allard for her detailed instructions
on how to use her wing profile tables and Kevin Heng for helpful discussions.
\end{acknowledgements}

\newpage

\appendix

\section{A fast method to calculate opacities from line lists}
\label{sect:line_calc}
If one wants to calculate the line opacities of a given molecule on a given grid of wave number points one must, in principle, and if no line truncation is
applied, calculate the line profile of every line at every wave number grid point. Even if one precomputes the opacities and tabulates them for later use
it can still take a long time to calculate the total molecular cross-section as the calculations need to be carried out at high resolution. Using
a fiducial resolution of $R=10^6$ and considering a species with of the order of $10^8$ lines one easily ends up with $\sim 10^{14}$ line profile
evaluations at just one given pressure and temperature.

To speed up the calculations of line opacities the method explained below was developed and used
for our opacity database calculations. In essence we calculate the line cores of every line at high resolution,
while calculating the line wings far from the line core on a much coarser grid. Once the contribution of all
the lines to the coarse grid has been calculated it is interpolated back to the fine grid. In detail we proceed as follows:

Divide the total wave number grid into subgrids of 10,000 grid points. Then start to go through
all these subgrids, which will be indexed by $m$:
\begin{itemize}
\item Calculate all line opacities of lines that are lying within the subgrid $m$ at all of its 10,000 wave number grid points.
\item Then iterate over all other subgrids (which are indexed by n). For the external lines in a given external subgrid with $n\ne m$
do the following:
\begin{enumerate}
\item If $\gamma_{G}>\gamma_{L}$ (i.e. Gaussian width larger than Lorentz width):
If the distance of the line to the subgrid border of $m$ is smaller than $f_{G}\gamma_{G}$, where $f_{G}$ is a factor that needs to be specified, then this line gets calculated at all of $m$'s 10,000 subgrid points. Otherwise go to step 2.
\item If $\gamma_{L}>\gamma_{G}$ or the distance to the subgrid border of $m$ is larger than $f_{G}\gamma_{G}$: \\
Consider the Lorentzprofile $\gamma/(\gamma^2+(x-x_0)^2)$. Far away from the line center its functional form is roughly $\gamma/(x-x_0)^2$. The relative deviation $\alpha$ from this form is
\begin{align}
\nonumber \alpha & = \frac{\gamma^2+(x-x_0)^2}{\gamma}\left[\frac{\gamma}{(x-x_0)^2}-\frac{\gamma}{\gamma^2+(x-x_0)^2}\right] \\
& = \frac{\gamma^2}{(x-x_0)^2}
\end{align}
I.e., if we want a maximum deviation of less than $\alpha$ from the above form, then we need that
\beq
|x-x_0|>\frac{\gamma}{\sqrt{\alpha}}.
\eeq
If this is not fulfilled, then the line opacity is just calculated at all of $m$'s 10,000 subgrid points.
It it is fulfilled go to step 3.
\item For all lines within a given external subgrid $n$ that fulfill the above inequation: Calculate the line strengths on a coarse subgrid of 10 points within the original subgrid $m$ and add the results for all these lines up, then interpolate back to the 10,000 original grid points in $m$, using a powerlaw interpolation and a coordinate transformation (a simple shift).
\item Move on to the next external subgrid $n+1$, go back to step 1.
\end{enumerate}
\item Move on to the next subgrid $m$.
\end{itemize}
\begin{figure}[htb!]
\centering
\includegraphics[width=0.485\textwidth]{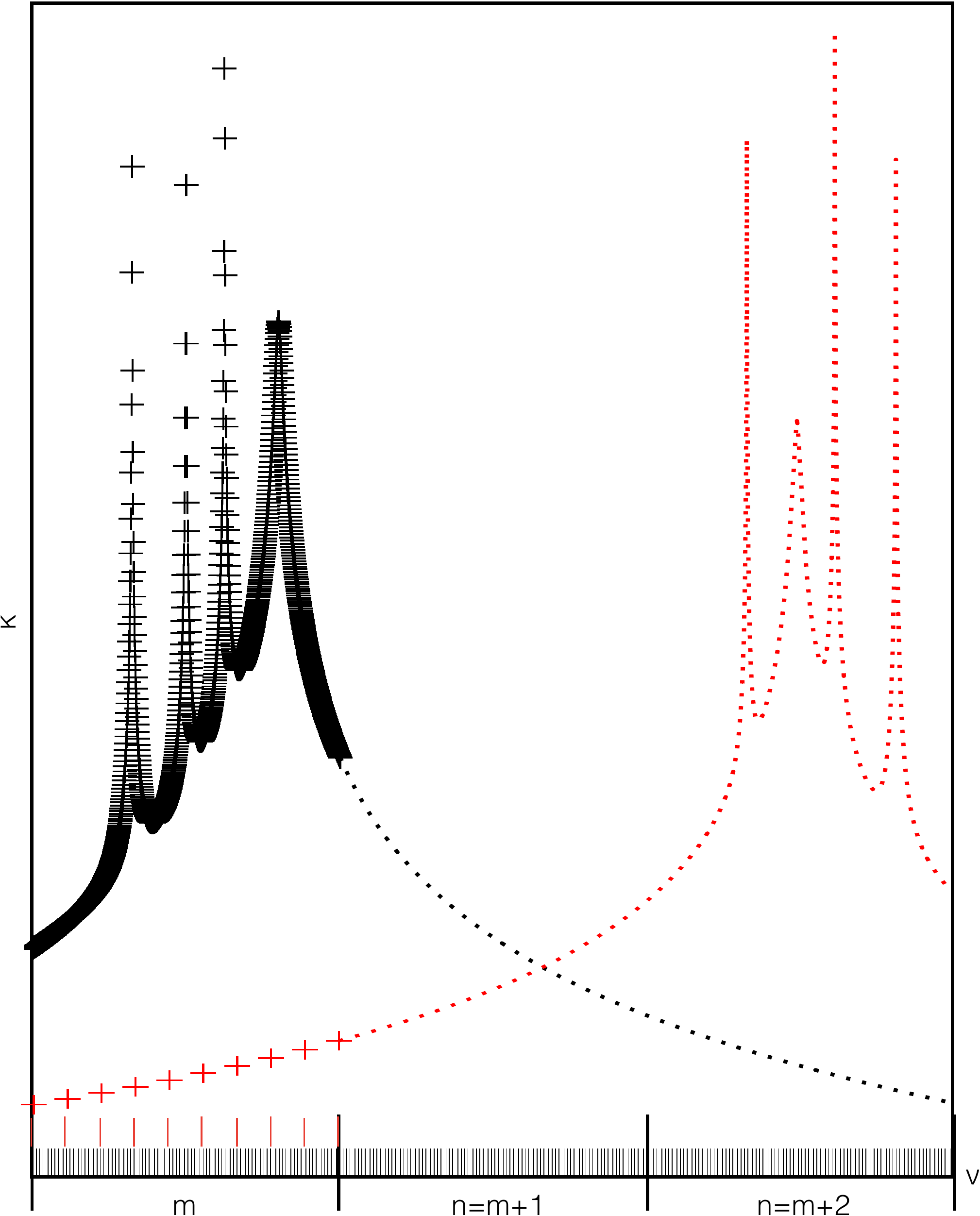}
\caption{Schematic drawing of the algorithm used to calculate the opacities. For the lines within subgrid $m$ the opacities get calculated on the
fine fiducial grid. If a subgrid $n\ne m$ is sufficiently far away (in this example n=m+2) the opacity of the lines in $n$ are calculated on a
coarse grid in $m$. The summed opacity values of the lines in grid $n$ are then interpolated on the fine grid in $m$
using a coordinate shifted powerlaw interpolation.}
\label{fig:lineacccalc}
\end{figure}
The reason to interpolate back to the 10,000 subgrid points of $m$ for all external subgrids $n$ seperatly is the following. \\
For a single line, far away from its line center, the line shape is roughly
\beq
\phi(x) = \frac{\gamma}{(x-x_0)^2} .
\eeq
Thus for all transitions $k_n$ of strength $S_{k_n}$ which are contained in a subgrid $n$ the total continuum line strength (i.e. the wing strength of
a line far away from its line core) will be
\beq
\sigma_{\rm tot,n}(x) = \sum_{k_n}\frac{S_{k_n}\gamma_{k_n}}{(x-x_{k_n})^2}.
\eeq
Seen from subgrid $m$ all lines $k_n$ within a given subgrid $n$ have roughly the same line center position (namely within subgrid $n$), thus one can
do the coordinate transformation $y_n=x-\bar{x}_n$, where $\bar{x}_n$ is the position of subgrid $n$ (e.g. the wave number at its center). This would yield
\beq
\sigma_{\rm tot,n}(y) = \sum_{k_n}\frac{S_{k_n}\gamma_{k_n}}{(y_n+\bar{x}_n-x_{k_n})^2} \approx \frac{1}{y_n^2}\sum_{k_n}S_{ij_n}\gamma_{k_n} \ .
\eeq
Thus for every subgrid $n$ one can do a coordinate transformation to $y_n$ and finds that the coarse 10-point subgrid continuum of subgrid $n$ seen
in subgrid $m$ should roughly behave like a powerlaw function with a powerlaw slope of $\sim-2$. This explains why a powerlaw interpolation in the coordinate $y_n$ is the best thing to do when interpolating the coarse continuum of the lines in subgrid n back to the fine grid in subgrid $m$. It should also be stressed that it is better to do an interpolation that finds the effective powerlaw slope, rather than taking it to be $-2$ and using  $\sum_{k_n}S_{k_n}\gamma_{k_n}$, as there will be slight deviations from this $-2$ powerlaw shape, as one knows that the line centers in subgrid $n$ are close to $\bar{x}_n$, but not exactly at $\bar{x}_n$. An interpolation will mitigate this problem by finding a slightly different powerlaw shape and an overall coefficient for the function that slightly deviates from $\sum_{k_n}S_{k_n}\gamma_{k_n}$. For continua produced by lines that are outside of the {\it total} grid we use a linear interpolation, as we don't actually check where these lines are sitting. In Figure \ref{fig:lineacccalc} one can see a schematic drawing explaining
the acceleration method introduced in this section.

All our opacity calculations were carried out using the accelerated method on a grid with a point spacing of $\lambda / \Delta \lambda = 10^6$. Additionally we performed a ``classic'' calculation on a reduced grid constructed by using every 1,000th fiducial wave number grid point. On this reduced grid we did not use the aforementioned acceleration method but calculated every line contribution at every wave number point. The high resolution result of the accelerated method was only kept if the maximum relative deviation at the points coinciding with the 1,000 times coarser test grid was smaller than 1 \%. If it was bigger, $f_G$ was increased and $\alpha$ was decreased and the calculation was repeated.

\section{Correlated-k: Going from $\mathcal{O}\left(C^N\right)$ to $\mathcal{O}\left(N\right)$}
\label{sect:corr-kapp}
In the following we describe our method of combining the opacity tables of multiple species.
Furthermore its implementation at different grid resolutions is explained.
For a general review of the correlated-k method see, e.g., \citet{marley2014}.

\subsection{The ``classical'' $\mathcal{O}\left(C^N\right)$ case}
\label{sect:classical_method}
The commonly utilized method to combine the k-tables of multiple species is numerically quite expensive,
as it is of order $\mathcal{O}\left(N_g^{N_{\rm sp}}\right)$, where $N_g$ is the number of grid points used in $g$-space ($g$ is the cumulative
opacity distribution function, see below) and $N_{\rm sp}$
is the number of species. In this traditional method, the computation of the total opacity $\kappa_{\rm tot}$ works as follows:
In a spectral region of the frequency interval $[\nu,\nu+\Delta\nu]$
the transmission of light $T$  through a layer of thickness $\Delta P$ which contains 2 spectrally active species is
\beq
T = \int_\nu^{\nu+\Delta\nu}{\rm exp}\left[-\frac{X_1\kappa_1(\nu)+X_2\kappa_2(\nu)}{a}\Delta P\right]\frac{{\rm d}\nu'}{\Delta \nu} \ ,
\eeq
where $X_i$ and $\kappa_i$ are the mass fractions and opacities of the two species, $a$ is the gravitational acceleration
in the atmosphere and $\Delta P$ is the atmospheric layer thickness in units of pressure.
For simplicity it is assumed that $X_i$ and $\kappa_i$ are constant within the atmospheric layer.
If one assumes the opacities of species 1 and 2 to be uncorrelated, i.e.
\beq
f_{\rm tot}(\kappa_1,\kappa_2)=f_1(\kappa_1)\cdot f_2(\kappa_2) \ ,
\eeq
where $f$ are the opacity distribution functions, one can rewrite the transmission $T$ as
\beq
T = \left[\int_\nu^{\nu+\Delta\nu}e^{-X_1\kappa_1(\nu)\Delta P/a}\frac{{\rm d}\nu'}{\Delta \nu}\right]\cdot \left[ \int_\nu^{\nu+\Delta\nu}e^{-X_2\kappa_2(\nu)\Delta P/a}\frac{{\rm d}\nu'}{\Delta \nu}\right] \ .
\eeq
An opacity distribution function within a frequency interval $[\nu,\nu+\Delta\nu]$ is defined by $f(\kappa){\rm d}\kappa$ being the fraction of
the opacity values within $[\nu,\nu+\Delta\nu]$ which lie between $\kappa$ and $\kappa+{\rm d}\kappa$. 
Going from frequency space to $g$-space, where g is the cumulative opacity distribution function (${\rm d}g = f(\kappa){\rm d}\kappa$), and approximating the integrals with sums yields
\beq
T \approx \sum_{i=1}^{N_g}\sum_{j=1}^{N_g}{\rm exp}\left[-\frac{X_1\kappa_{1,i}+X_2\kappa_{2,j}}{a}\Delta P\right]\Delta g_i \Delta g_j \ .
\label{equ:T_approx}
\eeq
The combined total k-table of species 1 and 2 therefore has the opacity values
\beq
\kappa_{{\rm tot},ij} = X_1\kappa_{1,i}+X_2\kappa_{2,j}
\label{equ:ck_class_1}
\eeq
which have to be weighted with
\beq
\Delta g_{ij} = \Delta g_i \Delta g_j .
\label{equ:ck_class_2}
\eeq
As is commonly pointed out the number of operations that need to be carried out in order to combine the k-tables of multiple species is thus $\mathcal{O}(N_g^{N_{\rm sp}})$, which can make the consideration of multiple species computationally expensive \citep[see, e.g.,][]{marley2014,lacis_oinas1991}.
\subsection{The $\mathcal{O}\left(N\right)$ case}
\subsubsection{Algorithm used at a bin size of $\lambda/\Delta\lambda=1000$ (R1000 method)}
\label{sect:1000_comb}
In order to combine the individual k-tables of all species of interest for finding the total k-table of an atmospheric layer we use a method
that is computationally less expensive. Similarly to the ``classical'' approach, the method makes use of the assumption that the opacities
are not correlated.
The main idea is to iteratively combine the opacities of two species: The opacity of a real species and the effective opacity of a ``help''-species.
If the opacities of all species are uncorrelated, then the combined opacity of two species is not correlated with the opacity of any other
remaining species. Furthermore the combined opacity of the two combined species can be treated as belonging to a new single species,
which is the ``help''-species.

We thus proceed in the following way: For every species, within every $\Delta \nu$ bin, we save the opacity distribution $\kappa(g)$ on
a grid of 30 points. The 30-point grid consists of two 15-point Gaussian grids ranging from 0 to 0.9 and from 0.9 to 1, respectively.\footnote{
This is not the same grid on which the radiative transport will be carried out on. The radiative transport grid consists of 20 points. The 30-point
Gaussian grid is only used for the combination of the k-tables.} Now, when starting to construct the total opacity, the first two species 1 and 2 get combined
according to equations (\ref{equ:ck_class_1}) and (\ref{equ:ck_class_2}). This results in $30\times 30=900$ new values $\kappa_{1+2,ij}$ which need
to be sorted by size. Using
the cumulative sum of the associated weights $\Delta g_i \Delta g_j$, where $\Delta g_i$ and $\Delta g_j$ are the respective Gauss-grid weights,
we interpolate the result back to the original 30-point Gauss-grid. This newly obtained opacity $\kappa_{1+2}$ is then iteratively combined with
the remaining species' opacities and results in the final opacity distribution $\kappa_{\rm tot}(g)$. In a procedural notation the method can thus be
described as
\begin{verbatim}
Total opacity = X_1 * kappa(g) of species 1
                           
For all remaining species (i = 2 to N_sp) {

  Total opacity =   combine(Total opacity,
       X_i*kappa(g) of species i)
       
  re-bin Total opacity to nominal g-grid

}
\end{verbatim}
For notational convenience the method for combining the opacities as introduced in this section will be called
``R1000 method'' in the following sections. The number of points used for combining two species' opacities
will be called $N_{\rm R1e3}$. As explained in this section, the nominal value of $N_{\rm R1e3}$ when working at a
resolution of $\lambda/\Delta \lambda = 1000$ is $N_{\rm R1e3}=30$.

\subsubsection{Algorithm used at a bin size of $\lambda/\Delta\lambda=10$ and $50$}
In order to correctly describe the opacity distributions at small $\lambda/\Delta\lambda$ many $g$-grid points would
need to be used, as especially at low pressures the opacity tables $\kappa(g)$ tend to be very sharply
peaked at $g$ values very close to 1. Therefore the R1000 method would become numerically inefficient and
cannot be used.

However, once more the idea is to combine two species iteratively in order to obtain the total opacity.
Moreover, we again make use of the assumption that the opacities are not correlated.
In the $\lambda/\Delta\lambda=10$ and $50$ cases the spectral bins are 100 or 20 times larger than in the
$\lambda/\Delta\lambda=1000$ case. They therefore  include many lines, and the assumption of
uncorrelatedness should be valid to an even higher degree than in the $\lambda/\Delta\lambda=1000$ case.

The idea to obtain the total opacity is the following:
In principle the combination of two species could be accomplished by randomly sampling the
2 individual opacity distributions and taking the sum of the sampled values as a set of the combined
opacity. In a numerically simplified version one could discretize the opacity distributions by providing
a pre-sampled set of $N$ opacity values and their corresponding weights $\Delta g$.

The random sampling could then be approximated by randomly drawing values from the opacity sets
of each species and adding them, taking into account their weights at the same time.
If one would sample continuous values from a distribution, it is possible to sample values from within
a given interval multiple times. Thus, if a discretized opacity value has been drawn from the opacity
set it must in principle not be excluded from being drawn in any of the next sampling steps.

The discretization is carried out in the following way in our method:
For every species we divide the $\kappa(g)$ table of every species into two sets.
The first set contains $\kappa(g)$ values with $g<g_{\rm bord}$. The $g$-coordinates are located
at the centers of grid cells defined by $N_p+1$ grid borders spaced equidistantly between $g=0$ and $g=g_{\rm bord}$.
The second set contains $\kappa(g)$ values with $g \ge g_{\rm bord}$.
These $g$ values are located at the centers of $N_p$ grid cells defined by $N_p+1$ grid borders spaced equidistantly
between $g=g_{\rm bord}$ and $g=1$. We chose $g_{\rm bord}=0.985$ and $N_p=128$.
The $N_p$ values of a species with $g<g_{\rm bord}$ will in the following be called $\kappa_{{\rm low}}$ and the
$N_p$ values with $g \ge g_{\rm bord}$ will be called $\kappa_{{\rm high}}$.
Additionally, for every species, we save the lowest and highest
opacity value within the frequency bin, corresponding to the $g=0$ and $g=1$ opacity values.
$\kappa_{{\rm low}}$ describes the low $g$, continuum properties of the species' opacity, while $\kappa_{{\rm high}}$
describes the high $g$, line core properties of the species' opacity.

Returning to sampling values from 2 species, the probability of
sampling and combining 2 values stemming from the respective $g<g_{\rm bord}$-regions is
$g_{\rm bord}^2$. The probability for combining 2 values from the $g<g_{\rm bord}$-region of species 1
and the $g>g_{\rm bord}$-region of species 2 is $g_{\rm bord}\cdot(1-g_{\rm bord})$ etc...

To speed up sampling, we now assume that once an opacity value of a given species has been drawn,
it cannot be drawn again (we will return to the validity of this approach below).

In order to approximate the sampling process of the combined opacity distribution function of two species, we then construct a $4N_p\times 2$ matrix
$K$ containing the various possible combinations of $\kappa_{{\rm low}}$ and $\kappa_{{\rm high}}$ of both species,
weighted by how common these combinations would be in a random sampling process of both species' opacities.
When sampling points from species 1 and combining them with sampled points from species 2 the assumption that
a given value can not be redrawn allows for a simple shuffling in the sampling process: 
\beq
K =
 \begin{pmatrix}
X_1\cdot{\rm shuffle}(\kappa_{1,{\rm low}})+X_2\cdot{\rm shuffle}(\kappa_{2,{\rm low}}) & \frac{g_{\rm bord}^2}{N_p}\\
X_1\cdot{\rm shuffle}(\kappa_{1,{\rm low}})+X_2\cdot{\rm shuffle}(\kappa_{2,{\rm high}}) & \frac{g_{\rm bord}\left(1-g_{\rm bord}\right)}{N_p}\\
X_1\cdot{\rm shuffle}(\kappa_{1,{\rm high}})+X_2\cdot{\rm shuffle}(\kappa_{2,{\rm low}}) & \frac{g_{\rm bord}\left(1-g_{\rm bord}\right)}{N_p}\\
X_1\cdot{\rm shuffle}(\kappa_{1,{\rm high}})+X_2\cdot{\rm shuffle}(\kappa_{2,{\rm high}}) & \frac{\left(1-g_{\rm bord}\right)^2}{N_p}
 \end{pmatrix}.
\eeq
The first column of $K$ represents the sampled values of the new combined opacity, the second column gives the weight of each sampled value, similar to the $\Delta g_1\Delta g_2$ weights in the classical method described in Section \ref{sect:classical_method}. We then sort the lines of the matrix $K$
by the values in the first column. After this we construct a vector $y$ of length $4N_p$ with the entries (starting at $m=2$)
\beq
y_{m} = y_{m-1}+\frac{k_{(m-1),2}+ k_{m,2}}{2}
\eeq
and $y_1 = k_{1,2}/2$. The second column of $K$ is then replaced with $y$. After this, the first column of $K$ contains the newly sampled $\kappa_{\rm tot}(g)$
values of the combined opacity of species 1 and 2, the second column contains the corresponding $g$ values. Using
$\kappa_{\rm tot}(0) = X_1\kappa_1(0)+X_2\kappa_2(0)$ and $\kappa_{\rm tot}(1) = X_1\kappa_1(1)+X_2\kappa_2(1)$ the total opacity can then be interpolated to
the $N_p$ low-$g$ and $N_p$ high-$g$ values to yield the final result. The resulting opacity is then ready for being combined with the opacity of the next species.
In order to shuffle the opacities we use the Knuth-shuffle algorithm, which is of order $\mathcal{O}(N_p)$.

\begin{figure}[htb!]
\centering
\includegraphics[width=0.485\textwidth]{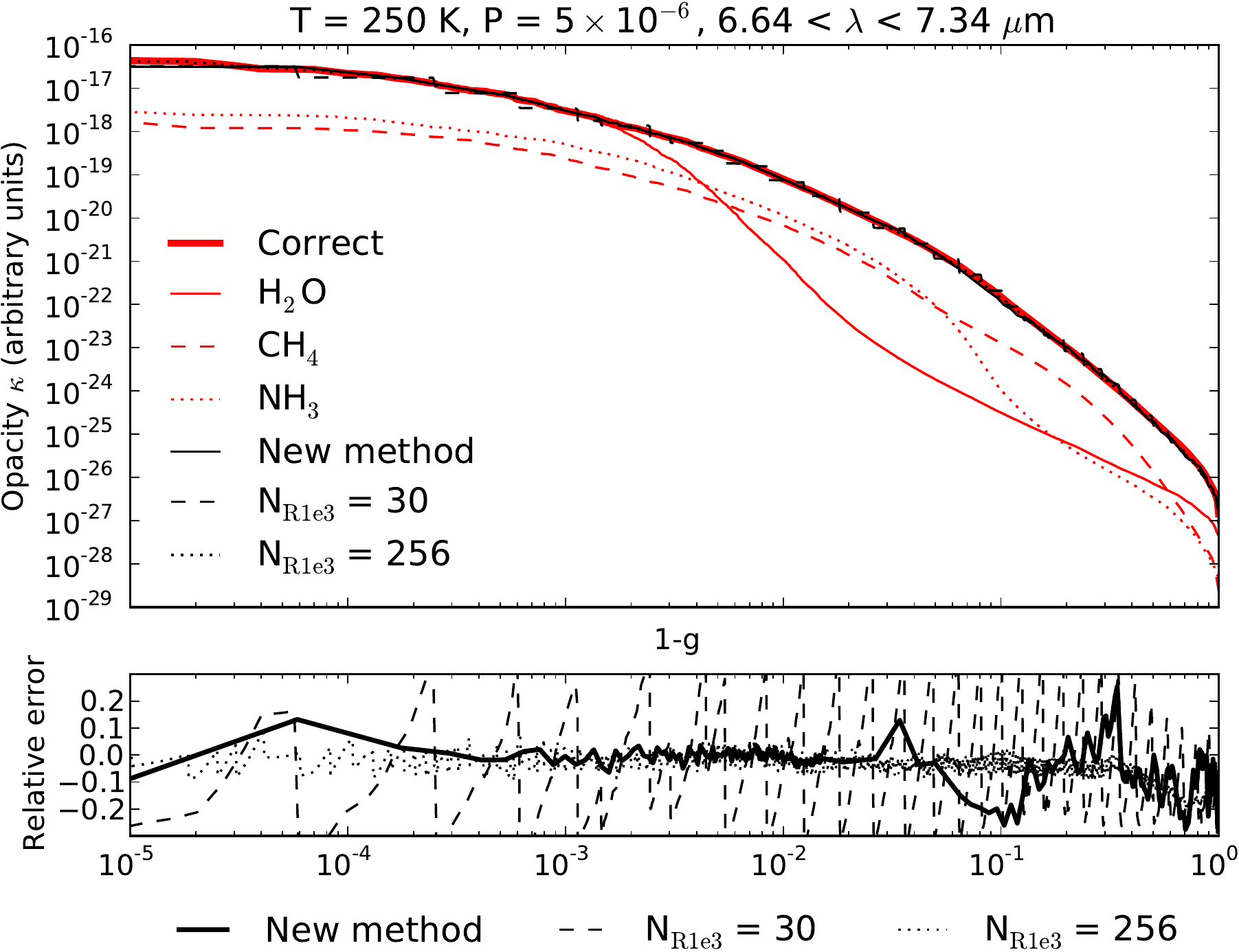}
\caption{Comparison of the different methods to combine the $\kappa(g)$ tables of different species. {\it Upper panel:}
Opacity of water (red solid line), methane (red dashed line) and ammonia (red dotted line) as a function of $g$.
The total $\kappa(g)$ obtained from adding the opacities in frequency space is shown as a red thick solid line. The results
when using the R1000 method with N$_{\rm R1e3}$ = 30 points and N$_{\rm R1e3}$ = 256 points are shown as black dashed
and dotted lines, respectively. The result when using the new method introduced in this section is shown as a black solid line.
{\it Lower panel:} Relative error of the three methods compared to the correct solution:
N$_{\rm R1e3}$ = 30 points (dashed line), N$_{\rm R1e3}$ = 256 points (dotted line), new method (thick solid line).}
\label{fig:ck_comb_test}
\end{figure}

The assumption of not being able to draw a given opacity value more than once is obviously not correct.
However, it has been found to not affect the quality of our results. From the above we see that in every combination
step one needs to sort $4N_p$ values. In the R1000 method we would have the same
computational costs when storing $N_{\rm R1e3}=2\sqrt{N_p}$ opacity points per species, losing resolution when comparing to the $2N_p$
points we use in the method introduced here.
Furthermore, the results of the R1000 method, at the same computational cost,
turn out to be much worse, both when comparing to the actual shape of the wanted total opacity distribution
as well as when comparing
\beq
\frac{1}{\Delta \nu}\int_{\nu}^{\nu+\Delta\nu}\kappa_{\nu'}{\rm d}\nu' \approx \sum_i \kappa_i \Delta g_i
\label{equ:calc_mean_opa}
\eeq
for both methods.\footnote{We will need to evaluate Eq. (\ref{equ:calc_mean_opa}) when computing the Planck mean opacity
in the temperature calculation.} The error of our method is in the range of \%, whereas the error of the R1000 method at the same
computational cost is in the range of tens of \%. Comparing the results of the new method with results of the R1000 method
when taking $N_{\rm R1e3}=2N_p$, i.e. the same number of points in both cases, yields slightly better results for the R1000 method.
However the numerical costs for the R1000 method are $\mathcal{O}(4N_p^2)$, while they are $\mathcal{O}(4N_p)$ in the new method
presented here.
The reason for the R1000 method at the fiducial resolution $N_{\rm R1e3}=30$ to fail here is that we consider 20-100 more points per
wavelength bin. This requires a higher resolution when trying to resolve the actual opacity distribution function.

In Figure \ref{fig:ck_comb_test} one can see an example calculation from combining the opacities of water, methane and ammonia
in the wavelength range going from 6.64 to 7.34 $\mu$m. The $\kappa(g)$ distributions of the individual species are shown in the
plot. All species are contributing approximately equally strong to the total opacity
in this example and have lines in the wavelength region of interest.
Therefore this case represents something like a worst-case scenario, as our method is the most
accurate when one species dominates or the other species only contribute via a their line continua.
We plot the correct total $\kappa(g)$ distribution, obtained when adding the opacities in frequency space first,
as well as the results obtained from using the R1000 method and the result from using the new method introduced in this section.
The $g$-grid used for the R1000 method was chosen to have $g$ values following a distribution $\propto {\rm d log}\kappa / {\rm d}g$
in order to trace strong changes in the opacity distributions.
One sees that our new method is never worse in accuracy than the R1000 method which even has a little higher computational cost
($N_{\rm R1e3}=30$), and usually has an relative error which is an order of magnitude smaller.
The error of the $N_{\rm R1e3}=256$ results is an order of magnitude smaller than the $N_{\rm R1e3}=30$ result.

Finally we note that our spectral calculations using the above efficient method at $\lambda/\Delta\lambda = 10$ do not deviate by more than 5 \%
(and usually less) in wavelength regions of appreciable flux when comparing to the rebinned $\lambda/\Delta\lambda = 10^6$ line-by-line
calculations (see Section \ref{sect:ck-rt_test}, Figure \ref{fig:spec_test}). This is a deviation commonly
stated for correlated-k \citep[see, e.g.,][]{fu_liou1992,lacis_oinas1991}.
The strength of the new method reported here is to be numerically efficient, while conserving the opacity information at a high level of detail.
\section{Use of the Variable Eddington Factor method to find the temperature}
\label{sect:VEF-method-app}
\subsection{Basic equations}
For the moment equation based approach of solving for the $PT$-structure we first need the equation of radiative transport, neglecting scattering processes for now:
\beq
\mathbf{n}\cdot \nabla I_{\rm \nu}(\mathbf{x},\mathbf{n}) = -\alpha_{\nu}(\mathbf{x})\left[I_{\rm \nu}(\mathbf{x},\mathbf{n})-S_{\rm \nu}(\mathbf{x},\mathbf{n})\right].
\eeq
In our case, the source function $S_\nu$ is simply the Planck function,
\beq
S_{\nu}(\mathbf{x},\mathbf{n}) = B_\nu(T),
\eeq
where $T=T(\mathbf{x})$. \\
We now make the plane-parallel assumption, which states that any spatially varying quantity can only vary in the vertical direction $z$. We chose $z$ to increase towards the upper layers of the atmosphere. The equation of radiative transport then transforms to
\beq
\mu\frac{{\rm d}}{{\rm d}z}I_{\rm \nu}(z,\mu,\phi) = -\alpha_{\nu}(z)\left[I_{\rm \nu}(z,\mu,\phi)-S_{\rm \nu}(z,\mu,\phi)\right],
\label{equ:pl_parallel_RT}
\eeq
where $\mu={\rm cos}(\theta)$ and $\theta$ being the angle between the vertical and the direction of the ray. $\phi$ is the polar angle around the $z$-axis. Note that $S$ is independent of both $\mu$ and $\phi$ when it is equal to the Planck function. \\
The zeroth, first and second radiative moments are defined as
\beq
\left[J_\nu(\mathbf{x}),\mathbf{H}_\nu(\mathbf{x}),\hat{K}_\nu(\mathbf{x})\right] = \frac{1}{4\pi}\oint I_\nu(\mathbf{x},\mathbf{n}) \left[1,\mathbf{n},\mathbf{n}\mathbf{n}\right]{\rm d}\Omega .
\eeq
In plane-parallel geometry and rotational symmetry around the $z$-axis (i.e. no $\phi$-dependence), only the $z$-component of the first moment $\mathbf{H}$ and only the $zz$-component of second moment $\hat{K}$ are unequal to 0 and one can define
\begin{align}
H(z) & = H_z(z), \\
K(z) & = K_{zz}(z),
\end{align}
where the $\nu$ subscript has been omitted. The definition of the three plane-parallel moments then is
\beq
\left[J_\nu(z),H_\nu(z),K_\nu(z)\right] = \frac{1}{2}\int_{-1}^{1} I_\nu(z,\mu) \left[1,\mu,\mu^2\right]{\rm d}\mu .
\label{equ:imp1}
\eeq
For radiation emanating from a small solid angle $\Delta \Omega_*$ (while keeping the $z$-only spatial dependancy) one finds that
\beq
\left[J_\nu(z),H_\nu(z),K_\nu(z)\right] = \frac{\Delta\Omega_*}{4\pi}I_{*,\nu}(z,\mu_*) \left[1,\mu_*,\mu_*^2\right],
\label{equ:imp2}
\eeq
where $\mu_*={\rm cos}(\theta_*)$ and $\theta_*$ being the angle between the vertical vector and the vector pointing in direction $\Omega_*$.
$H_\nu(z)$ and $K_\nu(z)$ are, once more, the $z$- and $zz$-component of $\mathbf{H}$ and $\hat{K}$.
If the radiation emanates from a star of radius $R_*$ at distance $d$, where $d\gg R_*$, then
\beq
\Delta \Omega_*=\pi\left(\frac{R_*}{d}\right)^2.
\label{equ:omega}
\eeq
Integration of Eq. (\ref{equ:pl_parallel_RT}) over the whole solid angle yields
\beq
\frac{{\rm d}}{{\rm d}z}H_\nu(z) = -\alpha_{\nu}(z)\left[J_{\rm \nu}(z)-B_{\rm \nu}(z)\right],
\label{equ:imp3}
\eeq
where we used that the source function is supposed to be the Planck function.
Note that this equation holds independently of the fact whether there is a $\phi$-dependance in the radiation field or not as long as the definition $H(z) = H_z(z)$ is used. It can thus also be used for the radiation emanating from a small solid angle. Multiplying Eq. (\ref{equ:pl_parallel_RT}) by $\mu$ and integrating over the whole solid angle again yields
\beq
\frac{{\rm d}}{{\rm d}z}K_\nu(z) = -\alpha_{\nu}(z)H_{\rm \nu}(z),
\label{equ:imp4}
\eeq
where the isotropy of the Planck function was used. Equations (\ref{equ:imp1}), (\ref{equ:imp2}), (\ref{equ:imp3}) and (\ref{equ:imp4}) are the equations of interest for the task of finding the $PT$ structure.
\subsection{Solution of the $PT$-structure problem}
The method explained below is based on the method used for protoplanetary disks introduced in \citet{dullemond2002}.

A useful spatial coordinate for the $PT$-structure calculation is the pressure $P$, rather than the height $z$: At the top of the atmosphere we have $z\rightarrow \infty$ and the starting point of $z=0$ can be chosen arbitrarily. In contrast to the pressure, where we have a well defined value at the top of the atmosphere, namely $P=0$. Furthermore the use of the pressure instead of some arbitrary height $z$ in the atmosphere makes the density drop out of all equations of interest. To eliminate the vertical height $z$ from the equations, we use the equation of hydrostatic equilibrium
\beq
{\rm d}P = - \rho g {\rm d}z,
\label{equ:hse}
\eeq
where $\rho$ is the density and $g$ is the gravitational acceleration, which is taken to be constant throughout the atmosphere. Furthermore we will use that
\beq
\alpha_\nu(z) = \rho \kappa_\nu(z),
\label{equ:alpha_kappa}
\eeq
where $\kappa_\nu$ is the monochromatic opacity. \\
The optical depth $\tau_\nu$ then relates to the height $z$ as
\beq
{\rm d}\tau_\nu =  -\rho \kappa_\nu {\rm d}z.
\label{equ:tau}
\eeq
This yields
\beq
{\rm d}\tau_\nu = \frac{\kappa_\nu}{g}{\rm d}P.
\label{equ:tauP}
\eeq
\subsubsection{Incident stellar irradiation}
For simplicity we assume the stellar irradiation field to be a black body in this section.
In the implemented version of our code we are able to use either a blackbody or an actual stellar spectrum.
In the latter case $H_{\nu,*}$ gets replaced with the stellar spectrum appropriate for a main sequence star
at a given effective temperature.

First we start with the stellar light shining at the atmosphere of the planet.
The stellar effective temperature shall be $T_*$. If one then defines the irradiation temperature as
\beq
T_{\rm irr} = \left(\frac{R_*}{d}\right)^{1/2}T_*
\eeq
then equations (\ref{equ:imp2}) and (\ref{equ:omega}), together with $I_{\nu,*}(P=0)=B_{\nu}(T_*)$ yield for the frequency integrated first moment in z-
direction that
\beq
H_*(P=0) = -\frac{\mu_*\sigma T_{\rm irr}^4}{4\pi},
\eeq
where it was used that
\beq
\int_{0}^{\infty}B_{\nu}(T_*) {\rm d}\nu = \frac{\sigma}{\pi} T_*^4.
\eeq
The negative sign implies that the radiation enters the planet, rather than leaving it. Furthermore we make the so-called {\it two stream approximation}, assuming that the stellar irradiation is in the optical wavelengths, while the radiation field inside the planet is in the IR-wavelengths due to the lower temperature of the planetary atmosphere. Any emission processes in the atmosphere at the stellar irradiation wavelengths are thus neglected and the corresponding source term in the equations of interest are neglected, leading to
\beq
\frac{{\rm d}}{{\rm d}\tau_\nu}H_{\nu,*} = J_{\nu,*}
\label{equ:temp_find}
\eeq
and
\beq
\frac{{\rm d}}{{\rm d}\tau_\nu}K_{\nu,*} = H_{\nu,*},
\eeq
where equations (\ref{equ:imp3}), (\ref{equ:imp4}) and (\ref{equ:tau}) were used.
Using Eq. (\ref{equ:imp2}) to see that $K_{\nu,*} = \mu_*^2J_{\nu,*}$ yields
\beq
\frac{{\rm d}^2}{{\rm d}\tau_\nu^2} H_{\nu,*} = \frac{1}{\mu_*^2}H_{\nu,*}.
\eeq
For attenuation in the atmosphere one then finds that
\beq
H_*(P) = \int_{0}^{\infty}H_{\nu,*}(P=0)e^{-\tau_\nu/\mu_*}{\rm d}\nu
\label{equ:imp_s_2}
\eeq
with
\beq
\tau_\nu = \frac{1}{g}\int_0^{P}\kappa_\nu(P'){\rm d}P'.
\label{equ:imp_s_3}
\eeq
The important equations from this section are equations (\ref{equ:imp_s_2}) and (\ref{equ:imp_s_3}).

In the case of taking the dayside or global average of the stellar radiation we
assume the stellar irradiation to be isotropic.
In this case we use that
\beq
I_*(\nu,P=0) = -4H_*(\nu,P=0) \ ,
\eeq
which follows from
\begin{align}
\nonumber H_*(\nu,P=0)  & = \frac{1}{2}\int_{-1}^{1}\mu I_*(\nu,P=0) {\rm d}\mu  \\
&= \frac{1}{2}\int_{-1}^{0}\mu I_*(\nu,P=0) {\rm d}\mu \ ,
\end{align}
where we used that the stellar light only shines downward in the last line.
Further assuming that at the top of the atmosphere $I_*$ is independent of $\mu$ (isotropy)
leads to the desired result.
We then carry out a full angle and frequency dependent radiative transport calculation for the
stellar intensity $I_*$, assuming only attenuation. From this we can calculate the stellar flux
$H_*$ in every layer.

\subsubsection{Planetary radiation field}
The total net flux leaving the planet is supposed to be $\sigma T_{\rm int}^4$. As the planet receives $-\mu_* \sigma T_{\rm irr}^4$, the wavelength integrated flux coming from within the planet at IR wavelengths must be
\beq
H(P=0) = \frac{\sigma T_{\rm int}^4}{4\pi} + \frac{\mu_*\sigma T_{\rm irr}^4}{4\pi}.
\eeq
The total flux is
\beq
H_{\rm tot} = H + H_*,
\label{equ:htot1}
\eeq
i.e.
\beq
H_{\rm tot}(P=0) = \frac{\sigma T_{\rm int}^4}{4\pi}.
\label{equ:htot2}
\eeq
as required. As there are no sinks or sources of energy for the radiation field in the steady state equilibrium case we know that
\beq
\frac{{\rm d}H_{\rm tot}}{{\rm d}P} = 0.
\eeq
Together with equations (\ref{equ:imp_s_2}), (\ref{equ:htot1}) and (\ref{equ:htot2}) this yields
\beq
H(P) = \frac{\sigma T_{\rm int}^4}{4\pi} - \int_{0}^{\infty}H_{\nu,*}(P=0)e^{-\tau_\nu/\mu_*}{\rm d}\nu.
\label{equ:find_H}
\eeq
For the solution one uses the wavelength dependent opacities of the previous full radiative transfer step to calculate the attenuation of the stellar light. \\
As the next step we define the $J_\nu$-averaged {\it Eddington factor} $f$ as
\begin{align}
\nonumber f & = \frac{1}{J}\int_{0}^{\infty}f_\nu J_\nu{\rm d}\nu \\
  & = \frac{K}{J},
\end{align}
where $K$ and $J$ are the wavelength integrated moments of zeroth and first order and $f_\nu = K_\nu/J_\nu$.
Eq. (\ref{equ:imp4}) then yields, together with $f$ and equations (\ref{equ:hse}) and (\ref{equ:alpha_kappa}):
\beq
\frac{{\rm d}}{{\rm d}P}\left(fJ\right) = \frac{1}{g}\int_0^\infty \kappa_\nu H_{\rm \nu}{\rm d} \nu
\eeq
and finally
\beq
\frac{{\rm d}}{{\rm d}P}\left(fJ\right) = \frac{1}{g}\kappa_H H,
\label{equ:integ_J}
\eeq
where $\kappa_H$ is the $H_\nu$ averaged opacity.
For the solution of J one takes
\beq
J(P=0) = \frac{1}{\psi} H(P=0),
\eeq
and uses $\psi$, $f(P)$ and $\kappa_H(P)$ of the previous full RT step and the results of Eq. (\ref{equ:find_H}) to integrate Eq. (\ref{equ:integ_J}) from $P=0$ to the pressure of interest.
\subsubsection{Finding the temperature}
Once one has obtained $J(P)$ of the planetary radiation field one can use the wavelength integrated version of Eq. (\ref{equ:imp3}), noting that $H_{\rm tot}={\rm constant}$ vertically. This yields
\beq
\frac{\sigma}{\pi}T^4\kappa_P(T)-\kappa_JJ-\int_{0}^\infty \kappa_\nu J_{\nu,*}{\rm d}\nu=0,
\eeq
where $\kappa_J$ is coming from the previous full radiative transfer step.
As one finds from a similar analysis as performed for $H_{\nu,*}$ that
\beq
J_{\nu,*}(P) = J_{\nu,*}(P=0)e^{-\tau_\nu/\mu_*}
\eeq
and as Eq. (\ref{equ:temp_find}) gives that
\beq
J_{\nu,*}(P=0) = -\frac{1}{\mu_*} H_{\nu,*}(P=0)
\eeq
one finally gets
\beq
\frac{\sigma}{\pi}T^4\kappa_P(T)-\kappa_JJ+\int_{0}^\infty \kappa_\nu \frac{1}{\mu_*} H_{\nu,*}(P=0)e^{-\tau_\nu/\mu_*}{\rm d}\nu=0,
\label{equ:temp_iter_eq}
\eeq
which has to be solved for $T$ to find the temperature at pressure $P$ for the next iteration step.
In the code this was done by applying a zbrent root finding algorithm taken from numerical recipes \citep{press1992}. Furthermore, every 20th iteration step we evolve
the temperature structure by using a Ng-accelerationn \citep{ng1974} applied on $T^4$. \\
From the corr-k full radiative transfer step we thus need the opacities of the previous iteration step for calculating the attenuation of the stellar light, the $J_\nu$-averaged Eddington factor, $\psi$, $\kappa_H$ and $\kappa_J$ as well as $\kappa_P$ to find the pressure temperature structure.

\subsection{Treatment of convection}
\label{subsect:treatment_of_convection}
After the radiative structure of the atmosphere has converged we switch on convection in our code. 

During the moment solution of the temperature, the radiative temperature profile is solved from top to bottom (starting at low $P_0$, typically $P_0=10^{-14}$ bar). We check in each layer $i$ whether it should be convective or not by comparing the effective radiative temperature gradient
\beq
\nabla_{\rm rad} = \left(\frac{T_i-T_{i-1}}{P_i-P_{i-1}}\right)\cdot\left(\frac{P_i+P_{i-1}}{T_i+T_{i-1}}\right)
\eeq
with $\nabla_{\rm ad} = (\Gamma_2 -1)/\Gamma_2$, where
\beq
\Gamma_{2} = \left[1-\frac{P}{c_P\rho T}\frac{\chi_T}{\chi_\rho}\right]^{-1} \ ,
\eeq
with $P$ being the pressure, $T$ the temperature, $\rho$ the density, $c_P$ the specific heat capacity,
$\chi_T = (\partial {\rm log}P/\partial {\rm log}T)_{\rho}$ and $\chi_\rho = (\partial {\rm log}P/\partial {\rm log}\rho)_{T}$
 \citep[see, e.g.,][]{hansen2004}. All required quantities can be obtained from the equilibrium chemistry code \emph{CEA}. 
We evaluate $\Gamma_2$ as $\Gamma_2 = (\Gamma_{2,i}+\Gamma_{2,i-1})/2$ on our grid.

We employ the Schwarzschild criterion, such that if $\nabla_{\rm rad} > \nabla_{\rm ad}$, we adjust the temperature
in layer $i$ to be
\beq
T_{i} = T_{i-1}\cdot\frac{P_{i-1}+P_i(2\Gamma_2-1)}{P_{i}+P_{i-1}(2\Gamma_2-1)}  \ .
\eeq

As the energy in a convective layer is not transported by radiation anymore, the integration of $J$ via Eq. (\ref{equ:integ_J}) is not possible in this
layer. However, in order to be able to discriminate between radiative and convective energy transport in layers lying below a current convective
layer we need to compare to the radiative temperature in deeper layers. For this we need to continue to computation of $J$ down to deeper layers.

We thus chose the approach that in a convective layer $i$ during the $n$-th iteration $J^n_i=\alpha^{n-1}_iB^n(T_i)$, with $\alpha^{n-1}_i=\tilde{J}^{n-1}_i/B^{n-1}(T^{n-1}_i)$, with $\tilde{J}^{n-1}_i$ being the mean intensity taken from the full angle and frequency dependent radiative transfer step of the previous iteration. The superscripts indicate the iteration number from which the respective quantity is used. We chose this approach as in the case of very efficient convection ($\nabla_{\rm layer} \rightarrow \nabla_{\rm ad}$) the atmospheric layer should be optically thick, i.e. $J_i\rightarrow B(T_i)$. When going to the next layer $i+1$ we radiatively
integrate $J$  to this next layer using Eq. (\ref{equ:integ_J}) and compare the resulting $\nabla_{\rm rad}$ with $\nabla_{\rm ad}$
again.

As in \citet{marley1996,burrows1997} we only allow a limited number atmospheric layers to be changed to convective
energy transport every iteration. This is done to allow the atmospheric structure to adapt to the introduction of convective layers. In
\citet{marley1996,burrows1997} only 1 layer per iteration is allowed to change. We allowed for the change of 2 layers per iteration,
because sometimes a layer on the brink to being convectively unstable will switch back and forth between being radiative or convective,
preventing the overall convergence of the atmospheric structure.
\bibliography{mybib}{}

\end{document}